\documentclass[12pt,notitlepage,fleqn]{article}
\usepackage{graphicx}
\usepackage{amsmath}
\usepackage{indentfirst}
\usepackage{amsfonts}
\usepackage{amssymb}
\begin{document}

\begin{center}
{\LARGE The Meson Spectrum of the BCC Quark Model }

(A Modification of the Quark Model)

\ \ \ \ \ \ \ \ \ \ \ \ 

{\normalsize Jiao Lin Xu }

{\small The Center for Simulational Physics, }

{\small The Department of Physics and Astronomy,}

{\small University of Georgia, Athens, GA 30602, U. S. A.}

E- mail: {\small \ jxu@hal.physast.uga.edu}

{\tiny \ \ \ \ \ \ \ \ \ \ }{\small \ \ \ \ }

{\normalsize Xin Yu}

{\small Chiron Corp.}

{\small Seattle, WA}

{\small E-mail: Xin-Yu@Chiron.Com\ \ \ \ \ \ \ \ }

\ \ \ \ \ \ \ \ \ \ \ \ \ \ \ \ \ \ \ \ \ \ \ \ \ \ \ 

\textbf{Abstract}
\end{center}

{\small Using the quark spectrum of the BCC Quark Model \cite{BCC MODEL} and
the phenomenological formula for the binding energies of the mesons,\ not only
have we deduced the intrinsic quantum numbers (I, S, C, b, and Q) of all
mesons as was done\ with the Quark Model \cite{QuarkModel}, but also we
deduced the meson mass spectrum\ in agreement with experimental
results\ \cite{Particle(2000)}\ that we could not deduce using the Quark
Model. The experimental meson spectrum gives some evidence of the existence of
the new quarks q}$_{_{S}^{\ast}}${\small (1391),\ q}$_{_{S}^{\ast}}%
${\small (2551), and\ q}$_{_{C}^{\ast}}${\small (6591)}$...,$ {\small which
are predicted by the BCC Quark Model.} {\small The meson }$\chi${\small (1600)
[2}$^{+}(2^{++})]$ {\small with I = 2 (predicted by the BCC Quark
Model--T(1603)) has already been discovered \cite{Meson (I=2)}. If this is
finally confirmed, it will provide a strong support for the BCC Quark Model.
We propose a search for the mesons \ D(5996), D}$_{S}${\small (6151), B(9504),
B}$_{S}${\small (9659), B}$_{C}${\small (11031)},{\small \ }$\eta
${\small (5926)}, $\eta${\small (17837), }$\psi${\small (25596), }$\Upsilon
${\small (17805), }$\Upsilon${\small (29597), T(960), T(1282), T(1603), and
T(1924). }

\section{Introduction}

According to the Quark Model \cite{QuarkModel}, a meson is composed of a quark
and an antiquark. Although the Quark Model assumes many elementary quarks (6
flavors, 3 colors--a total of 18 elementary quarks), in fact, those quarks are
still not enough to explain the full experimental meson spectrum, as
demonstrated in lists\ (\ref{Meson-A})\ and (\ref{Meson-B}) \cite{MESON-2000} below:\ \ %

\begin{equation}%
\begin{tabular}
[c]{|l|l|l|l|l|l|l|}\hline
{\small N}$^{2s+1}${\small L}$_{J}$ & {\small J}$^{PC}$ & {\small u}%
$\overline{d}${\small ,u}$\overline{u}${\small ,d}$\overline{d}$ &
{\small \ \ \ \ \ \ \ u}$\overline{u}${\small ,d}$\overline{d},$%
{\small s}$\overline{s}$ & {\small \ \ c}$\overline{c}$ & {\small b}%
$\overline{b}$ & $\overline{s}${\small u,}$\overline{s}${\small d}\\\hline
&  & {\small \ \ I = 1} & {\small \ \ \ \ \ \ \ \ \ I = 0} & {\small I=0} &
{\small I=0} & {\small I=1/2}\\\hline
{\small 1}$^{1}${\small \ \ \ S}$_{0}$ & {\small 0}$^{-+}$ & {\small \ \ \ \ }%
$\pi${\small (139)} & $\eta${\small (549),}$\eta^{\prime}${\small (958)} &
{\small \ }$\eta_{c}${\small (2980)} &  & {\small K(494)}\\\hline
{\small 1}$^{3}${\small \ \ \ S}$_{1}$ & {\small 1}$^{--}$ & {\small \ \ \ \ }%
$\rho${\small (770)} & $\omega${\small (782),}$\phi${\small (1020)} &
{\small J/}$\psi${\small (3097)} & $\Upsilon${\small (9460)} & {\small K}%
$^{\ast}${\small (892)}\\\hline
{\small 1}$^{1}${\small \ \ \ p}$_{1}$ & {\small 1}$^{+-}$ & {\small b}$_{1}%
${\small (1235)} & {\small h}$_{1}${\small (1170), h}$_{1}${\small (1380)} &
{\small h}$_{c}${\small (1p)} &  & {\small K}$_{1B}^{\dagger}${\small (1270)}%
\\\hline
{\small 1}$^{3}${\small \ \ \ p}$_{0}$ & {\small 0}$^{++}$ & {\small a}$_{0}%
${\small (1450)}$^{\ast}$ & {\small f}$_{0}${\small (1370)}$^{\ast}%
,${\small f}$_{0}${\small (1710)}$^{\ast}$ & $\chi_{c0}${\small (1p)} &
$\chi_{b0}${\small (1p)} & {\small K}$_{0}^{\ast}${\small (1430)}\\\hline
{\small 1}$^{3}${\small \ \ \ p}$_{1}$ & {\small 1}$^{++}$ & {\small a}$_{1}%
${\small (1260)} & {\small f}$_{1}${\small (1285), f}$_{1}${\small (1420)} &
$\chi_{c1}${\small (1p)} & $\chi_{b1}${\small (1p)} & {\small K}%
$_{1A}^{\dagger}$\\\hline
{\small 1}$^{3}${\small \ \ \ p}$_{2}$ & {\small 2}$^{++}$ & {\small a}$_{2}%
${\small (1320)} & {\small f}$_{2}${\small (1270), f''}$_{2}${\small (1525)} &
$\chi_{c2}${\small (1p)} & $\chi_{b2}${\small (1p)} & {\small K}$_{2}^{\ast}%
${\small (1430)}\\\hline
{\small 1}$^{1}${\small \ \ \ D}$_{2}$ & {\small 2}$^{-+}$ & $\pi_{2}%
${\small (1670)} & $\eta_{2}${\small (1645), }$\eta_{2}${\small (1870)} &  &
& {\small K}$_{2}${\small (1770)}\\\hline
{\small 1}$^{3}${\small \ \ \ D}$_{1}$ & {\small 1}$^{--}$ & $\rho
${\small (1700)} & $\omega${\small (1650)} & $\psi${\small (3770)} &  &
{\small K}$^{\ast}${\small (1680)}$^{\dagger}$\\\hline
{\small 1}$^{3}${\small \ \ \ D}$_{2}$ & {\small 2}$^{--}$ &  &  &  &  &
{\small K}$_{2}${\small (1820)}\\\hline
{\small 1}$^{3}${\small \ \ \ D}$_{3}$ & {\small 3}$^{--}$ & $\rho_{3}%
${\small (1690)} & $\omega_{3}${\small (1670),}$\phi_{3}${\small (1850)} &  &
& {\small K}$_{3}^{\ast}${\small (1780)}\\\hline
{\small 1}$^{3}${\small \ \ \ F}$_{4}$ & {\small 4}$^{++}$ & $\alpha_{4}%
${\small (2040)} & {\small f}$_{4}${\small (2050),f}$_{4}${\small (2220)} &  &
& {\small K}$_{4}^{\ast}${\small (2045)}\\\hline
{\small 2}$^{1}${\small \ \ \ S}$_{0}$ & {\small 0}$^{-+}$ & $\pi
${\small (1300)} & $\eta${\small (1295), }$\eta${\small (1440)} & $\eta_{c}%
${\small (2S)} &  & {\small K(1460)}\\\hline
{\small 2}$^{3}${\small \ \ \ S}$_{1}$ & {\small 1}$^{--}$ & $\rho
${\small (1450)} & $\omega${\small (1420),}$\phi${\small (1680)} & $\psi
${\small (2S)} & $\Upsilon(2s)$ & {\small K}$^{\ast}${\small (1410)}\\\hline
{\small 2}$^{3}${\small \ \ \ P}$_{2}$ & {\small 2}$^{++}$ &  & $f_{2}%
${\small (1810),f}$_{2}${\small (2010)} &  & $\chi_{b2}${\small (2p)} &
{\small K}$_{2}^{\ast}${\small (1980)}\\\hline
{\small 3}$^{1}${\small \ \ \ S}$_{0}$ & {\small 0}$^{-+}$ & $\pi
${\small (1800)} & $\eta${\small (1760)} &  &  & {\small K(1830)}\\\hline
&  &  &  &  &  & \\\hline
\end{tabular}
\label{Meson-A}%
\end{equation}%

\begin{equation}%
\begin{tabular}
[c]{|l|l|l|l|l|l|l|}\hline
N$^{2s+1}$L$_{J}$ & J$^{PC}$ & c$\overline{u}$,c$\overline{d}$ &
c$\overline{s}$ & $\overline{b}u,\overline{b}d$ & $\overline{b}s$ &
$\overline{b}c$\\\hline
&  &  I=1/2 & I = 0 & I = 1/2 & I = 0 & I=0\\\hline
1$^{1}$ \ \ S$_{0}$ & 0$^{-+}$ & D(1869) & D$_{s}(1969)$ & B(5279) &
B$_{s}(5370)$ & B$_{c}(6400)$\\\hline
1$^{3}$ \ \ S$_{1}$ & 1$^{--}$ & D$^{\ast}$(2010) & D$_{s}^{\ast}$(2112) &
B$^{\ast}$(5325) & B$_{S}^{\ast}$(5850) & \\\hline
1$^{1}$ \ \ p$_{1}$ & 1$^{-+}$ & D$_{1}$(2420) & D$_{s1}$(2536) &  &  &
\\\hline
1$^{3}$ \ \ p$_{0}$ & 0$^{++}$ &  &  &  &  & \\\hline
1$^{3}$ \ \ p$_{1}$ & 1$^{++}$ &  &  &  &  & \\\hline
1$^{3}$ \ \ p$_{2}$ & 2$^{++}$ & D$_{2}^{\ast}$(2460) &  &  &  & \\\hline
\end{tabular}
\label{Meson-B}%
\end{equation}

\ \ 

A. In list (1), in order to explain the masses of the light unflavored mesons
($\eta$, $\omega$, $\phi$, h, and f) with I = 0, the Quark Model has to give
up the principle that a meson is made of a single quark and a single
antiquark, and allow a meson to be a mixture of three quark-antiquark pairs
(u$\overline{u}$, d$\overline{d}$, and s$\overline{s})$. (Note: the meson is
not a superposition of three quark-antiquark pairs (u$\overline{u}$,
d$\overline{d}$, and s$\overline{s})$ because the three pairs are independent
elementary particle pairs in the Quark Model). For example:%

\begin{equation}%
\begin{tabular}
[c]{|l|}\hline
$\eta(549)=\eta_{8}\cos\theta_{p}-\eta_{1}\sin\theta_{p},$\\\hline
$\eta^{\prime}(958)=\eta_{8}\sin\theta_{p}+\eta_{1}\cos\theta_{p},$\\\hline
$\eta_{1}=(u\overline{u}+d\overline{d}+s\overline{s})/\sqrt{3},$\\\hline
$\eta_{8}=(u\overline{u}+d\overline{d}-2s\overline{s})/\sqrt{6}.$\\\hline
\end{tabular}
\label{Mixture Moson}%
\end{equation}
This ``mixture'' violates the principle that a meson is made of a quark and an
antiquark. It also makes the model need too many parameters \cite{Marshah} by
introducing the parameter $\theta_{p}$. There are more than twenty such
mixture mesons ($\eta$(549), $\eta^{\prime}$(958), $\phi$(1020) h$_{1}%
$(1170)...$\phi_{3}$(1850), f$_{4}$(2220)) (see (\ref{Meson-A})). However, the
BCC Quark Model, as a modification of the Quark Model, needs only two
elementary quarks, u (3-colors) and d (3-colors), and deduces an excited (from
the vacuum) quark spectrum (including S, C, b, I, Q, and M) \cite{BCC MODEL}.
This quark spectrum alone is enough to structure the full meson spectrum (a
meson is made of a quark and an antiquark). The BCC Quark Model does not need
the mixture. For example:%

\begin{equation}%
\begin{tabular}
[c]{|l|}\hline
q$_{S}^{\ast}$(1111)$\overline{q_{S}^{\ast}(1111)}$=$\eta$(549) [$\bullet\eta
$(547)] (u$\overline{u}$,d$\overline{d},$s$\overline{s}$ mixture in
{\small Quark Model)},\\\hline
q$_{S}^{\ast}($1391)$\overline{q_{S}^{\ast}(1391)}$=$\eta$(952)[$\bullet
\eta^{\prime}$(958)] (u$\overline{u}$,d$\overline{d},$s$\overline{s}$ mixture
in {\small Quark Model}),\\\hline
q$_{N}^{\ast}$(931)$\overline{q_{N}^{\ast}(1471)}$ = $\eta($1021)
[$\bullet\phi($1020)] (u$\overline{u}$,d$\overline{d},$s$\overline{s}$ mixture
in {\small Quark Model)}.\\\hline
\end{tabular}
\label{Meson Example (I=0)}%
\end{equation}

B. In list (1), there are more than 15 K mesons (K(494), K$^{\ast}$(892),
K$_{1}$(1270)... K$_{4}^{\ast}$(2045)... K$_{4}$(2500)) (see (\ref{Meson-A})),
but only one strange quark (s) and two quarks (u and d) can form the K mesons
in the Quark Model. Thus, the Quark Model has to assume that the K mesons are
all $\overline{s}$u and $\overline{s}$d with different angular momenta (J) and
parities (P). These types of assumptions bring trouble. For example, the Quark
Model assumes that $\eta_{C}($2969) ($\Gamma=13.2$ Mev) and J/$\psi$(3097)
($\Gamma=87Kev)$ are c$\overline{c}$ mesons \cite{MESON-2000}. However, the
ground state $\eta_{C}($2969) has a shorter life time (13.2 Mev) than the
excited state J/$\psi$(3097) ($\Gamma=87Kev)$. This is contrary to a physical
law--the ground state has the longest lifetime. Since there is a series of the
quarks $q_{S}^{\ast}(1111),$ $q_{S}^{\ast}(1391),$ $q_{S}^{\ast}(2011)...$ in
the BCC Quark Model {\small \cite{BCC MODEL}}. Therefore, the BCC Quark Model
can present different K mesons using different quark pairs. For example:\ \ \ \ \ \ \ \ \ \ \ \ \ \ \ \ \ \ %

\begin{equation}%
\begin{tabular}
[c]{|l|}\hline
q$_{N}^{\ast}($931)$\overline{q_{S}^{\ast}(1111)}$ =$K(494)$ [$\bullet
K(494)$)]$\leftrightarrow\overline{s}u,$ $\overline{s}d$ (Quark
Model),\\\hline
q$_{N}^{\ast}($931)$\overline{q_{S}^{\ast}(1391)}$ =$K(899)$ [$\bullet
K(892)$)]$\leftrightarrow\overline{s}u,$ $\overline{s}d.$ (Quark
Model),\\\hline
q$_{N}^{\ast}($931)$\overline{q_{S}^{\ast}(2551)}$=$K(1804)$ [$\bullet
$K(1820)]$\leftrightarrow\overline{s}u,$ $\overline{s}d$ (Quark
Model).\\\hline
\end{tabular}
\label{K Mesons}%
\end{equation}

Mesons and quark pairs are essentially one-to-one correspondence in the BCC
Quark Model. Although sometimes there is a meson that is a superposition of
two quark-antiquark pairs, it is not a mixture of different elementary quark
pairs, but\textbf{\ }only a superposition of different excited states of the
same elementary quarks, u and d, in the BCC Quark Model.

C. Although the Quark Model can deduce the quantum numbers of the meson
spectrum, it cannot deduce the mass spectrum for mesons. It can only give some
mass relations (Gell-Mann--Okubo) in an Octet \cite{Gell-Mann--Okubo}.

D. The binding energy of a quark and an antiquark in a meson has not been
discussed clearly in the Quark Model. There is not a formula for the binding
energy in the Quark Model. According to the quark masses \cite{Quark
Mass-2000} of the Quark Model, m$_{u}$ =1.5 to 5 Mev and m$_{d}$ = 3 to 9 Mev,
$\pi^{+}(u\overline{d})$ and $\pi^{-}(d\overline{u})$ have the binding energy%
\begin{equation}
\text{E(}\pi^{\pm}\text{)}_{bind}\text{= M}_{\pi^{\pm}}\text{- (m}%
_{u}\text{+m}_{\overline{d}}\text{) = [139-(4.5 to 14)] Mev
$>$%
0}.\label{Pai  Binding Energy}%
\end{equation}
From E = MC$^{2}$, the mesons $\pi^{\pm}$ cannot be formed, because if the two
quarks (u and $\overline{d}$) separate into two individual quarks, the system
will have a much lower energy than the energy of the meson $\pi^{\pm}.$ This
(\ref{Pai Binding Energy}) means that the interactive force between\textbf{\ }%
the two quarks (u and $\overline{d}$) is repulsive. Similarly, the proton
(uud) has a binding energy
\begin{equation}
\text{E(p)}_{bind}\text{ = M}_{proton}\text{- (m}_{u}\text{+m}_{u}%
\text{+m}_{d}\text{) = [939-(6 to 19)] Mev
$>$%
0.}\label{E(P)bind}%
\end{equation}
This (\ref{E(P)bind}) means that the interactive forces among\textbf{\ }the
three quarks are repulsive with respect to one another. Thus, the three quarks
cannot make a stable proton. However, the proton is absolutely a stable
particle. The A-bomb and H-bomb tests have already shown that the formula E =
MC$^{2}$ is completely right. The small masses of the quarks (in the Quark
Model) not only will destroy the confinement theory \cite{CONFINEMENT}, but
also are not enough to construct the stable baryons (\ref{E(P)bind}) and the
mesons (\ref{Pai Binding Energy}).

Therefore, the Quark Model needs modification and development. The BCC Quark
Model is a good modification and development of the Quark Model. It needs only
two elementary quarks, u (3 colors) and d (3 colors), and deduces an excited
(from vacuum) quark spectrum (including S, C, b, I, Q, and M) \cite{NetXu
(Quark)}. The quark spectrum is enough to construct the full meson spectrum
using the principle that a meson is made of a single quark and a single
antiquark. The masses of the quarks (in the BCC Quark Model) are large enough
to construct the stable mesons and baryons. The binding energies of the mesons
are essentially a constant (-1723Mev).

This paper is organized as follows: The quantum numbers of the mesons are
presented in Section II. The probability that a quark and an antiquark form a
meson is discussed in Section III. The binding energies of the mesons are
introduced in Section IV. A comparison of the results of the BCC Quark Model
and experimental results is listed in Section V. The evidence of some new
quarks is given in Section VI. The predictions and experimental verifications
of the BCC Quark Model are stated in Section VII. The conclusions are in
Section VIII.

\newpage

\section{The Quantum Numbers of the Mesons}

First, we list the quark spectrum of the BCC Quark Model in
(\ref{Quark-Spectrum})\ and (\ref{Quark Quantum Number}) \cite{NetXu (Quark)} :%

\begin{equation}%
\begin{tabular}
[c]{|l|l|l|l|}\hline
d q(m) & \ \ \ \ \ \ \ \ d q(m) & d q(m) & d q(m)\\\hline
1 \textbf{q}$_{N}^{\ast}($\textbf{931)}$^{\#}$ & \textbf{\ \ \ \ \ \ \ \ 1\ q}%
$_{S}^{\ast}($\textbf{1111)}$^{\#}$ & 1 \textbf{q}$_{C}^{\ast}$\textbf{(2271)}%
$^{\#}$ & 1 \textbf{q}$_{b}^{\ast}$\textbf{(5531)}$^{\#}$\\\hline
(u$_{931}^{\ast}$, d$_{931}^{\ast}$) &
\begin{tabular}
[c]{ll}%
1{\small q}$_{S}^{\ast}(${\small 1391)} & 2{\small q}$_{\Xi}^{\ast}%
(${\small 1291)}%
\end{tabular}
& 1{\small q}$_{C}^{\ast}${\small (2441)} & 1{\small q}$_{b}^{\ast}%
${\small (9951)}\\\hline
1{\small q}$_{N}^{\ast}(${\small 1201)} &
\begin{tabular}
[c]{ll}%
1{\small q}$_{S}^{\ast}(${\small 2011)} & 3{\small q}$_{\Xi}^{\ast}%
(${\small 1471)}%
\end{tabular}
& 1{\small q}$_{C}^{\ast}${\small (2961)} & 1q$_{b}^{\ast}$(15811)\\\hline
1{\small q}$_{N}^{\ast}(${\small 1471)} &
\begin{tabular}
[c]{ll}%
1{\small q}$_{S}^{\ast}(${\small 2451)} & 2{\small q}$_{\Xi}^{\ast}%
(${\small 1651)}%
\end{tabular}
& 1{\small q}$_{C}^{\ast}${\small (6591)} & \ \ \ \ \ \ \ ...\\\hline
2{\small q}$_{N}^{\ast}(${\small 1831)} &
\begin{tabular}
[c]{ll}%
1{\small q}$_{S}^{\ast}(${\small 2551)} & 2{\small q}$_{\Xi}^{\ast}%
(${\small 1831)}%
\end{tabular}
& 1{\small q}$_{C}^{\ast}${\small (13791)} & \\\hline
4{\small q}$_{N}^{\ast}(${\small 1921)} &
\begin{tabular}
[c]{ll}%
3{\small q}$_{S}^{\ast}(${\small 2641)} & 1{\small q}$_{\Xi}^{\ast}%
(${\small 1921)}%
\end{tabular}
& \ \ \ \ \ \ \ ... & \\\hline
2{\small q}$_{N}^{\ast}(${\small 2191)} &
\begin{tabular}
[c]{ll}%
1{\small q}$_{S}^{\ast}(${\small 2731)} & 2{\small q}$_{\Xi}^{\ast}%
(${\small 2011)}%
\end{tabular}
& 1{\small q}$_{\Sigma_{C}}^{\ast}(${\small 2441)} & \\\hline
2{\small q}$_{N}^{\ast}(${\small 2551)}$^{\dagger}$ &
\begin{tabular}
[c]{ll}%
\ \ \ \ \ ...\ \ \ \ \ \  & 2{\small q}$_{\Xi}^{\ast}(${\small 2191)}%
\end{tabular}
& 1{\small q}$_{\Sigma_{C}}^{\ast}(${\small 2531)} & \\\hline
3{\small q}$_{N}^{\ast}${\small (2641)} &
\begin{tabular}
[c]{ll}%
1{\small q}$_{\Sigma}^{\ast}(${\small 1201)} & {\small 2q}$_{\Xi}^{\ast}%
(${\small 2371)}%
\end{tabular}
& 1{\small q}$_{\Sigma_{C}}^{\ast}(${\small 2961)} & \\\hline
2{\small q}$_{N}^{\ast}(${\small 2731)} &
\begin{tabular}
[c]{ll}%
3{\small q}$_{\Sigma}^{\ast}(${\small 1651)} & {\small 3q}$_{\Xi}^{\ast}%
(${\small 2551)}%
\end{tabular}
& \ \ \ \ \ \ \ ... & \\\hline
\ \ \ \ \ \ ... &
\begin{tabular}
[c]{ll}%
5{\small q}$_{\Sigma}^{\ast}(${\small 1921)} & {\small 8q}$_{\Xi}^{\ast}%
(${\small 2731)}%
\end{tabular}
&  & \\\hline
1q$_{\Delta}^{\ast}(${\small 1291)} &
\begin{tabular}
[c]{ll}%
2{\small q}$_{\Sigma}^{\ast}(${\small 2011)} & \ \ \ \ \ \ ...\ \ \ \ \ \
\end{tabular}
& 1{\small q}$_{\Xi_{C}}^{\ast}${\small (2541)} & \\\hline
2q$_{\Delta}^{\ast}(${\small 1651)} &
\begin{tabular}
[c]{ll}%
1{\small q}$_{\Sigma}^{\ast}(${\small 2371)} & 1{\small q}$_{\Omega}^{\ast}%
${\small (1651)}%
\end{tabular}
& 1{\small q}$_{\Xi_{C}}^{\ast}${\small (3161)} & \\\hline
1q$_{\Delta}^{\ast}(${\small 2011)} &
\begin{tabular}
[c]{ll}%
3{\small q}$_{\Sigma}^{\ast}(${\small 2551)} & 1{\small q}$_{\Omega}^{\ast}%
${\small (2451)}%
\end{tabular}
& \ \ \ \ \ \ \ ... & \\\hline
1q$_{\Delta}^{\ast}(${\small 2371)} &
\begin{tabular}
[c]{ll}%
2{\small q}$_{\Sigma}^{\ast}(${\small 2641)} & 1{\small q}$_{\Omega}^{\ast}%
${\small (3071)}%
\end{tabular}
&  & \\\hline
4q$_{\Delta}^{\ast}(${\small 2731)} &
\begin{tabular}
[c]{ll}%
{\small 3q}$_{\Sigma}^{\ast}(${\small 2731)} & 1{\small q}$_{\Omega}^{\ast}%
${\small (3711)}%
\end{tabular}
& 1{\small q}$_{\Omega_{C}}^{\ast}${\small (2651)} & \\\hline
{\small 3q}$_{\Delta}^{\ast}(${\small 3091)} &
\begin{tabular}
[c]{ll}%
{\small 5q}$_{\Sigma}^{\ast}(${\small 3091)} & \ \ \ \ \ \ ... \ \ \ \
\end{tabular}
& 1{\small q}$_{\Omega_{C}}^{\ast}${\small (3471)} & \\\hline
\ \ \ \ \ \ ... & \ \ \ \ \ \ ... & \ \ \ \ \ \ \ \ ... & \\\hline
\end{tabular}
\label{Quark-Spectrum}%
\end{equation}

{\footnotesize Where d is degeneracy.\ }

{\footnotesize Since q}$_{\Sigma}^{\ast}(${\footnotesize 1291) has asymmetric
n numbers \{[n}$_{1}${\footnotesize \ = (0, 0, 2), n}$_{2}${\footnotesize \ =
(-1,0,1), n}$_{3}${\footnotesize \ = (0, -1, 1)]}$\Rightarrow$%
{\footnotesize \ }$\Lambda${\footnotesize (1391) [n}$_{1}${\footnotesize \ =
(0, 0, 2)] and }$\Delta${\footnotesize (1291) [n}$_{1}${\footnotesize \ = (1,
0, 1), n}$_{2}${\footnotesize \ = (-1,0,1), n}$_{3}${\footnotesize = (0,1,1),
n}$_{4}${\footnotesize \ = (0, -1, 1)]\}(see Appendix I), we have removed it
from the above list. }

{\footnotesize \# The five ground quarks q}$_{N}^{\ast}(${\footnotesize 931)
[q}$_{{\small u}}^{\ast}${\footnotesize (931), q}$_{{\small d}}^{\ast}%
${\footnotesize (931)], q}$_{S}^{\ast}(${\footnotesize 1111), q}$_{C}^{\ast}%
${\footnotesize (2271), and q}$_{b}^{\ast}${\footnotesize (5531) are the five
quarks (u, d, s, c, and b) of the Quark Model. They have the same quantum
numbers, but different masses. The mass differences of the same kind of quarks
are essentially a constant (m}$_{q_{N}^{\ast}}${\footnotesize -m}$_{u,d}%
${\footnotesize =935, m}$_{q_{s}}${\footnotesize -m}$_{s}${\footnotesize =997,
m}$_{q_{c}}${\footnotesize -m}$_{c}${\footnotesize =1030, m}$_{q_{b}^{\ast}}%
${\footnotesize -m}$_{b}${\footnotesize =1340), which roughly is the mass of a
proton. If we use only the five ground quarks to form the mesons, we get the
results of the Quark Model.}

$\dagger${\footnotesize 2q}$_{N}^{\ast}(${\footnotesize 2551)}$^{\dagger}%
$$\equiv${\footnotesize q}$_{N}^{\ast}(${\footnotesize 2541)}$+$%
{\footnotesize q}$_{N}^{\ast}(${\footnotesize 2551)}%

\begin{equation}%
\begin{tabular}
[c]{l}%
\ \ \ \ \ \ \ \ \ \ \ \ \ \ \ \ \ \ \ \ \ \ \ \ \ The Quantum Numbers of the
Quarks\\%
\begin{tabular}
[c]{|l|l|l|l|l|l|l|l|l|l|l|l|}\hline
{\small q}$^{\ast Q}$ & {\small q}$_{N}^{\ast,\frac{2}{3}}$ & {\small q}%
$_{N}^{\ast\frac{-1}{3}}$ & {\small q}$_{\Delta}^{\ast\frac{5}{3}}$ &
{\small q}$_{\Delta}^{\ast\frac{2}{3}}$ & {\small q}$_{\Delta}^{\ast\frac
{-1}{3}}$ & {\small q}$_{\Delta}^{\ast\frac{-4}{3}}$ & {\small q}$_{S}%
^{\ast\frac{-1}{3}}$ & {\small \ q}$_{C}^{\ast\frac{2}{3}}$ & {\small \ q}%
$_{b}^{\ast\frac{-1}{3}}$ & {\small q}$_{\Xi}^{\ast\frac{-1}{3}}$ &
{\small q}$_{\Xi}^{\ast\frac{-4}{3}}$\\\hline
S & \ \ 0 & \ \ 0 & \ \ 0 & \ \ 0 & \ \ 0 & \ \ 0 & -1 & \ \ 0 & \ \ 0 & -2 &
-2\\\hline
C & \ \ 0 & \ \ 0 & \ \ 0 & \ \ 0 & \ \ 0 & \ \ 0 & \ \ 0 & \ \ 1 & \ \ 0 &
\ \ 0 & \ \ 0\\\hline
b & \ \ 0 & \ \ 0 & \ \ 0 & \ 0 & \ \ 0 & \ \ 0 & \ \ 0 & \ \ 0 & -1 & \ \ 0 &
\ \ 0\\\hline
I & 1/2 & 1/2 & 3/2 & 3/2 & 3/2 & 3/2 & \ \ 0 & \ \ 0 & \ \ 0 & 1/2 &
1/2\\\hline
I$_{Z}$ & 1/2 & -1/2 & 3/2 & 1/2 & -1/2 & -3/2 & \ \ 0 & \ \ 0 & \ \ 0 & 1/2 &
-1/2\\\hline
Q & 2/3 & -1/3 & 5/3 & 2/3 & -1/3 & -4/3 & -1/3 & 2/3 & -1/3 & -1/3 &
4/3\\\hline
\end{tabular}
\\
\ \ \ \ \ \ \ ------------------------------------------------------------------------------------\\
\ \ \ \
\begin{tabular}
[c]{|l|l|l|l|l|l|l|l|l|l|l|}\hline
{\small q}$^{\ast Q}$ & {\small q}$_{\Sigma}^{\ast\frac{2}{3}}$ &
{\small q}$_{\Sigma}^{\ast\frac{-1}{3}}$ & {\small q}$_{\Sigma}^{\ast\frac
{-4}{3}}$ & {\small q}$_{\Omega}^{\ast\frac{-4}{3}}$ & {\small q}$_{\Sigma
_{C}}^{\ast\frac{5}{3}}$ & {\small q}$_{\Sigma_{C}}^{\ast\frac{2}{3}}$ &
{\small q}$_{\Sigma_{C}}^{\ast\frac{-1}{3}}$ & {\small q}$_{\Xi_{C}}%
^{\ast\frac{2}{3}}$ & {\small q}$_{\Xi_{C}}^{\ast\frac{-1}{3}}$ &
{\small q}$_{\Omega_{C}}^{\ast\frac{-1}{3}}$\\\hline
S & \ -1 \  & \ \ -1 & \ -1 & -3 & \ \ 0 & \ \ 0 & \ \ 0 & -1 & -1 &
-2\\\hline
C & \ \ 0 & \ \ 0 & \ \ 0 & \ \ 0 & \ \ 1 & \ \ 1 & \ \ 1 & \ \ 1 & \ \ 1 &
\ \ 1\\\hline
b & \ \ 0 & \ \ 0 & \ \ 0 & \ \ 0 & \ \ 0 & \ \ 0 & \ \ 0 & \ \ 0 & \ \ 0 &
\ \ 0\\\hline
I & \ \ 1 & \ \ 1 & \ \ 1 & \ \ 0 & \ 1 & \ \ 1 & \ \ 1 & 1/2 & 1/2 &
\ \ 0\\\hline
I$_{Z}$ & \ \ 1 & \ \ 0 & -1 & \ \ 0 & \ \ 1 & \ \ 0 & -1 & 1/2 & -1/2 &
\ \ 0\\\hline
Q & 2/3 & -1/3 & -4/3 & -4/3 & 5/3 & 2/3 & -1/3 & 2/3 & -1/3 & -1/3\\\hline
\end{tabular}
\end{tabular}
\label{Quark Quantum Number}%
\end{equation}

Then, we list the formula of the quantum number sum laws of the quark pairs:%

\begin{equation}%
\begin{tabular}
[c]{l}%
$B_{q\overline{q}}=B_{q}+B_{\overline{q}}$, $S_{q\overline{q}}=S_{q}%
+S_{\overline{q}}$, $C_{q\overline{q}}=C_{q}+C_{\overline{q}}$,\\
$b_{q\overline{q}}=b_{q}+b_{\overline{q}}$, $Q_{q\overline{q}}=Q_{q}%
+Q_{\overline{q}}$, $\overline{I}_{q\overline{q}}=\overline{I}_{q}%
+\overline{I}_{\overline{q}}$.
\end{tabular}
\label{SumLaw}%
\end{equation}

Using the quark quantum numbers (\ref{Quark Quantum Number}) and the sum law
(\ref{SumLaw}), we can find the quantum numbers (S, C, b, I, Q,...) of the
quark pairs (q$_{i}^{\ast}\overline{q_{j}^{\ast}})$. We list S, C, b, I, Q of
the quark pairs (q$_{i}^{\ast}\overline{q_{j}^{\ast}})$ as follows:
\begin{equation}
\
\begin{tabular}
[c]{|l|l|l|l|l|l|l|l|l|l|}\hline
q$_{i}^{\ast}\overline{q_{j}^{\ast}}$ & {\small S} & {\small C} & {\small b} &
{\small I; Q; Meson} & q$_{i}^{\ast}\overline{q_{j}^{\ast}}$ & S & C & b &
\ {\small I; Q; Meson}\\\hline
{\small q}$_{N}^{\ast}\overline{q_{N}^{\ast}}$ & {\small 0} & {\small 0} &
{\small 0} &
\begin{tabular}
[c]{l}%
{\small 0; 0; }$\eta$\\
{\small 1; 1,0,-1; }$\pi$%
\end{tabular}
& {\small q}$_{C}^{\ast}\overline{q_{b}^{\ast}}$ & {\small 0} & {\small 1} &
{\small 1} & \ {\small 0; 1; B}$_{C}$\\\hline
{\small q}$_{S}^{\ast}\overline{q_{S}^{\ast}}$ & {\small 0} & {\small 0} &
{\small 0} & \ {\small 0; 0; }$\eta$ & {\small q}$_{C}^{\ast}\overline
{q_{\Sigma}^{\ast}}$ & {\small 1} & {\small 1} & {\small 0} & \ {\small 1; 2,
1, 0; D}$_{\Sigma}$\\\hline
{\small q}$_{C}^{\ast}\overline{q_{C}^{\ast}}$ & {\small 0} & {\small 0} &
{\small 0} & \ {\small 0; 0; \ }$\psi$ & {\small q}$_{C}^{\ast}\overline
{q_{\Xi}^{\ast}}$ & {\small 2} & {\small 1} & {\small 0} & $\ \frac{1}{2}%
${\small ; 2, 1; D}$_{\Xi}$\\\hline
{\small q}$_{b}^{\ast}\overline{q_{b}^{\ast}}$ & {\small 0} & {\small 0} &
{\small 0} & \ {\small 0; 0; \ }$\Upsilon$ & {\small q}$_{C}^{\ast}%
\overline{q_{\Omega}^{\ast}}$ & {\small 3} & {\small 1} & {\small 0} &
\ {\small 0; 2; D}$_{\Omega}$\\\hline
{\small q}$_{\Sigma}^{\ast}\overline{q_{\Sigma}^{\ast}}$ & {\small 0} &
{\small 0} & {\small 0} &
\begin{tabular}
[c]{l}%
{\small 0; 0; }$\eta$\\
{\small 1; 1,0,\ -1; }$\pi$\\
{\small 2; }$\pm${\small 2,}$\pm${\small 1,0; T}%
\end{tabular}
& q$_{\Delta}^{\ast}\overline{q_{\Delta}^{\ast}}$ & {\small 0} & {\small 0} &
{\small 0} &
\begin{tabular}
[c]{l}%
{\small 0; 0; }$\eta${\small \ }\\
{\small 1; 1, 0, -1; }$\pi$\\
{\small 2; }$\pm${\small 2, }$\pm${\small 1, 0; T}\\
{\small 3; }$\pm${\small 3,}$\pm${\small 2,}$\pm${\small 1,0; W}%
\end{tabular}
\\\hline
{\small q}$_{\Xi}^{\ast}\overline{q_{\Xi}^{\ast}}$ & {\small 0} & {\small 0} &
{\small 0} &
\begin{tabular}
[c]{l}%
{\small 0; 0; }$\eta$\\
{\small 1; 1, 0, -1; }$\pi$%
\end{tabular}
& {\small q}$_{b}^{\ast}\overline{q_{\Sigma}^{\ast}}$ & {\small 1} &
{\small 0} & {\small -1} & \ {\small 1;1,0,-1;B}$_{\Sigma}$\\\hline
{\small q}$_{\Omega}^{\ast}\overline{q_{\Omega}^{\ast}}$ & {\small 0} &
{\small 0} & {\small 0} & \ {\small 0; 0; }$\eta$ & {\small q}$_{b}^{\ast
}\overline{q_{\Xi}^{\ast}}$ & 2 & 0 & -1 & $\ \frac{1}{2}${\small ; 1,0;
B}$_{\Xi}$\\\hline
{\small q}$_{N}^{\ast}\overline{q_{S}^{\ast}}$ & {\small 1} & {\small 0} &
{\small 0} & $\ \frac{1}{2}${\small ; 1, 0; K} & {\small q}$_{b}^{\ast
}\overline{q_{\Omega}^{\ast}}$ & 3 & 0 & -1 & \ 0; 1; B$_{\Omega}$\\\hline
$\overline{q_{N}^{\ast}}q_{C}^{\ast}$ & {\small 0} & {\small 1} & {\small 0} &
$\ \frac{1}{2}${\small ; 1, 0; }D & {\small q}$_{\Sigma}^{\ast}\overline
{q_{\Omega}^{\ast}}$ & 2 & 0 & 0 & \ {\small 1;2,1,0; }$\pi_{\Xi}$\\\hline
{\small q}$_{N}^{\ast}\overline{q_{b}^{\ast}}$ & {\small 0} & {\small 0} &
{\small 1} & $\ \frac{1}{2}${\small ; 1, 0; B} & {\small q}$_{\Xi}^{\ast
}\overline{q_{\Omega}^{\ast}}$ & 1 & 0 & 0 & $\ \frac{1}{2}${\small ;1,0;
K}\\\hline
{\small q}$_{N}^{\ast}\overline{q_{\Sigma}^{\ast}}$ & {\small 1} & {\small 0}%
& {\small 0} &
\begin{tabular}
[c]{l}%
$\frac{1}{2}${\small ; 1, 0; K}\\
$\frac{3}{2}${\small ; 2, }$\pm${\small 1, 0;F}%
\end{tabular}
& {\small q}$_{\Sigma}^{\ast}\overline{q_{\Xi}^{\ast}}$ & 1 & 0 & 0 & $%
\begin{tabular}
[c]{l}%
$\frac{1}{2}${\small ; 1, 0; K}\\
$\frac{3}{2}${\small ; 2, }$\pm${\small 1, 0;F}%
\end{tabular}
$\\\hline
{\small q}$_{N}^{\ast}\overline{q_{\Xi}^{\ast}}$ & {\small 2} & {\small 0} &
{\small 0} &
\begin{tabular}
[c]{l}%
{\small 0; 1; }$\eta_{\Xi}$\\
{\small 1; 2, 1, 0; }$\pi_{\Xi}$%
\end{tabular}
& q$_{\Delta}^{\ast}\overline{q_{N}^{\ast}}$ & {\small 0} & {\small 0} &
{\small 0} & $%
\begin{tabular}
[c]{l}%
{\small 1; 1, 0, -1; }$\pi$\\
{\small 2; \ }$\pm${\small 2, }$\pm${\small 1, 0; T}%
\end{tabular}
$\\\hline
{\small q}$_{N}^{\ast}\overline{q_{\Omega}^{\ast}}$ & {\small 3} & {\small 0}%
& {\small 0} & $\ \frac{1}{2}${\small ; 2, 1; K}$_{\Omega}$ & q$_{\Delta
}^{\ast}\overline{q_{S}^{\ast}}$ & {\small 1} & {\small 0} & {\small 0} &
$\ \frac{3}{2}${\small ; 2, 1, 0, -1; F}\\\hline
$q_{C}^{\ast}\overline{q_{S}^{\ast}}$ & {\small 1} & {\small 1} & {\small 0} &
\ {\small 0; 1; }$D_{S}$ & $\overline{q_{\Delta}^{\ast}}q_{C}^{\ast}$ &
{\small 0} & {\small 1} & {\small 0} & $\ \frac{3}{2}${\small ; 2, 1 ,0, -1;
F}$_{C}$\\\hline
$q_{b}^{\ast}\overline{q_{S}^{\ast}}$ & {\small -1} & {\small 0} & {\small 1}%
& \ {\small 0; 0; }$B_{S}$ & q$_{\Delta}^{\ast}\overline{q_{b}^{\ast}}$ &
{\small 0} & {\small 0} & {\small 1} & $\ \frac{3}{2}${\small ; 2, 1, 0, -1;
F}$_{b}$\\\hline
{\small q}$_{S}^{\ast}\overline{q_{\Sigma}^{\ast}}$ & {\small 0} & {\small 0}%
& {\small 0} & \ {\small 1;1,0,-1; }$\pi$ & q$_{\Delta}^{\ast}\overline
{q_{\Sigma}^{\ast}}$ & {\small 1} & {\small 0} & {\small 0} &
\begin{tabular}
[c]{l}%
$\frac{1}{2}${\small ; 1, 0; K}\\
$\frac{3}{2}${\small ; 2, 1, 0, -1; F}\\
$\frac{5}{2}${\small ; 3,}$\pm${\small 2,}$\pm${\small 1,0; S}%
\end{tabular}
\\\hline
{\small q}$_{S}^{\ast}\overline{q_{\Xi}^{\ast}}$ & {\small 1} & {\small 0} &
{\small 0} & $\frac{1}{2}${\small ; 1, 0; K} & q$_{\Delta}^{\ast}%
\overline{q_{\Xi}^{\ast}}$ & {\small 2} & {\small 0} & {\small 0} & $%
\begin{tabular}
[c]{l}%
{\small 1; 2, 1, 0; }$\pi_{\Xi}$\\
{\small 2; 3, 2, }$\pm${\small 1, 0; T}$_{\Xi}$%
\end{tabular}
$\\\hline
{\small q}$_{S}^{\ast}\overline{q_{\Omega}^{\ast}}$ & {\small 2} & {\small 0}%
& {\small 0} & \ {\small 0; 1;}$\eta_{\Xi}$ & q$_{\Delta}^{\ast}%
\overline{q_{\Omega}^{\ast}}$ & {\small 3} & {\small 0} & {\small 0} &
$\ \frac{3}{2}${\small ; 3, 2, 1, 0; F}$_{\Omega}$\\\hline
\end{tabular}
\label{Quark Pair}%
\end{equation}

{\small Since the quarks (q}$_{\Sigma_{C}}^{\ast}${\small , q}$_{\Xi_{C}%
}^{\ast}${\small , and\ q}$_{\Omega_{C}}^{\ast}${\small ) cannot form a meson
(see next section (14) and (24)), they are omitted.}

Although we have deduced the quantum numbers of the quark pairs (q$_{i}^{\ast
}\overline{q_{j}^{\ast}}$), there is only a possibility that the quark pair
(q$_{i}^{\ast}\overline{q_{j}^{\ast}}$) may form\textbf{\ }a meson with these
quantum numbers. However, whether a quark\textbf{\ }pair (q$_{i}^{\ast
}\overline{q_{j}^{\ast}}$) really forms a meson (which can be observed by
experiment) depends\textbf{\ }on a probability of formation.

\ \ \ \ \ \ \ \ \ \ \ \ \ \ \ \ \ \ \ \ 

\section{The Probability That a Quark and an Antiquark \textbf{Form }a Meson}

Since the quarks are born from different symmetry axes and symmetry points
with different quantum numbers and masses, the probability that a quark and an
antiquark form a meson is not the same for different quark pairs. We need to
find the laws for the probability.

\subsection{The Probability of Producing a Quark From the Vacuum}

First, we shall find the probability of producing a quark, q$^{\ast}$, from
the vacuum.

A1. According to the BCC Quark Model \cite{BCC MODEL} (see Fig. 1 of Appendix
I), the q$_{\Delta}^{\ast}$ and q$_{N}^{\ast}$ quarks with S = 0 are born on
the $\Delta$-axis (8 symmetry operations) and the D-axis; the q$_{C}^{\ast}%
$\ quarks are born on the $\Delta$-axis (single energy bands from $\Delta S$ =
+ 1) at the point $\Gamma$ (see Fig. 5 (b));\ the q$_{S}^{\ast}$ quarks
$($q$_{S}^{\ast}(1391),$q$_{S}^{\ast}(4271)...)$ are born on the $\Delta$-axis
(single energy bands from $\Delta S$ = - 1) at the point H (see Fig. 5 (b)).
The\ q$_{S}^{\ast}$ and $q_{\Sigma}^{\ast}$\ quarks with S = -1 are born on
the $\Lambda$-axis (6 symmetry operations) (see Fig. 2 (b)) and the F-axis
(see Fig. 4 (a)). The q$_{\Xi}$\ quarks are born on the $\Sigma$-axis (4
symmetry operations) (see Fig. 3 (a)) and the G-axis (4 symmetry operations)
(see Fig. 4 (b)). The q$_{S}^{\ast}$\ quarks (q$_{S}^{\ast}$(1111),q$_{S}%
^{\ast}$(2551)) and the q$_{b}^{\ast}$ quarks (q$_{b}^{\ast}$(5531),
q$_{b}^{\ast}$(9951), q$_{b}^{\ast}$(15811)) are born on the $\Sigma$-axis
(single energy bands from $\Delta S$ = + 1) at the point N (see Fig. 5 (c));
and the q$_{\Omega}^{\ast}$\ quarks (q$_{\Omega}^{\ast}(1651)$, q$_{\Omega
}^{\ast}(7211)...)$ are born on the $\Sigma$-axis (single energy bands from
$\Delta S$ = - 1) at the point $\Gamma$ (see Fig. 5 (c)). Thus, we assume that
the probability (P$_{q^{\ast}}$) of producing a quark (q$^{\ast}$) is
\begin{equation}
P_{q^{\ast}}=C_{1}\exp\text{[O(q}^{\ast}\text{)] = }C_{1}\text{exp(}8+2\times
S_{G})\label{P(q)}%
\end{equation}
where S$_{G}$($\equiv$S+C+b) is the generalized strange number. O(q$^{\ast}%
$)=8+2$\times$S$_{G}$ (probability\textbf{\ }unit):
\begin{equation}%
\begin{tabular}
[c]{|l|l|l|l|l|l|}\hline
quark & q$_{C}^{\ast}$ & q$_{N}^{\ast},$q$_{\Delta}^{\ast}$ & q$_{S,}^{\ast}%
$q$_{b,}^{\ast}$q$_{\Sigma}^{\ast}$ & q$_{\Xi}^{\ast}$ & q$_{\Omega}^{\ast}%
$,\\\hline
S$_{G}$=S+C+b & +1 & 0 & -1 & -2 & -3,\\\hline
O(q$^{\ast}$)(Unit) & 10 & 8 & 6 & 4 & 2.\\\hline
\end{tabular}
\label{O(q)}%
\end{equation}
Thus, from (\ref{O(q)}), the unflavored\textbf{\ }quarks (q$_{N}^{\ast}$ and
q$_{\Delta}^{\ast})$ have O$(q^{\ast})$ = 8 units; the charmed quarks
(q$_{C}^{\ast}$) has O(q$_{C}^{\ast}$)= 10 (8+2 which is from $\Delta_{S}=$
+1) units; the strange quarks (q$_{S}^{\ast}$ and $q_{\Sigma}^{\ast}$) have
O$(q^{\ast})=$ 6 units; the q$_{\Xi}^{\ast}$ quarks have O$(q_{\Xi}^{\ast})=$
4 units; the q$_{b}^{\ast}$ quarks have O$(q_{b}^{\ast})=$ 6 (4+2 which is
from $\Delta_{S}=+1)$ units; the q$_{\Omega}^{\ast}$ quarks have O$(q_{\Omega
}^{\ast})=$ 2 (4-2 which is from $\Delta s=-1)$ units.

A2. The quarks {\small q}$_{\Sigma_{C}}^{\ast}$, {\small q}$_{\Xi_{C}}^{\ast}%
$, and\ {\small q}$_{\Omega_{C}}^{\ast}$ are born from `` second division''
\cite{NetXu(Appen B)}. In the first division, we got O({\small q}$_{\Sigma
}^{\ast}$) = 6, O({\small q}$_{\Xi}^{\ast}$) = 4, and O({\small q}$_{\Omega
}^{\ast}$) = 2 from (\ref{O(q)}). The second division will lower the
probabilities (O({\small q}$_{\Sigma}^{\ast}$), O({\small q}$_{\Xi}^{\ast}$),
and O({\small q}$_{\Omega}^{\ast}$)). For simplification, we assume that the
second division deduces half O(q$^{\ast}$) values:%

\begin{equation}
\text{O({\small q}}_{\Sigma_{C}}^{\ast}\text{) = 3, O({\small q}}_{\Xi_{C}%
}^{\ast}\text{) = 2, and O({\small q}}_{\Omega_{C}}^{\ast}\text{) = 1.}
\label{O(q) - 2 Division}%
\end{equation}

A3. The quarks which originate from the single energy bands of the $\Delta
$-axis (q$_{S}^{\ast}$(1391), q$_{C}^{\ast}$(2271), q$_{S}^{\ast}$(4271),
q$_{C}^{\ast}$(6591), q$_{S}^{\ast}$(10031)...) and the quarks which originate
from the single energy bands of the $\Sigma$-axis (q$_{S}^{\ast}$(1111),
q$_{\Omega}^{\ast}$(1651), q$_{S}^{\ast}$(2551), q$_{\Omega}^{\ast}$(3711),
q$_{b}^{\ast}$(5531), q$_{\Omega}^{\ast}$(7211), q$_{b}^{\ast}$(9951)...) have
double O(q$^{\ast}$). Using (\ref{O(q)}), we have%
\begin{equation}%
\begin{tabular}
[c]{|l|l|}\hline%
\begin{tabular}
[c]{l}%
${\small \Delta}$\\
{\small S= 0}%
\end{tabular}
&
\begin{tabular}
[c]{llll}%
$\Delta${\small S=+1} & {\small C=+1} & {\small q}$_{C}^{\ast}${\small (2271),
q}$_{C}^{\ast}${\small (6591), q}$_{C}^{\ast}${\small (13791)...} &
O{\small (q$_{C}^{\ast}$)=20}\\
$\Delta${\small S=-1} & {\small S=-1} & {\small q}$_{S}^{\ast}${\small (1391),
q}$_{S}^{\ast}${\small (4271), q}$_{S}^{\ast}${\small (10031)...} &
O{\small (q$_{S}^{\ast}$)=12}%
\end{tabular}
\\\hline%
\begin{tabular}
[c]{l}%
$\Sigma$\\
{\small S=-2}%
\end{tabular}
&
\begin{tabular}
[c]{llll}%
$\Delta${\small S=+1} & {\small S}$_{G}${\small =-1} & {\small q}$_{S}^{\ast}%
${\small (1111), q}$_{S}^{\ast}${\small (2551), q}$_{b}^{\ast}$%
{\small (5531)...} & O{\small (q$^{\ast}$)=12}\\
$\Delta${\small S=-1} & {\small S=-3} & {\small q}$_{\Omega}^{\ast}%
${\small (1651), q}$_{\Omega}^{\ast}${\small (3711), q}$_{\Omega}^{\ast}%
${\small (7211)...} & O{\small (q$_{\Omega}^{\ast}$)=4}%
\end{tabular}
\\\hline
\end{tabular}
\label{DoubleO(q)}%
\end{equation}

Thus, for q$_{S}^{\ast}$(1391), q$_{S}^{\ast}$(4271), q$_{S}^{\ast}$(10031),
q$_{S}^{\ast}$(1111), q$_{S}^{\ast}$(2551), q$_{b}^{\ast}$(5531), and
q$_{b}^{\ast}$(9951), O(q$^{\ast}$)= 2$\times$6 = 12 units$;$ for q$_{C}%
^{\ast}(2271)$, q$_{C}^{\ast}(6591)$, and q$_{C}$(13800){\small , O}%
(q$_{C}^{\ast}$) = 2$\times$10 = 20 units; for q$_{\Omega}^{\ast}$(1651),
q$_{\Omega}^{\ast}$(3711), and q$_{\Omega}^{\ast}$(7211), O(q$_{\Omega}^{\ast
}$) = 2$\times$2 = 4 units.

A4. The ground states have double O(q$^{\ast}$)\textbf{\ }values: for
q$_{S}^{\ast}(1111)$, O(q$_{S}^{\ast}$(1111)) = 2$\times$12 = 24 units; for
q$_{C}^{\ast}(2271)$, O(q$_{C}^{\ast}$(2271)) = 2$\times$20 = 40 units; for
q$_{b}^{\ast}$(5531), O(q$_{b}^{\ast}$(5531)) = 2$\times$12 = 24; for (the
general ground state) q$_{N}^{\ast}(931)$ originates from the center point
$\Gamma$ at which three $\Delta$-axes meet together (there are 48 symmetric
operations), O(q$_{N}^{\ast}$(931)) = 2$\times$(3$\times$8) = 48 units.

A5. Generally, a high energy quark will have a smaller probability of
producing the quark from the vacuum. For simplification, we assume the
following: for the high energy quark q$_{C}^{\ast}(2271)$, O(q$_{C}^{\ast
}(2271)$)=40 $\rightarrow$30; for the high energy quark q$_{b}^{\ast}(5531)$,
O(q$_{b}^{\ast}(5531)$)= 24$\rightarrow$18; and for the quarks q$_{N}^{\ast}%
$(931) and q$_{S}^{\ast}$(1111) with low energies, (q$^{\ast}$) values will
not be changed, O(q$_{N}^{\ast}$(931)) = 48, O(q$_{S}^{\ast}$(1111))=24. We do
not consider any other quarks in this paper. $\ $

\subsection{The Probability that a Pair of Quarks ($q_{i}^{\ast}%
\overline{q_{j}^{\ast}})$ is Formed}

B1. From Eq. (\ref{P(q)}), we can deduce the probability (P$_{q_{i}^{\ast
}\overline{q_{j}^{\ast}}})$ that a quark (q$_{i}^{\ast}$) and an antiquark
($\overline{q_{j}^{\ast}}$) form a quark and an antiquark pair ($q_{i}^{\ast
}\overline{q_{j}^{\ast}}$) as follows:
\begin{equation}
\text{P}_{q_{i}^{\ast}\overline{q_{j}^{\ast}}}\varpropto\text{P}_{q_{i}^{\ast
}}\times\text{P}_{\overline{q_{j}^{\ast}}}\varpropto\text{C}_{1}%
\text{e}^{O(q_{i}^{\ast})}\times\text{C}_{1}\text{e}^{O(\overline{q_{j}^{\ast
}})}\text{=C}\times\exp\text{[O(}q_{i}^{\ast}\text{)+O(}\overline{q_{j}^{\ast
}}\text{)].} \label{P(qq)}%
\end{equation}

B2. A quark and its own antiquark have double $[O(q_{i}^{\ast})+O(\overline
{q_{j}^{\ast}})]$:%

\begin{align}
\text{P}_{q_{i}^{\ast}\overline{q_{j}^{\ast}}}\text{{}}  &  =\text{C}%
\times\exp\text{\{(1+}\delta_{ij}\text{)[O(}q_{i}^{\ast}\text{)+O(}%
\overline{q_{j}^{\ast}}\text{)]\}}\label{P[O(qiqj)]}\\
&  =\text{C}\times\exp\text{[O(}q_{i}^{\ast}\overline{q_{j}^{\ast}}\text{)],}%
\end{align}
where
\begin{equation}
\text{O(}q_{i}^{\ast}\overline{q_{j}^{\ast}}\text{)=(1+}\delta_{ij}%
\text{)[O(}q_{i}^{\ast}\text{)+O(}\overline{q_{j}^{\ast}}\text{)].}
\label{O(qiqj)}%
\end{equation}
For example, for q$_{N}^{\ast}$(931)$\overline{q_{N}^{\ast}(931)},$
$O(q_{i}^{\ast}\overline{q_{j}^{\ast}})=$2$\times$[(48)+(48)]=192 units; for
q$_{S}^{\ast}$(1111)$\overline{q_{S}^{\ast}(1111)},$ $O(q_{i}^{\ast}%
\overline{q_{j}^{\ast}})=$2$\times$[(24)+(24)]= 96 units; for q$_{N}^{\ast}%
$(1831)$\overline{q_{N}^{\ast}(1831)}$, $O(q_{i}^{\ast}\overline{q_{j}^{\ast}%
})=$ 2$\times$[(8)+(8)]=32 units.

B3. A quark and an antiquark of the same kind (with the same I, S, C, b, and
Q, but different masses) have one and a half $[O(q_{i}^{\ast})+O(\overline
{q_{j}^{\ast}})]$:%

\begin{equation}
\text{O(}q_{i}^{\ast}\text{(m}_{k}\text{)}\overline{q_{j}^{\ast}(m_{l}%
}\text{))=[1+0.5}\delta_{ij}(1+\delta_{kl}\text{)][O(}q_{i}^{\ast}%
\text{(m}_{k}\text{))+O(}\overline{q_{j}^{\ast}(m_{l}}\text{)].}%
\end{equation}
For example, for q$_{N}^{\ast}$(931)$\overline{q_{N}^{\ast}(1831)},$
$O(q_{i}^{\ast}\overline{q_{j}^{\ast}})=$1.5$\times$[48+8]=84 units; for
q$_{S}^{\ast}$(1111)$\overline{q_{S}^{\ast}(1391)},$ $O(q_{i}^{\ast}%
\overline{q_{j}^{\ast}})=$1.5$\times$[24+12]= 54 units; for q$_{C}^{\ast}%
$(2271)$\overline{q_{C}^{\ast}(2441)}$, $O(q_{i}^{\ast}\overline{q_{j}^{\ast}%
})=$ 1.5$\times$[(30)+10]=60 units.

B4. A quark pair ($q_{i}^{\ast}\overline{q_{j}^{\ast}}$) in which the quark
($q_{i}^{\ast})$ has different I, S, C, and b from the antiquark ($q_{j}%
^{\ast})$ will have a lower probability (P$_{q_{i}^{\ast}\overline{q_{j}%
^{\ast}}})$:

\qquad%
\begin{equation}
\text{P}_{q_{i}^{\ast}\overline{q_{j}^{\ast}}}\text{{}= C}\exp\text{(1-O(}%
\Delta\text{I)) [O(}q_{i}^{\ast}\text{(m}_{k}\text{)+O(}\overline{q_{j}^{\ast
}(m_{l})}\text{)]=Cexp\{O[}q_{i}^{\ast}\text{(m}_{k}\text{)(}\overline
{q_{j}^{\ast}(m_{l})}\text{]\}} \label{P(qiqj)}%
\end{equation}
where%

\begin{equation}
\text{O[}q_{i}^{\ast}\text{(m}_{k}\text{)(}\overline{q_{j}^{\ast}(m_{l}%
)}\text{] = (1-O(}\Delta\text{I)) [O(}q_{i}^{\ast}\text{(m}_{k}\text{)+O(}%
\overline{q_{j}^{\ast}(m_{l})}\text{]} \label{O[Qi(mk)Qj(ml)]}%
\end{equation}
O($\Delta$I) $\equiv\Delta$G$\times\Delta$I$\times\frac{I_{q}+I_{\overline{q}%
}-1/2}{2(I_{q}+I_{\overline{q}})+1}$, $\Delta$G=$\left|  \text{S}_{i}%
\text{-S}_{j}\right|  $+$\left|  \text{C}_{i}\text{-C}_{j}\right|  $+$\left|
\text{b}_{i}\text{-b}_{j}\right|  ,$ $\Delta$I= $\left|  \text{I}_{i}%
\text{-I}_{j}\right|  $. We list O($\Delta$I) values as follows:%

\begin{equation}%
\begin{tabular}
[c]{|l|l|l|l|}\hline
Quark Pair & O($\Delta$I) &  & O($\Delta$I)\\\hline
q$_{N}^{\ast}\overline{q_{S}^{\ast}}$, $q_{C}^{\ast}\overline{q_{N}^{\ast}}$,
$q_{N}^{\ast}\overline{q_{b}^{\ast}}$, q$_{N}^{\ast}\overline{q_{\Delta}%
^{\ast}}$ & \ \ 0 & $q_{N}^{\ast}\overline{q_{\Sigma}^{\ast}}$ & $1/8$\\\hline
$\overline{q_{S}^{\ast}}q_{S}^{\ast}$, $\overline{q_{S}^{\ast}}q_{C}^{\ast}$,
$\overline{q_{S}^{\ast}}q_{b}^{\ast}$, $\overline{q_{S}^{\ast}}q_{\Sigma
,}^{\ast}$ & \ \ 0 & $\overline{q_{S}^{\ast}}q_{\Delta}^{\ast}$ &
$3/8$\\\hline
q$_{C}^{\ast}\overline{q_{C}^{\ast}}$, $q_{C}^{\ast}\overline{q_{b}^{\ast}}$,
$q_{C}^{\ast}\overline{q_{\Omega}^{\ast}}$, $q_{C}^{\ast}\overline{q_{\Xi
}^{\ast}}$ & \ \ 0 & $q_{C}^{\ast}\overline{q_{\Delta}^{\ast}}$ &
$3/8$\\\hline
q$_{b}^{\ast}\overline{q_{b}^{\ast}}$, q$_{b}^{\ast}\overline{q_{\Omega}%
^{\ast}}$, q$_{b}^{\ast}\overline{q_{\Xi}^{\ast}}$ & \ \ 0 & $\overline
{q_{b}^{\ast}}q_{\Delta}^{\ast}$ & $3/8$\\\hline
$\overline{q_{\Sigma}^{\ast}}q_{S}^{\ast}$, $\overline{q_{\Sigma}^{\ast}%
}q_{\Sigma}^{\ast}$, $q_{\Delta}^{\ast}\overline{q_{\Delta}^{\ast}}$ & \ \ 0 &
$\overline{q_{\Sigma}^{\ast}}q_{\Delta}^{\ast}$ & $1/6$\\\hline
q$_{C}^{\ast}\overline{q_{\Sigma}^{\ast}}$ & 1/3 & $\overline{q_{b}^{\ast}%
}q_{\Sigma}^{\ast}$ & $1/3$\\\hline
\end{tabular}
\label{O(Dalta(I))}%
\end{equation}

B5. If a quark (antiquark) has a very small value of O(q$^{\ast}$), it will be
very difficult for it to form a meson with an antiquark (quark). For
simplification, we assume that if a quark q$^{\ast}$ (antiquark $\overline
{q^{\ast}}$) has
\begin{equation}
O(q^{\ast})(\text{or }O(\overline{q^{\ast}}))\leq4\text{ \ }units\text{,}
\label{O<4}%
\end{equation}
it cannot form a meson with any other antiquark $\overline{q^{\ast}}$ (quark
q$^{\ast}$). Thus, q$_{\Xi}^{\ast}$, q$_{\Omega}^{\ast}$, q$_{\Sigma_{C}%
}^{\ast}$, q$_{\Xi_{C}}^{\ast}$, and q$_{\Omega_{C}}^{\ast}$ cannot form a
meson with any antiquark. This assumption is based on today's experiments. In
the future, we might be able to find mesons that are made by q$_{\Xi}^{\ast}$,
q$_{\Omega}^{\ast}$, q$_{\Sigma_{C}}^{\ast}$, q$_{\Xi_{C}}^{\ast}$, and
q$_{\Omega_{C}}^{\ast}$.

Using (\ref{O[Qi(mk)Qj(ml)]}),(\ref{O(Dalta(I))}), and (\ref{O<4}), we can
deduce O[q$_{i}^{\ast}$(m$_{k}$)$\overline{q_{j}^{\ast}(m_{l})}$] values for
all quark pairs (q$_{i}^{\ast}$(m$_{k}$)$\overline{q_{j}^{\ast}(m_{l})}$):%
\begin{equation}
\
\begin{tabular}
[c]{|l|l|l|l|l|l|}\hline
{\small q}$_{N}^{\ast}${\small (931)}$\overline{q_{N}^{\ast}(931)}$ &
{\small 192} & {\small q}$_{N}^{\ast}${\small (931)}$\overline{q_{S}^{\ast
}(1111)}$ & {\small 72} & {\small q}$_{C}^{\ast}${\small (2271)}%
$\overline{q_{b}^{\ast}(5531)}$ & {\small 48}\\\hline
{\small q}$_{N}^{\ast}${\small (931)}$\overline{q_{N}^{\ast}(m)}$ &
{\small 84} & {\small q}$_{N}^{\ast}${\small (931)}$\overline{q_{S}^{\ast
}(\operatorname{Si}\text{ngle})}$ & {\small 60} & {\small q}$_{C}^{\ast}%
${\small (2271)}$\overline{q_{b}^{\ast}(\operatorname{Si}\text{ngle})}$ &
{\small 42}\\\hline
{\small q}$_{N}^{\ast}${\small (m}$_{i}${\small )}$\overline{q_{N}^{\ast
}(m_{i})}$ & {\small 32} & {\small q}$_{N}^{\ast}${\small (931)}%
$\overline{q_{S}^{\ast}(m)}$ & {\small 54} & {\small q}$_{C}^{\ast}%
${\small (2271)}$\overline{q_{\Sigma}^{\ast}(m)}$ & {\small 24}\\\hline
{\small q}$_{N}^{\ast}${\small (m}$_{i}${\small )}$\overline{q_{N}^{\ast
}(m_{j})}$ & {\small 24} & {\small q}$_{N}^{\ast}${\small (931)}%
$\overline{q_{C}^{\ast}(2271)}$ & {\small 78} & {\small q}$_{C}^{\ast}%
${\small (2271)}$\overline{q_{\Xi}^{\ast}(m)}$ & //\\\hline
q$_{S}^{\ast}$(1111)$\overline{q_{S}^{\ast}(1111)}$ & {\small 96} &
{\small q}$_{N}^{\ast}${\small (931)}$\overline{q_{C}^{\ast}(\operatorname{Si}%
\text{ngle})}$ & {\small 68} & {\small q}$_{C}^{\ast}${\small (2271)}%
$\overline{q_{\Omega}^{\ast}(m)}$ & //\\\hline
q$_{S}^{\ast}$(1111)$\overline{q_{S}^{\ast}(\operatorname{Si}\text{ngle})}$ &
{\small 54} & {\small q}$_{N}^{\ast}${\small (931)}$\overline{q_{C}^{\ast}%
(m)}$ & {\small 58} & {\small q}$_{C}^{\ast}${\small (2271)}$\overline
{q_{\Delta}^{\ast}(m_{j})}$ & 24\\\hline
q$_{S}^{\ast}$(1111)$\overline{q_{S}^{\ast}(m)}$ & {\small 45} &
{\small q}$_{N}^{\ast}${\small (931)}$\overline{q_{b}^{\ast}(5531)}$ & 66 &
q$_{b}^{\ast}$(5531)$\overline{q_{\Sigma}^{\ast}(m)}$ & 16\\\hline
{\small q}$_{S}^{\ast}${\small (m}$_{i}${\small )}$\overline{q_{S}^{\ast
}(\operatorname{Si}\text{ngle})}$ & 27 & {\small q}$_{N}^{\ast}$%
{\small (931)}$\overline{q_{b}^{\ast}(\operatorname{Si}\text{ngle})}$ & 60 &
{\small q}$_{b}^{\ast}${\small (5531)}$\overline{q_{\Xi}^{\ast}(m)}$ &
//\\\hline
{\small q}$_{S}^{\ast}${\small (m}$_{i}${\small )}$\overline{q_{S}^{\ast
}(m_{i})}$ & {\small 24} & {\small q}$_{N}^{\ast}${\small (931)}%
$\overline{q_{\Delta}^{\ast}(m)}$ & 56 & {\small q}$_{b}^{\ast}$%
{\small (5531)}$\overline{q_{\Omega}^{\ast}(m)}$ & //\\\hline
{\small q}$_{S}^{\ast}${\small (m}$_{i}${\small )}$\overline{q_{S}^{\ast
}(m_{j})}$ & {\small 18} & {\small q}$_{N}^{\ast}${\small (931)}%
$\overline{q_{\Sigma}^{\ast}(m)}$ & 47 & {\small q}$_{b}^{\ast}$%
{\small (5531)}$\overline{q_{\Delta}^{\ast}(m)}$ & 16\\\hline
{\small q}$_{C}^{\ast}${\small (2271)}$\overline{q_{C}^{\ast}(2271)}$ &
{\small 120} & {\small q}$_{N}^{\ast}${\small (931)}$\overline{q_{\Xi}^{\ast
}(m)}$ & // & {\small q}$_{\Sigma}^{\ast}${\small (m}$_{i}${\small )}%
$\overline{q_{\Sigma}^{\ast}(m_{i})}$ & 24\\\hline
{\small q}$_{C}^{\ast}${\small (2271)}$\overline{q_{C}^{\ast}%
(\operatorname{Si}\text{ngle})}$ & 75 & {\small q}$_{N}^{\ast}${\small (931)}%
$\overline{q_{\Omega}^{\ast}(m)}$ & // & {\small q}$_{\Sigma}^{\ast}%
${\small (m}$_{i}${\small )}$\overline{q_{\Sigma}^{\ast}(m_{j})}$ & 18\\\hline
{\small q}$_{C}^{\ast}${\small (2271)}$\overline{q_{C}^{\ast}(m)}$ &
{\small 60} & q$_{S}^{\ast}$(1111)$\overline{q_{C}^{\ast}(2271)}$ & 54 &
{\small q}$_{\Sigma}^{\ast}${\small (m}$_{i}${\small )}$\overline{q_{\Xi
}^{\ast}(m_{j})}$ & //\\\hline
{\small q}$_{C}^{\ast}${\small (m}$_{i}${\small )}$\overline{q_{C}^{\ast
}(\operatorname{Si}\text{ngle})}$ & 45 & q$_{S}^{\ast}$(1111)$\overline
{q_{C}^{\ast}(\operatorname{Si}ngle)}$ & 44 & {\small q}$_{\Sigma}^{\ast}%
${\small (m}$_{i}${\small )}$\overline{q_{\Omega}^{\ast}(m_{j})}$ & //\\\hline
{\small q}$_{C}^{\ast}${\small (m}$_{i}${\small )}$\overline{q_{C}^{\ast
}(m_{i})}$ & {\small 40} & q$_{S}^{\ast}$(1111)$\overline{q_{C}^{\ast}(m)}$ &
34 & {\small q}$_{\Sigma}^{\ast}${\small (m}$_{i}${\small )}$\overline
{q_{\Delta}^{\ast}(m_{j})}$ & {\small 12}\\\hline
{\small q}$_{C}^{\ast}${\small (m}$_{i}${\small )}$\overline{q_{C}^{\ast
}(m_{j})}$ & {\small 30} & q$_{S}^{\ast}$(1111)$\overline{q_{b}^{\ast}(5531)}$%
& 42 & {\small q}$_{\Xi}^{\ast}${\small (m}$_{i}${\small )}$\overline{q_{\Xi
}^{\ast}(m_{i})}$ & //\\\hline
{\small q}$_{b}^{\ast}${\small (5531)}$\overline{q_{b}^{\ast}(5531)}$ &
{\small 72} & q$_{S}^{\ast}$(1111)$\overline{q_{b}^{\ast}(\operatorname{Si}%
ngle)}$ & {\small 36} & {\small q}$_{\Xi}^{\ast}${\small (m}$_{i}$%
{\small )}$\overline{q_{\Xi}^{\ast}(m_{j})}$ & //\\\hline
{\small q}$_{b}^{\ast}${\small (5531)}$\overline{q_{b}^{\ast}(m_{i})}$ &
{\small 45} & q$_{S}^{\ast}$(1111)$\overline{q_{\Sigma}^{\ast}(m)}$ & 30 &
{\small q}$_{\Xi}^{\ast}${\small (m}$_{i}${\small )}$\overline{q_{\Omega
}^{\ast}(m_{j})}$ & //\\\hline
{\small q}$_{b}^{\ast}${\small (m}$_{i}${\small )}$\overline{q_{b}^{\ast
}(m_{i})}$ & {\small 48} & q$_{S}^{\ast}$(1111)$\overline{q_{\Xi}^{\ast}(m)}$%
& // & {\small q}$_{\Xi}^{\ast}${\small (m}$_{i}${\small )}$\overline
{q_{\Delta}^{\ast}(m_{j})}$ & //\\\hline
{\small q}$_{b}^{\ast}${\small (m}$_{i}${\small )}$\overline{q_{b}^{\ast
}(m_{j})}$ & {\small 36} & q$_{S}^{\ast}$(1111)$\overline{q_{\Omega}^{\ast
}(m)}$ & // & {\small q}$_{\Omega}^{\ast}${\small (m}$_{i}${\small )}%
$\overline{q_{\Omega}^{\ast}(m_{j})}$ & //\\\hline
q$_{\Delta}^{\ast}${\small (m}$_{i}${\small )}$\overline{q_{\Delta}^{\ast
}(m_{i})}$ & 32 & q$_{S}^{\ast}$(1111)$\overline{q_{\Delta}^{\ast}(m)}$ & 32 &
{\small q}$_{\Omega}^{\ast}${\small (m}$_{i}${\small )}$\overline{q_{\Delta
}^{\ast}(m_{i})}$ & //\\\hline
q$_{\Delta}^{\ast}${\small (m}$_{i}${\small )}$\overline{q_{\Delta}^{\ast
}(m_{j})}$ & 24 &  &  &  & \\\hline
\end{tabular}
\label{O(QiQj)-MAX}%
\end{equation}

$\ $

\subsection{The Probability (P(I$_{m}$)) of a Meson with Isospin I$_{m}$}

The above\ probabilities (\ref{O(QiQj)-MAX})\ are the maximum probability
(P$_{Maximum})$ of each pair. However, the real probability (P(I$_{meson}$))
of a meson being made by a quark and an antiquark will depend on the isospin
(I$_{meson}$) of the meson. The probability of a meson with a larger isospin
(I) will be smaller:
\begin{align}
P(I_{Mes})  &  =C_{2}\times\exp\{O(q_{i}^{\ast}\overline{q_{j}^{\ast}%
})[1-\frac{I_{Mes}-\left|  I_{q}-I_{\overline{q}}\right|  }{I_{q}%
+I_{\overline{q}}+1}]\}\nonumber\\
&  =C_{2}\times\exp[O(I_{q_{i}^{\ast}\overline{q}_{j}})], \label{P(I)}%
\end{align}
where
\begin{equation}
O(I_{q_{i}^{\ast}\overline{q}_{j}})=O(q_{i}^{\ast}\overline{q_{j}^{\ast}%
})[1-\frac{I_{Mes}-\left|  I_{q}-I_{\overline{q}}\right|  }{I_{q}%
+I_{\overline{q}}+1}]. \label{O(I)}%
\end{equation}
We list $O(I_{q_{i}^{\ast}\overline{q}_{j}})$ values as follows:
\begin{equation}%
\begin{tabular}
[c]{|l|l|l|l|}\hline
(I$_{q}$)$\otimes$(I$_{\overline{q}}$)=(I$_{M_{1}}$)$\oplus$(I$_{M_{2}}$)... &
\ \ \ O(I$_{M_{1}}$) & \ \ \ O(I$_{M_{2}}$) & \ \ \ \ \ O(I$_{M_{3}}$)\\\hline
{\small (0)}$\otimes${\small (0)=(0)} & {\small O(0)=}$O(q_{i}^{\ast}%
\overline{q_{j}^{\ast}})$ &  & \\\hline
{\small (0)}$\otimes${\small (}$\frac{1}{2}${\small )=(}$\frac{1}{2}%
${\small )} & {\small O(}$\frac{1}{2}${\small )=}$O(q_{i}^{\ast}%
\overline{q_{j}^{\ast}})$ &  & \\\hline
{\small (0)}$\otimes${\small (1)=(1)} & {\small O(}$1${\small )=}%
$O(q_{i}^{\ast}\overline{q_{j}^{\ast}})$ &  & \\\hline
{\small (0)}$\otimes${\small (}$\frac{3}{2}${\small )=(}$\frac{3}{2}%
${\small )} & {\small O(}$\frac{3}{2}${\small )=}$O(q_{i}^{\ast}%
\overline{q_{j}^{\ast}})$ &  & \\\hline
{\small (}$\frac{1}{2}${\small )}$\otimes${\small (}$\frac{1}{2}%
${\small )=(0)}$\oplus${\small (1)} & O{\small (0)=}$O(q_{i}^{\ast}%
\overline{q_{j}^{\ast}})$ & O{\small (1)=}$\frac{1}{2}O(q_{i}^{\ast}%
\overline{q_{j}^{\ast}})$ & \\\hline
{\small (}$\frac{1}{2}${\small )}$\otimes${\small (1)=(}$\frac{1}{2}%
${\small )}$\oplus${\small (}$\frac{3}{2}${\small )} & O{\small (}$\frac{1}%
{2}${\small )=}$O(q_{i}^{\ast}\overline{q_{j}^{\ast}})$ & O{\small (}$\frac
{3}{2}${\small )=}$\frac{3}{5}O(q_{i}^{\ast}\overline{q_{j}^{\ast}})$ &
\\\hline
{\small (}$\frac{1}{2}${\small )}$\otimes${\small (}$\frac{3}{2}%
${\small )=(1)}$\oplus${\small (2)} & O{\small (1)=}$O(q_{i}^{\ast}%
\overline{q_{j}^{\ast}})$ & O{\small (2)=}$\frac{2}{3}O(q_{i}^{\ast}%
\overline{q_{j}^{\ast}})$ & \\\hline
{\small (1)}$\otimes${\small (1)=(0)}$\oplus${\small (1)}$\oplus${\small (2)}%
& O{\small (0)=}$O(q_{i}^{\ast}\overline{q_{j}^{\ast}})$ & O{\small (1)=}%
$\frac{2}{3}O(q_{i}^{\ast}\overline{q_{j}^{\ast}})$ & O{\small (2)=}$\frac
{1}{3}O(q_{i}^{\ast}\overline{q_{j}^{\ast}})$\\\hline
{\small (1)}$\otimes${\small (}$\frac{3}{2}${\small )=(}$\frac{1}{2}%
${\small )}$\oplus${\small (}$\frac{3}{2}${\small )}$\oplus${\small (}%
$\frac{5}{2}${\small )} & O{\small (}$\frac{1}{2}${\small )=}$O(q_{i}^{\ast
}\overline{q_{j}^{\ast}})$ & O{\small (}$\frac{3}{2}${\small )=}$\frac{5}%
{7}O(q_{i}^{\ast}\overline{q_{j}^{\ast}})$ & O{\small (}$\frac{5}{2}%
${\small )=}$\frac{3}{7}O(q_{i}^{\ast}\overline{q_{j}^{\ast}})$\\\hline
{\small (}$\frac{3}{2}${\small )}$\otimes${\small (}$\frac{3}{2}%
${\small )=(0)}$\oplus(${\small 1)}$\oplus${\small (2)}$\oplus${\small (3)} &
O{\small (0)=}$O(q_{i}^{\ast}\overline{q_{j}^{\ast}})$ & O{\small (1)=}%
$\frac{3}{4}O(q_{i}^{\ast}\overline{q_{j}^{\ast}})$ & O{\small (2)=}$\frac
{1}{2}O(q_{i}^{\ast}\overline{q_{j}^{\ast}})$\\\hline
&  &  &  O{\small (3)=}$\frac{1}{4}O(q_{i}^{\ast}\overline{q_{j}^{\ast}}%
)$\\\hline
\end{tabular}
\label{O(M(I))}%
\end{equation}

\subsection{The Degeneracy $D_{q^{\ast}}($ $D_{\overline{q^{\ast}}})$ of the
Quarks (antiquarks)}

Considering the degeneracy ($D_{q^{\ast}}($ $D_{\overline{q^{\ast}}}))$ of the
quarks (antiquarks) (\ref{Quark-Spectrum}), the total\emph{\ }probability of
forming a meson will be

\qquad\qquad%
\begin{align}
P_{total}  &  =D_{q^{\ast}}\times D_{\overline{q^{\ast}}}\times P(I_{Meson}%
)\nonumber\\
&  =C_{3}\times C_{2}\times e^{d_{q^{\ast}}+d_{\overline{q^{\ast}}}}\times
e^{O(I_{Mes})}\nonumber\\
&  =C\times\exp[d_{q^{\ast}}+d_{\overline{q^{\ast}}}+O(I_{Mes})]\nonumber\\
&  =C\times\exp(\Phi) \label{P-TOTAL}%
\end{align}
where $d_{q^{\ast}}(d_{\overline{q^{\ast}}})$ is the degeneracy of the
quark(antiquark) q$^{\ast}$ ($\overline{q^{\ast}}),$ and $\Phi$ is defined as
\begin{equation}
\Phi\equiv d_{q^{\ast}}+d_{\overline{q^{\ast}}}+O(I_{Mes}). \label{Fai}%
\end{equation}
The formula (\ref{P-TOTAL}) is the probability of a meson ($q_{i}^{\ast
}\overline{q_{j}^{\ast}})$ with isospin I. The constant C is unknown.
The\textbf{\ }value $\Phi$ can be deduced from Eq. (\ref{Fai}). It is very
important to note that $\Phi$ ($d_{q^{\ast}}+d_{\overline{q^{\ast}}}%
+O(I_{Mes})$) is the result of the symmetry of the BCC\ quark lattice. Thus,
the probability formula (\ref{P-TOTAL}) is based on the foundation of the
symmetry of the BCC\ quark lattice.

\section{The Phenomenological Formula for the Binding Energies of the Mesons}

Using the sum laws (\ref{SumLaw}), we have deduced the quantum numbers of all
quark pairs (\ref{Quark Pair}). However, the masses (M$_{q_{i}^{\ast}%
\overline{q_{j}^{\ast}}}$) of the mesons (M(q$_{i}^{\ast}\overline{q_{j}%
^{\ast}}$)) cannot be found using the sum laws alone because we have to
consider the binding energy:
\begin{equation}
M(q_{i}^{\ast}\overline{q_{j}^{\ast}})=m_{q_{i}^{\ast}}+m_{\overline
{q_{j}^{\ast}}}+E_{binding}\text{.} \label{Meson Mass}%
\end{equation}
There is not a theoretical formula for the binding energies in the Quark
Model. Thus, we propose a unified phenomenological formula for the meson
binding energies now.

Although we do not know the exact formula of the binding energies E$_{B}$(i,
j) of q$_{i}^{\ast}\overline{q_{j}^{\ast}}$ inside a meson, we have found some
common characteristics of the binding energies.

First, according to the BCC Quark Model \cite{NetXu (Quark)}, all quarks (with
different I, S, C, b, Q, and mass) are different energy band excited states of
the two elementary quarks, u and d. The two elementary quarks, u and d, are
the different component (I$_{Z}$) states of the same quark (q). Thus, for
different mesons, the binding energies (between the quark and the antiquark in
a meson) are essentially the same (a constant-- (- 1723 Mev))-- the first term
in E$_{B}$(i, j) (see (\ref{E(i,j)})).

Second, the strange quarks, the charmed quarks, and the bottom quarks have
smaller binding energies than the\emph{\ }unflavored\textbf{\ }ones. Thus, we
will see a term, N($\left|  \text{S}_{i\text{ }or\text{ j}}\right|  $
+1.5$\left|  \text{C}_{i\text{ }or\text{ j}}\right|  $ +3$\left|  b_{i\text{
}or\text{ j}}\right|  $)$,$ in E$_{B}$(i, j) (see (\ref{E(i,j)})). This term
makes their binding energies\ smaller.

Third, the ground state quarks (q$_{N}^{\ast}$(931), q$_{S}^{\ast}$(1111),
q$_{C}^{\ast}(2271)$, q$_{b}^{\ast}(5531)$) and their antiquarks have larger
binding energies than non-ground state quarks. The term 2$\delta_{ng}$ in
E$_{B}$(i, j) (see (\ref{E(i,j)})) makes the ground states and their
antiquarks have 200 Mev more binding energies than the non-ground state
quarks. The quarks that are born from the single energy bands of the $\Delta
$-axis from $\Delta S=+1$(q$_{C}^{\ast}$(6591), $q_{C}^{\ast}$(13791)...) (see
list (\ref{DoubleO(q)})) are the brothers of the ground quark q$_{C}^{\ast
}(2271)$) \cite{NetXu (Quark)}, and the quarks that are born from the single
energy bands of the $\Sigma$-axis from $\Delta S=+1$ ( q$_{S}^{\ast}(2551)$,
q$_{b}^{\ast}$(9951), q$_{b}^{\ast}$(15811)...) (see list (\ref{DoubleO(q)}))
are the brothers of the ground quarks q$_{S}^{\ast}$(1111) and q$_{b}^{\ast
}(5531)$) \cite{NetXu (Quark)}. They can be dealt with as the ground states
($\delta_{ng}$=0) when we calculate their binding energies. The term
``brother'' means that they are born at the same symmetry point of the single
energy bands of the same symmetry axis.

Fourth, the binding energies of the quarks and their own antiquarks are larger
than other cases (see first term ``2'' inside [...](1-$\delta_{ij}$) of
E$_{B}$(i, j)$,$ the ``2'' makes the binding energy of the pairs 200 Mev more
than nonpair cases). The quarks that are born from the single energy bands of
the $\Delta$-axis with $\Delta$S= -1 (q$_{S}^{\ast}$(1391), q$_{S}^{\ast}%
$(4271), q$_{S}^{\ast}$(10031)...) (see list (\ref{DoubleO(q)})) and the
quarks that are born from the single energy bands of the $\Sigma$-axis with
$\Delta$S= -1 (q$_{\Omega}^{\ast}$(1651), q$_{\Omega}^{\ast}$(3711),
q$_{\Omega}^{\ast}$(7211)...) (see list (\ref{DoubleO(q)})) can be dealt with
as the ground states when we calculate their paired binding energies.

Fifth, the strange quark with a larger isospin has smaller binding energies
than other cases (see the term `-1.5(SI$_{q}$-$SI_{\widetilde{q}})$' of
f(I,S,C) inside [...] of E$_{B}$(i, j) (see (\ref{E(i,j)})).

Sixth, the quark pairs (q$_{i}^{\ast}\overline{q_{j}^{\ast}}$) in which the
quark (q$_{i}^{\ast}$) has different I, S, and C from the antiquark
($\overline{q_{j}^{\ast}}$) will have different (smaller) binding energies
(see the term \ $\Delta$I -$\delta$S -$\Delta$C\ in f(I,S,C) inside [...] of
E$_{B}$(i, j)) (see (\ref{E(i,j)}) and (\ref{F(I,S,C)})).

Finally, the quarks with larger masses have slightly larger binding energies
than the smaller quarks of the same kind [see term (1-$\widetilde{m}),$
$\widetilde{m}=m_{q}m_{\overline{q}}/m_{g}m_{\overline{g}}]$, especially, the
quark pairs (q$_{i}^{\ast}\overline{q_{i}^{\ast}}$) in which the quarks
(q$_{i}^{\ast}$) and the antiquarks ($\overline{q_{i}^{\ast}}$) are born on
the single energy binds of the $\Delta$-axis and the $\Sigma$-axis (see list
(\ref{DoubleO(q)})). The term (-$\delta_{SP}N_{i})$ makes the binding energy
larger. The term 25(G$_{q}$-$G_{\overline{q}}$-SI$_{q}$+$SI_{\overline{q}}$)
is a small adjustment.

Based on these characteristics, we propose the following phenomenological
binding energy formula: the binding energy E$_{B}$(i, j) of a quark
(q$_{i}^{\ast}$) and an antiquark ($\overline{q_{j}^{\ast}}$) in the meson
M(q$_{i}^{\ast}\overline{q_{j}^{\ast}}$) is\ \ \ \ \ %

\begin{equation}
\text{E}_{B}\text{(i, j)}=\text{-1723+}\text{100\{N+[2-}\left|  S\right|
\text{+}\delta_{ng}\text{f(I,S,C)](1-}\delta_{ij}\text{)+2}\delta
_{ng}\text{+(1-}\widetilde{m}\text{)-}\delta_{SP}\text{N}_{i}\text{\}+25A.}
\label{E(i,j)}%
\end{equation}
Where
\begin{equation}
N=\left|  S_{i\text{ }or\text{ j}}\right|  +1.5\left|  C_{i\text{ }or\text{
j}}\right|  +3\left|  b_{i\text{ }or\text{ j}}\right|  , \label{N Meaning}%
\end{equation}
S is the strange number, C is the charmed number, and b is the bottom number.
If q$_{i}^{\ast}$ and $\overline{q_{j}^{\ast}}$ have the same$\left|
S\right|  $, $\left|  C\right|  $, or$\left|  b\right|  $, N$_{i\text{
}or\text{ j}}$ = N$_{i}.$ For example, for q$_{S}^{\ast}$(1111)$\overline
{q_{S}^{\ast}(1111)}$, N$_{i\text{ }or\text{ j}}$= N$_{i}$ = $\left|
S\right|  =$1; for q$_{C}$(2271)$\overline{q_{C}^{\ast}(2271)}$, N$_{i\text{
}or\text{ j}}$ = 1.5$\left|  C\right|  $ = 1.5; for q$_{b}$(5540)$\overline
{q_{b}^{\ast}(5531)}$, N$_{i\text{ }or\text{ j}}$= 3$\left|  b_{i\text{
}or\text{ j}}\right|  $ = 3; for q$_{S}^{\ast}$(1111)$\overline{q_{S}^{\ast
}(2551)}$, N$_{i\text{ }or\text{ j}}$=$\left|  S\right|  $= 1; for
q$_{N}^{\ast}$(931)$\overline{q_{S}^{\ast}(1111)}$, N$_{i\text{ }or\text{ j}}%
$=1; for q$_{N}^{\ast}$(931)$\overline{q_{C}^{\ast}(2271)}$, N$_{i\text{
}or\text{ j}}$= 1.5$\left|  \overline{C}\right|  =$ 1.5; for q$_{N}^{\ast}%
$(931)$\overline{q_{b}^{\ast}(5531)}$, N$_{i\text{ }or\text{ j}}$= 3$\left|
\overline{b}\right|  =$3.
\begin{equation}
f(I,S,C)=-1.5(SI-\overline{SI})+\Delta I-\delta S-2.5\left|  C\right|  ,
\label{F(I,S,C)}%
\end{equation}
SI$_{q}$ is strange number S times($\times$) isospin I of q$_{i}^{\ast}$;
$\overline{SI}$ is strange number S times($\times$) isospin I of
$\overline{q_{j}^{\ast}}$. $\Delta$I=$\left|  \text{I}_{i}\text{-I}%
_{\overline{j}}\right|  ,$ $\delta$S=$\left|  \text{S}_{i}\text{-S}%
_{j}\right|  $, $\Delta$C=$\left|  \text{C}_{i}\text{-C}_{j}\right|  $.
$\delta_{ng}$ = 1 if either q or $\overline{q}$ is not a ground state (or a
quark that is born from the single band of the $\Delta$-axis and the $\Sigma
$-axis), otherwise $\delta_{ng}$ = 0. In (\ref{E(i,j)}),%
\begin{equation}
\widetilde{m}=m_{q_{i}^{\ast}}m_{\overline{q_{j}^{\ast}}}/m_{g_{i}%
}m_{\overline{g_{j}}}\text{,} \label{m value}%
\end{equation}
here $m_{g_{i}}$ ($m_{\overline{g_{j}}}$) is the mass of the ground state
(anti-ground state).$\ $There are only five ground quarks -- q$_{u}^{\ast}%
$(931) [O(q$_{u}^{\ast}$)=48], q$_{d}^{\ast}$(931) [O(q$_{d}^{\ast}$)=48],
q$_{S}^{\ast}$(1111) [O(q$_{S}^{\ast}$)=24], q$_{C}^{\ast}$(2271)
[O(q$_{C}^{\ast}$)=30], and q$_{b}^{\ast}$(5531) [O(q$_{b}^{\ast}$)=18]. For
the single energy bands of the $\Delta$-axis and the $\Sigma$-axis,
$\widetilde{m}$ = m$_{q}$(J$_{n}$)$\times$m$_{\overline{q^{\prime}}}%
$(J$_{n^{\prime}}$)/m$_{q}$(J$_{n-1}$)$\times$m$_{\overline{q^{\prime}}}%
$(J$_{n^{\prime}-1}$),\textbf{\ }J$_{n}$\ is an order number of the single
energy band with $\Delta S\neq0$ at the same symmetry point \cite{NetXu
(Quark)}. $\delta_{SP}$=1 if i = j and both quarks (q$_{i}^{\ast}$ and
$\overline{q_{i}^{\ast}})$ are born from the single binds (\ref{DoubleO(q)});
$\delta_{SP}$=0 otherwise.
\begin{equation}
A=G_{q}-G_{\overline{q}}-SI_{q}+SI_{\overline{q}}, \label{A Values}%
\end{equation}
where G = S+1.5C+3b.

The above formula looks very complex, but in fact, it is very simple and easy
to use. Usually, we simplify it into seven cases as follows:\qquad\qquad

I. For ground quark pairs, $\mathbf{\delta}_{ij}$=1, $\mathbf{\delta}_{ng}$=0,
$\delta_{SP}$=0, $\widetilde{m}$=1, G$_{q}$=-G$_{\overline{q}}$, SI$_{q}%
$=-SI$_{\overline{q}}$, from (\ref{E(i,j)})
\begin{equation}
\text{E}_{B}\text{(i, i) = -1723+100N}_{i}\text{+50(G}_{q}\text{-SI}%
_{q}\text{).} \label{Ground Pair}%
\end{equation}

II. For the quark pairs (q$_{i}^{\ast}\overline{q_{i}^{\ast}}$) born from the
single energy bands (\ref{DoubleO(q)}), $\mathbf{\delta}_{ij}$=1,
$\mathbf{\delta}_{ng}$= 0, G$_{q}$= -G$_{\overline{q}}$, SI$_{q}$=
-SI$_{\overline{q}}$, $\delta_{SP}$= 1, from (\ref{E(i,j)}),
\begin{equation}
\text{E}_{B}\text{(i, i) = -1623+100(-}\widetilde{m}\text{)+50(G}%
_{q}\text{-SI}_{q}\text{).} \label{Single Pair}%
\end{equation}
For the single energy bands $\widetilde{m}$ =m$_{q}(J_{n})\times
m_{\overline{q^{\prime}}}(J_{n^{\prime}})/$m$_{q}(J_{n-1})\times
m_{\overline{q^{\prime}}}(J_{n^{\prime}-1}),$\textbf{\ }$J_{n}$\ is an order
number of the single energy band with $\Delta S\neq0$ at the same symmetry
point \cite{NetXu (Quark)}.

III.\ For non-ground pairs\textbf{, }$\mathbf{\delta}_{ij}$ = 1,
$\mathbf{\delta}_{ng}$ = 1, $\delta_{SP}$ = 0,\ G$_{q}$= -G$_{\overline{q}}$,
SI$_{q}$= -SI$_{\overline{q}}$, from (\ref{E(i,j)}),\ \ \ \ \ \ \ \ \ \
\begin{equation}
\text{E}_{B}\text{(i, i) = -1423+100(N}_{i}\text{-}\widetilde{m}%
\text{)+50(G}_{q}\text{-SI}_{q}\text{).} \label{NoGround Pair}%
\end{equation}

IV. For quarks and antiquarks that are both ground quarks (q$_{i}^{\ast
}\overline{q_{j}^{\ast}},$ i $\neq$ j)$,\delta_{i}${}$_{j}$ = 0, $\delta_{ng}%
$=0, $\widetilde{m}$ = 1, $\delta_{SP}$ = 0, from (\ref{E(i,j)}),
\begin{equation}
\text{E}_{B}\text{(i, j) = -1523+100[1.5}\left|  C\right|  \text{+3}\left|
b\right|  \text{]+25(G}_{q}\text{-G}_{\overline{q}}\text{-SI}_{q}%
\text{+SI}_{\overline{q}}\text{).} \label{No-Pair Ground}%
\end{equation}

V. For the quarks (not pairs) that are born from the single energy bands with
$\Delta$S= +1 of the $\Delta$-axis and the $\Sigma$-axis, $\delta_{i}${}$_{j}$
= 0, $\delta_{ng}$= 0, $\widetilde{m}$ = m$_{q}$(J$_{n}$)$\times$%
m$_{\overline{q^{\prime}}}$(J$_{n^{\prime}}$)$/$m$_{q}$(J$_{n-1}$)$\times
$m$_{\overline{q^{\prime}}}$(J$_{n^{\prime}-1}$), $\delta_{SP}$ = 0, SI$_{q}%
$-SI$_{\overline{q}}$ = 0. From (\ref{E(i,j)}),
\begin{equation}
\text{E}_{B}\text{(i, j) = -1423+100[1.5}\left|  C\right|  \text{+3}\left|
b\right|  \text{-}\widetilde{m}\text{]+25(G}_{q}\text{-G}_{\overline{q}%
}\text{-SI}_{q}\text{+SI}_{\overline{q}}\text{).} \label{SINGLE No Pair}%
\end{equation}
\qquad

VI. For\ quarks and antiquarks are not both ground quarks q$_{i}^{\ast
}\overline{q_{j}^{\ast}}$ ( i $\neq$ j), $\mathbf{\delta}_{ij}$ = 0,
$\mathbf{\delta}_{ng}$ =1, $\delta_{SP}$ = 0. From (\ref{E(i,j)}),%

\begin{equation}
\text{E}_{B}\text{(i, j) = -1223+100[3}\left|  b\right|  \text{-1.5(SI-}%
\overline{SI}\text{)+}\Delta\text{I-}\delta\text{S-}\Delta\text{C-}%
\widetilde{m}\text{]+25(G}_{q}\text{-G}_{\overline{q}}\text{-SI}_{q}%
\text{+SI}_{\overline{q}}\text{).}%
\end{equation}

VII. For the quark (q$_{i}^{\ast}(m_{k})$) and the antiquark ($\overline
{q_{j}^{\ast}(m_{l})}$) that have the same quantum numbers (I, S, C, b, and
Q)$,$ but different masses, $\delta_{ng}$= 1, $\delta_{SP}$=0. Since they have
the same quantum numbers, the quantum numbers of E$_{B}$(i,j) should be the
same as those of the pair quarks (N). Because the masses of q$_{i}^{\ast}$ and
$\overline{q_{j}^{\ast}}$ are different, E$_{B}$(i,j) will take the
non-quantum number portion (`2') from non-pair term [2-$\left|  S\right|
$+$\delta_{ng}$f(S,I)] (1-$\delta_{ij}$). Thus, from (\ref{E(i,j)})%

\begin{equation}
\text{E}_{B}\text{(i, j)\ = -1223+100(N}_{i}\text{-}\widetilde{m}%
\text{)+25(G}_{q}\text{-G}_{\overline{q}}\text{-SI}_{q}\text{+SI}%
_{\overline{q}}\text{).}%
\end{equation}

For example, we apply the above formulas to find the following mesons
({\small O(}q$_{i}^{\ast}\overline{q_{j}^{\ast}})$ is from the list
(\ref{O(QiQj)-MAX}) and (\ref{O(M(I))}) ):

I. For ground quark pairs,\textbf{ }from (\ref{Ground Pair}), we have

\qquad E$_{B}$(i, i) = -1723+100N$_{i}$+50[($G_{q}$-GI$_{q}$].

For q$_{N}^{\ast}$(931)$\overline{q_{N}^{\ast}({\small 931})}$, \ \qquad E=
-1723+100(0)+0 = -1723;

for q$_{S}^{\ast}$(1111)$\overline{q_{S}^{\ast}({\small 1111})}$,
\ \ \ \ \ \ E= -1723+100(1)+50[-1-0] =-1673;

for q$_{C}^{\ast}$(2271)$\overline{q_{C}^{\ast}({\small 2271})}$, \ \ \ \ \ E=
-1723+100(1.5)+50(+1.5-0)\ = -1498;

for q$_{b}^{\ast}$(5531)$\overline{q_{b}^{\ast}({\small 5531})}$, \ \ \ \ \ E=
-1723+100(3.0)+50(-3) =-1573.%

\begin{equation}%
\begin{tabular}
[c]{|l|l|l|l|l|}\hline
q$_{i}^{\ast}${\small (m)} & q$_{N}^{\ast}$(931) & q$_{S}^{\ast}$(1111) &
q$_{C}^{\ast}$(2271) & q$_{b}^{\ast}$(5531)\\\hline
$\overline{q_{j}^{\ast}(m)}$ & $\overline{q_{N}^{\ast}({\small 931})}$ &
$\overline{q_{S}^{\ast}({\small 1111})}$ & $\overline{q_{C}^{\ast
}({\small 2271})}$ & $\overline{q_{b}^{\ast}({\small 5531})}$\\\hline
{\small E}$_{B}$ & {\small -1723} & {\small -1673} & {\small -1498} &
{\small -1573}\\\hline
{\small Theory} & $\pi$(139) & $\eta$(549) & J/$\Psi$(3044) & $\Upsilon
$(9489)\\\hline
{\small O(}q$_{i}^{\ast}\overline{q_{j}^{\ast}})$ & {\small 192} & {\small 96}%
& {\small 120} & {\small 72}\\\hline
{\small Exper.} & $\pi$(139) & $\eta$(547) & J/$\Psi$(3097) & $\Upsilon
$(9460)\\\hline
\end{tabular}
\end{equation}
.

II. The quark pairs born from the single energy bands, from (\ref{Single Pair}),

\qquad%
\begin{equation}
\text{E}_{B}\text{(i, i) = -1623 + 100(-}\widetilde{m}\text{) + 50(G}%
_{q}\text{-SI}_{q}\text{),}%
\end{equation}
for the single energy bands $\widetilde{m}$ =m$_{q}$(J$_{n}$)$\times
$m$_{\overline{q^{\prime}}}$(J$_{n^{\prime}}$)$/$m$_{q}$(J$_{n-1}$)$\times
$m$_{\overline{q^{\prime}}}$(J$_{n^{\prime}-1}$)$,$\textbf{\ }$J_{n}$\ is an
order number of the single energy band with $\Delta S\neq0$ at the same
symmetry point \cite{NetXu (Section V)}.

1). The quark pairs on the single energy bands of the $\Delta$ axis ($\Delta
S=+1)$,

\qquad E = -1623+100(- $\widetilde{m})+50(1.5)=-1548-100$ $\widetilde{m},$

\qquad for q$_{C}^{\ast}$(6591)$\overline{q_{C}(6591)}$, $\qquad\widetilde{m}%
$= (6591$\times$6591)/(2271$\times$2271) = 8.42,

\qquad for q$_{C}$(13791)$\overline{q_{C}(13791)}$, $\ \ \ \widetilde{m}$=
(13791$\times$13791)$/$(6591$\times$6591) =4.38.%

\begin{equation}%
\begin{tabular}
[c]{|l|l|l|l|}\hline
q$_{i}^{\ast}${\small (m)}$\overline{q_{j}^{\ast}(m)}$ & q$_{C}^{\ast}%
$(2271)$\overline{q_{C}^{\ast}({\small 2271})}$ & q$_{C}^{\ast}$%
{\small (6591)}$\overline{q_{C}^{\ast}({\small 6591})}$ & q$_{C}%
$({\small 13791})$\overline{q_{C}^{\ast}({\small 13791})}$\\\hline
E & {\small -1498} & {\small -1548-842=-2390} & {\small --1548-438=
-1986}\\\hline
Theory & $J/\Psi(3044)$ & $\psi(10792)$ & $\psi(25596)$\\\hline
O(q$_{i}^{\ast}\overline{q_{j}^{\ast}})$ & 120 & 80 & 80\\\hline
Exper. & $J/\Psi(3097)$ & $\Upsilon(10355)$ & ?\\\hline
\end{tabular}
\end{equation}

2). The quark pairs on the single energy bands of the $\Sigma$ axis ($\Delta S=+1),$

\qquad\ for \ {\small q}$_{S}${\small (2560)}$\overline{q_{S}^{\ast}(2551)}%
$\qquad\ \ E = -1623+50(-1)-100 $\widetilde{m}$ = -1673 -527,

\qquad\qquad\qquad\qquad\qquad\qquad\qquad\qquad$(\widetilde{m}$
{\small =(2551}$\times${\small 2551)/(1111}$\times${\small 1111) = 5.22);}

\qquad\ for \ q$_{b}^{\ast}$(9951)$\overline{q_{b}^{\ast}(9951)}$\qquad\ \ E =
-1623+50(-3)-100 $\widetilde{m}$ = -1773 -324,

\qquad\qquad\qquad\qquad\qquad\qquad\qquad\qquad${\small (}\widetilde{m}$
{\small = (9951}${\small \times}${\small 9951)/(5531}${\small \times}%
${\small 5531) = 3.24);}\qquad\qquad

\qquad\ for \ q$_{b}^{\ast}$(15811)$\overline{q_{b}^{\ast}(15811)}$
\ \ \ \ \ E = -1623+50(-3)-100 $\widetilde{m}$ = -1773 -252,

\qquad\qquad\qquad\qquad\qquad\qquad\qquad\qquad$(\widetilde{m}${\small =
\ (15811}${\small \times}${\small 15811)}/{\small (9951}$\times${\small 9951)
= 2.52).}\qquad%

\begin{equation}%
\begin{tabular}
[c]{|l|l|l|l|l|l|}\hline
q$_{i}^{\ast}${\small (m)} & q$_{S}^{\ast}$(1111) & q$_{S}^{\ast}%
$({\small 2551}) & q$_{b}^{\ast}$(5531) & q$_{b}^{\ast}$(9951) & q$_{b}^{\ast
}$(15811)\\\hline
$\overline{q_{j}^{\ast}(m)}$ & $\overline{q_{S}^{\ast}({\small 1111})}$ &
$\overline{q_{S}^{\ast}({\small 2551})}$ & $\overline{q_{b}^{\ast
}({\small 5531})}$ & $\overline{q_{b}^{\ast}({\small 9951})}$ & $\overline
{q_{b}^{\ast}({\small 15811})}$\\\hline
E & {\small -1673} & -2200 & {\small -1573} & -2097 & -2025\\\hline
Theory & $\eta$(549) & $\eta$(2902) & $\Upsilon$(9489) & $\Upsilon$(17805) &
$\Upsilon$(29597)\\\hline
O(q$_{i}^{\ast}\overline{q_{j}^{\ast}})$ & 96 & 48 & 72 & 48 & 48\\\hline
Exper. & $\eta$(547) & $\eta_{C}$(2980) & $\Upsilon$(9460) & ? & ?\\\hline
\end{tabular}
\end{equation}

3). The quark pairs on the single energy bands of the $\Delta$ axis ($\Delta S=-1),$

\qquad E$_{B}$(i, i) = -1623 + 100(-$\widetilde{m}$) + 50(G$_{q}$-SI$_{q}$) =
- 1673-100$\widetilde{m}.$

\ \ \qquad For q$_{S}^{\ast}$(1391)$\overline{q_{S}^{\ast}{\small (1391)}}%
${\small , \ \ \ \ }$\widetilde{m}$=(1391$\times$1391)/(1111$\times$1111) = 1.567,

\ \ \qquad for q$_{S}^{\ast}$(4271)$\overline{q_{S}^{\ast}{\small (4271)}}%
${\small , \ \ \ \ }$\widetilde{m}$=(4271$\times$4271)/(1391$\times$1391)= 9.43,

\qquad\ \ for q$_{S}^{\ast}$(10031)$\overline{q_{S}^{\ast}{\small (10031)}}%
${\small , }$\widetilde{m}$=(10031$\times$10031)/(4271$\times$4271) = 5.52.
\begin{equation}%
\begin{tabular}
[c]{|l|l|l|l|l|}\hline
q$_{i}^{\ast}${\small (m)} & q$_{S}^{\ast}$(1111) & q$_{S}^{\ast}$(1391) &
q$_{S}^{\ast}$(4271) & q$_{S}^{\ast}$(10031)\\\hline
$\overline{q_{j}^{\ast}(m)}$ & $\overline{q_{S}^{\ast}({\small 1111})}$ &
$\overline{q_{S}^{\ast}{\small (1391)}}$ & $\overline{q_{S}^{\ast
}{\small (4271)}}$ & $\overline{q_{S}^{\ast}{\small (10031)}}$\\\hline
{\small E} & {\small -1691} & {\small -1830} & {\small -2616} & {\small -2225}%
\\\hline
{\small Theory} & $\eta${\small (549)} & $\eta${\small (952)} & $\eta
${\small (5926)} & $\eta${\small (17837)}\\\hline
{\small O(}q$_{i}^{\ast}\overline{q_{j}^{\ast}}${\small )} & {\small 96} &
{\small 48} & {\small 48} & {\small 48}\\\hline
{\small Exper.} & $\eta${\small (547)} & $\eta^{\prime}${\small (958)} & ? &
?\\\hline
\end{tabular}
\end{equation}

\ \ \ \ \ \ \ \ \ \ \ \ \ \ \ \ \ \ \ \ \ \ \ \ \ \ \ \ 

\ IV. For quarks and antiquarks that are both the ground states (i $\neq j)$,
from (\ref{No-Pair Ground})

\qquad E$_{B}$(i, j)= -1523+100[1.5$\left|  C\right|  $+3$\left|  b\right|
$]+25(G$_{q}$-$G_{\overline{q}}$-SI$_{q}$+$SI_{\overline{q}}$) \ \ \ \ \ \ \ \ \ \ \ \ \ \ \ \ \ \ \ \ \ \ \ \ \ \ \ \ \ \ \ 

\ 1). For \ q(m)$\overline{q_{N}^{\ast}({\small 931})}$. $\overline{SI}$ = 0,
G$_{\overline{q}}$ = 0, \qquad\qquad\qquad\ \ 

\qquad for q$_{S}^{\ast}$(1111)$\overline{q_{N}^{\ast}({\small 931})}$,
\qquad\ E = -1523+0+25(-1) = -1548;

\qquad for q$_{C}^{\ast}$(2271)$\overline{q_{N}^{\ast}({\small 931})}$,
\ \ \ \ \ \ \ \ E = -1523+150+25(1.5) = -1336;

\qquad for q$_{b}^{\ast}$(5531)$\overline{q_{N}^{\ast}({\small 931})}$,
\qquad\ E = -1523+300+25(-3) = -1298.%

\begin{equation}%
\begin{tabular}
[c]{|l|l|l|l|}\hline
q$_{i}^{\ast}${\small (m)} & q$_{S}^{\ast}$(1111) & q$_{C}^{\ast}$(2271) &
q$_{b}^{\ast}$(5531)\\\hline
$\overline{q_{j}^{\ast}(m)}$ & $\overline{q_{N}^{\ast}({\small 931})}$ &
$\overline{q_{N}^{\ast}({\small 931})}$ & $\overline{q_{N}^{\ast}%
({\small 931})}$\\\hline
E$_{bind}$ & -1548 & -1336 & -1298\\\hline
Theory & K(494) & D(1866) & B(5164)\\\hline
O(q$_{i}^{\ast}\overline{q_{j}^{\ast}})$ & 72 & 78 & 66\\\hline
Exper. & K(494) & D(1869) & B(5279)\\\hline
\end{tabular}
\end{equation}

\qquad\qquad\qquad

\ \ 2. For \ q({\small m})$\overline{q_{S}^{\ast}({\small 1111})}$.
$\overline{SI}=0,$ $G_{\overline{q}}=1,$ $SI_{\overline{q}}=0,$

\qquad for q$_{C}^{\ast}$(2271)$\overline{q}_{S}$(1111), \ E =
-1523+150+25(1.5-1) = -1360.5;

\qquad for q$_{b}^{\ast}$(5531)$\overline{q}_{S}$(1111), \ E =
-1523+300+25(-3-1) = -1323.\ 

\qquad%

\begin{equation}%
\begin{tabular}
[c]{|l|l|l|}\hline
q$_{i}^{\ast}${\small (m)} & q$_{C}^{\ast}$(2271) & q$_{b}^{\ast}%
$(5531)\\\hline
$\overline{q_{j}^{\ast}(m)}$ & $\overline{q_{S}^{\ast}({\small 1111})}$ &
$\overline{q_{S}^{\ast}({\small 1111})}$\\\hline
E$_{bind}$ & -1361 & -1323\\\hline
Theory & {\small D}$_{S}(${\small 2022)} & {\small B}$_{S}(${\small 5319)}%
\\\hline
O(q$_{i}^{\ast}\overline{q_{j}^{\ast}})$ & 54 & 42\\\hline
Exper. & {\small D}$_{S}^{\pm}(${\small 1969)} & {\small B}$_{S}%
(5369)$\\\hline
\end{tabular}
\end{equation}
\ \ \ \ \ \ \ \ \ \ \ \ \ \ \ \ \ \ 

3. For q({\small m})$\overline{q_{C}^{\ast}(2271)},$

\qquad E$_{B}$(i,j) = -1523+100[1.5$\left|  C\right|  $+3$\left|  b\right|
$]+25(G$_{q}$-$G_{\overline{q}}$-SI$_{q}$+$SI_{\overline{q}}$) = -1111.
\begin{equation}%
\begin{tabular}
[c]{|l|l|}\hline
q$_{i}^{\ast}${\small (m)}$\overline{q_{j}^{\ast}(m)}$ & q$_{b}^{\ast}%
$(5531)$\overline{q_{C}^{\ast}(2271)}$\\\hline
E$_{bind}$ & -1111\\\hline
Theory & {\small B}$_{C}${\small (6691)(I=0, Q=-1)}\\\hline
O(q$_{i}^{\ast}\overline{q_{j}^{\ast}})$ & 38\\\hline
Exper. & B$_{C}(6400)$\\\hline
\end{tabular}
\end{equation}

For a full list of the meson mass spectrum, please see Appendix II.

From the quark spectrum (\ref{Quark-Spectrum})\ and (\ref{Quark Quantum
Number}), using sum laws (\ref{SumLaw}), we have found the intrinsic quantum
numbers (I, S, C, b, and Q) of the mesons (\ref{Quark Pair}); with the
phenomenological binding energy formula (\ref{E(i,j)}), we have deduced the
masses of the mesons; in terms of the probability formulas (\ref{O(QiQj)-MAX})
and (\ref{P-TOTAL}), we have also got the probabilities that a quark and an
antiquark form a meson. Therefore, we deduce the meson spectrum (I, S, C, b,
Q, Mass, and the possibility P$_{total}$ (\ref{P-TOTAL})). The theoretical
meson spectrum is listed in Appendix II.

\newpage

\section{Comparing Results}

We compare the theoretical results to the experimental results
\cite{Particle(2000)} using the seven tables found below. In the comparison,
we will use the following conventions:\ \ 

(1). We do not take into account the angular momenta of the experimental
results. We assume that the small differences of the masses in the same group
of the mesons originate from their different angular momenta. If we ignore
this effect, their masses should be essentially the same. The mesons of the
same kind (the same I, S, C, b, and Q) with roughly the same masses but
different angular momentums and parities form a group. We use the average of
the masses to represent the mass of the group mesons.

(2). We use the meson name to represent the intrinsic quantum numbers
(\ref{Quark Pair}).

(3). If there are many possible mesons that correspond to a group of
experimental observed mesons, we use the meson with the maximum discovered
probability to represent the group.\ \ \ \ \ \ \ \ \ \ \ \ \ \ \ \ 

(4). The mesons with O(I$_{Meson}$) (see (\ref{O(I)}))
\begin{equation}
O(I_{Meson})<24 \label{O(I)<24}%
\end{equation}
are omitted since they can not be discovered now (because of technical limitations).

(5). For the quark masses of the Quark Model \cite{Quark Mass-2000}, we take
the average values:%
\begin{align}
\text{m}_{u}\text{{}}  &  =\text{ 3Mev (1.5 to 5 Mev), }\nonumber\\
\text{m}_{d}\text{{}}  &  =\text{ 6Mev (3 to 9 Mev), }\nonumber\\
\text{m}_{s}\text{{}}  &  =\text{ 123Mev (75 to 170 Mev), }\nonumber\\
\text{m}_{c}\text{{}}  &  =\text{ 1250Mev (1150 to 1350 Mev), }\nonumber\\
\text{m}_{b}\text{{}}  &  =\text{ 4200Mev (4000 to 4400 Mev).}
\label{QUARK MASS of  THE QUARK MODEL}%
\end{align}

\newpage

\qquad

\ \ \ \ \ \ Table I. \ Light Unflavored Mesons ({\small S = C = b = 0,}
{\small I = 0, Q =} {\small 0\ )}

$%
\begin{tabular}
[c]{|l|l|l|}\hline
The\ BCC Quark Model & Exper. & \ \ \ \ \ The\ Quark Model\\\hline
\ q$_{i}^{\ast}$(m$_{k}$)$\overline{q_{j}^{\ast}(m_{l})}=\eta(m)$ &
\ M{\tiny eson}(m) & \ \ \ \ \ \ \ \ \ q$_{i}^{\ast}$(m$_{k}$)$\overline
{q_{j}^{\ast}(m_{l})}$\\\hline
{\small q}$_{S}^{\ast}${\small (1111)}$\overline{{\small q}_{S}^{\ast
}{\small (1111)}}${\small =}$\eta${\small (549)} & $\bullet\eta${\small (547)}%
& {\small u(3)}$\overline{{\small u(3)}}${\small ,d(6)}$\overline
{{\small d(6)}}${\small ,s(123)}$\overline{{\small s(123)}}$\\\hline
{\small q}$_{N}^{\ast}${\small (931)}$\overline{{\small q}_{N}^{\ast
}{\small (1201)}}${\small =}$\eta${\small (780)} & $\bullet\omega
${\small (782)} & {\small u(3)}$\overline{{\small u(3)}}${\small ,d(6)}%
$\overline{{\small d(6)}}${\small ,s(123)}$\overline{{\small s(123)}}$\\\hline
{\small q}$_{N}^{\ast}${\small (1210)}$\overline{{\small q}_{N}^{\ast
}{\small (1201)}}${\small =}$\eta${\small (813)} & {\small f}$_{0}%
${\small ({\tiny 400-1200})} & {\small u(3)}$\overline{{\small u(3)}}%
${\small ,d(6)}$\overline{{\small d(6)}}${\small ,s(123)}$\overline
{{\small s(123)}}$\\\hline
{\small q}$_{S}^{\ast}${\small (1391)}$\overline{{\small q}_{S}^{\ast
}{\small (1391)}}${\small =}$\eta${\small (952)} & $\bullet\eta^{\prime}%
${\small (958)} & {\small u(3)}$\overline{{\small u(3)}}${\small ,d(6)}%
$\overline{{\small d(6)}}${\small ,s(123)}$\overline{{\small s(123)}}$\\\hline
{\small q}$_{\Delta}^{\ast}${\small (1291)}$\overline{{\small q}_{\Delta
}^{\ast}{\small (1291)}}${\small =}$\eta${\small (967)} & $\bullet$%
{\small f}$_{0}${\small (980)} & {\small u(3)}$\overline{{\small u(3)}}%
${\small ,d(6)}$\overline{{\small d(6)}}${\small ,s(123)}$\overline
{{\small s(123)}}$\\\hline
{\small q}$_{N}^{\ast}${\small (931)}$\overline{{\small q}_{N}^{\ast
}{\small (1471)}}${\small =}$\eta${\small (1021)} & $\bullet\phi
${\small (1020)} & {\small u(3)}$\overline{{\small u(3)}}${\small ,d(6)}%
$\overline{{\small d(6)}}${\small ,s(123)}$\overline{{\small s(123)}}$\\\hline
{\small q}$_{S}^{\ast}${\small (1111)}$\overline{{\small q}_{S}^{\ast
}{\small (1391)}}${\small =}$\eta${\small (1204)} & $\bullet${\small h}$_{1}%
${\small (1170)} & {\small u(3)}$\overline{{\small u(3)}}${\small ,d(6)}%
$\overline{{\small d(6)}}${\small ,s(123)}$\overline{{\small s(123)}}$\\\hline
{\small q}$_{N}^{\ast}${\small (1471)}$\overline{{\small q}_{N}^{\ast
}{\small (1471)}}${\small =}$\eta${\small (1269)} & $\bullet\overline{\eta}%
${\small (1283)} & {\small u(3)}$\overline{{\small u(3)}}${\small ,d(6)}%
$\overline{{\small d(6)}}${\small ,s(123)}$\overline{{\small s(123)}}$\\\hline
{\small q}$_{N}^{\ast}${\small (931)}$\overline{{\small q}_{N}^{\ast
}{\small (1831)}}${\small =}$\eta${\small (1342)} & $\bullet\overline{\eta}%
${\small (1375)} & {\small u(3)}$\overline{{\small u(3)}}${\small ,d(6)}%
$\overline{{\small d(6)}}${\small ,s(123)}$\overline{{\small s(123)}}$\\\hline
{\small q}$_{N}^{\ast}${\small (931)}$\overline{{\small q}_{N}^{\ast
}{\small (1921)}}${\small =}$\eta${\small (1423)} & $\bullet\overline{\eta}%
${\small (1428)} & {\small u(3)}$\overline{{\small u(3)}}${\small ,d(6)}%
$\overline{{\small d(6)}}${\small ,s(123)}$\overline{{\small s(123)}}$\\\hline
q$_{\Delta}^{\ast}${\small (1651)}$\overline{{\small q}_{\Delta}^{\ast
}{\small (1651)}}${\small =}$\eta${\small (1565)} & $\bullet\overline{\eta}%
${\small (1525)} & {\small u(3)}$\overline{{\small u(3)}}${\small ,d(6)}%
$\overline{{\small d(6)}}${\small ,s(123)}$\overline{{\small s(123)}}$\\\hline
{\small q}$_{N}^{\ast}${\small (931)}$\overline{{\small q}_{N}^{\ast
}{\small (2191)}}${\small =}$\eta${\small (1664)} & $\bullet\overline{\eta}%
${\small (1656)} & {\small u(3)}$\overline{{\small u(3)}}${\small ,d(6)}%
$\overline{{\small d(6)}}${\small ,s(123)}$\overline{{\small s(123)}}$\\\hline
{\small q}$_{S}^{\ast}${\small (1111)}$\overline{{\small q}_{S}^{\ast
}{\small (2011)}}${\small =}$\eta${\small (1768)} & $\bullet\overline{\eta}%
${\small (1760)} & {\small u(3)}$\overline{{\small u(3)}}${\small ,d(6)}%
$\overline{{\small d(6)}}${\small ,s(123)}$\overline{{\small s(123)}}$\\\hline
{\small q}$_{N}^{\ast}${\small (1831)}$\overline{{\small q}_{N}^{\ast
}{\small (1831)}}${\small =}$\eta${\small (1852)} & $\bullet\overline{\eta}%
${\small (1860)} & {\small u(3)}$\overline{{\small u(3)}}${\small ,d(6)}%
$\overline{{\small d(6)}}${\small ,s(123)}$\overline{{\small s(123)}}$\\\hline
{\small q}$_{N}^{\ast}${\small (931)}$\overline{{\small q}_{N}^{\ast
}{\small (2551)}}${\small =}$\eta${\small (1985)} & $\bullet\overline{\eta}%
${\small (1930)} & {\small u(3)}$\overline{{\small u(3)}}${\small ,d(6)}%
$\overline{{\small d(6)}}${\small ,s(123)}$\overline{{\small s(123)}}$\\\hline
{\small q}$_{N}^{\ast}${\small (931)}$\overline{{\small q}_{N}^{\ast
}{\small (2641)}}${\small =}$\eta${\small (2065)} & $\bullet\overline{\eta}%
${\small (2030)} & {\small u(3)}$\overline{{\small u(3)}}${\small ,d(6)}%
$\overline{{\small d(6)}}${\small ,s(123)}$\overline{{\small s(123)}}$\\\hline
{\small q}$_{N}^{\ast}${\small (931)}$\overline{{\small q}_{N}^{\ast
}{\small (2731)}}${\small =}$\eta${\small (2146)} & $\ \overline{\eta}%
${\small (2199)} & {\small u(3)}$\overline{{\small u(3)}}${\small ,d(6)}%
$\overline{{\small d(6)}}${\small ,s(123)}$\overline{{\small s(123)}}$\\\hline
{\small q}$_{S}^{\ast}${\small (1111)}$\overline{{\small q}_{S}^{\ast
}{\small (2641)}}${\small =}$\eta${\small (2341)} & $\bullet\overline{\eta}%
${\small (2320)} & {\small u(3)}$\overline{{\small u(3)}}${\small ,d(6)}%
$\overline{{\small d(6)}}${\small ,s(123)}$\overline{{\small s(123)}}$\\\hline
{\small q}$_{S}^{\ast}${\small (1111)}$\overline{{\small q}_{S}^{\ast
}{\small (2731)}}${\small =}$\eta${\small (2423)} & f$_{6}${\small (2510)} &
{\small u(3)}$\overline{{\small u(3)}}${\small ,d(6)}$\overline{{\small d(6)}%
}${\small ,s(123)}$\overline{{\small s(123)}}$\\\hline
{\small q}$_{S}^{\ast}${\small (2551)}$\overline{{\small q}_{S}^{\ast
}{\small (2551)}}${\small =}$\eta${\small (2902)} & $\eta_{C}${\small (2980)}%
& {\small u(3)}$\overline{{\small u(3)}}${\small ,d(6)}$\overline
{{\small d(6)}}${\small ,s(123)}$\overline{{\small s(123)}}$\\\hline
{\small q}$_{\Delta}^{\ast}${\small (2731)}$\overline{{\small q}_{\Delta
}^{\ast}{\small (2731)}}${\small =}$\eta${\small (3178)} & $\ \chi
${\small (3250)} & {\small u(3)}$\overline{{\small u(3)}}${\small ,d(6)}%
$\overline{{\small d(6)}}${\small ,s(123)}$\overline{{\small s(123)}}$\\\hline
{\small q}$_{S}^{\ast}${\small (4271)}$\overline{{\small q}_{S}^{\ast
}{\small (4271)}}${\small =}$\eta${\small (5926)} & \ \ \ \ \ ? &
\ \ \ \ \ \ \ \ \ \ \ \ \ \ \ \ \ \ ?\\\hline
{\small q}$_{S}^{\ast}${\small (10031)}$\overline{{\small q}_{S}^{\ast
}{\small (10031)}}${\small =}$\eta${\small (17837)} & \ \ \ \ \ ? &
\ \ \ \ \ \ \ \ \ \ \ \ \ \ \ \ \ \ ?\\\hline
\ \ \ \ \ \ \ \ \ \ \ \ \ \ \ \ \ \ ... & \ \ \ \ \ ... &
\ \ \ \ \ \ \ \ \ \ \ \ \ \ \ \ \ ...\\\hline
\end{tabular}
$

\ \ \ \ \ \ \ \ \ \ \ \ \ \ \ \ \ \ \ \ \ \ \ \ \ \ \ \newpage

\qquad\qquad

\ \ Table II. Light Unflavored Mesons ({\small S = C = b = 0,} {\small I = 1,
Q = }$\pm${\small 1, 0)}

\ $%
\begin{tabular}
[c]{|l|l|l|}\hline
The BCC Quark Model & Exper. & Quark Model\\\hline
\ q$_{i}^{\ast}$(m$_{k}$)$\overline{q_{j}^{\ast}(m_{l})}=\pi(m)$ &
{\small Meson(m)} & \ \ \ \ \ \ \ \ \ q$_{i}^{\ast}$(m$_{k}$)$\overline
{q_{j}^{\ast}(m_{l})}$\\\hline
{\small q}$_{N}^{\ast}${\small (931)}$\overline{q_{N}^{\ast}{\small (931)}}%
${\small =}$\pi${\small (139)} & $\bullet\pi${\small (139)} & {\small u(3)}%
$\overline{d{\small (6)}}${\small ,u(3)}$\overline{u{\small (3)}}%
${\small ,d(6)}$\overline{d{\small (6)}}$\\\hline
{\small q}$_{N}^{\ast}${\small (931)}$\overline{q_{N}^{\ast}{\small (1201)}}%
${\small =}$\pi${\small (780)} & $\bullet\rho${\small (770)} & {\small u(3)}%
$\overline{d{\small (6)}}${\small ,u(3)}$\overline{u{\small (3)}}%
${\small ,d(6)}$\overline{d{\small (6)}}$\\\hline
{\small q}$_{N}^{\ast}${\small (931)}$\overline{q_{\Delta}^{\ast
}{\small (1291)}}${\small =}$\pi${\small (960)} & $\bullet${\small a}$_{0}%
${\small (980)} & {\small u(3)}$\overline{d{\small (6)}}${\small ,u(3)}%
$\overline{u{\small (3)}}${\small ,d(6)}$\overline{d{\small (6)}}$\\\hline
{\small q}$_{N}^{\ast}${\small (931)}$\overline{q_{\Delta}^{\ast
}{\small (1651)}}${\small =}$\pi${\small (1282)} & $\bullet\overline{\pi}%
${\small (1248)} & {\small u(3)}$\overline{d{\small (6)}}${\small ,u(3)}%
$\overline{u{\small (3)}}${\small ,d(6)}$\overline{d{\small (6)}}$\\\hline
{\small q}$_{N}^{\ast}${\small (931)}$\overline{q_{N}^{\ast}{\small (1831)}}%
${\small =}$\pi${\small (1342)} & $\bullet\overline{\pi}${\small (1340)} &
{\small u(3)}$\overline{d{\small (6)}}${\small ,u(3)}$\overline{u{\small (3)}%
}${\small ,d(6)}$\overline{d{\small (6)}}$\\\hline
{\small q}$_{N}^{\ast}${\small (931)}$\overline{q_{N}^{\ast}{\small (1921)}}%
${\small =}$\pi${\small (1423)} & $\bullet\overline{\pi}${\small (1450)} &
{\small u(3)}$\overline{d{\small (6)}}${\small ,u(3)}$\overline{u{\small (3)}%
}${\small ,d(6)}$\overline{d{\small (6)}}$\\\hline
{\small q}$_{N}^{\ast}${\small (931)}$\overline{q_{\Delta}^{\ast
}{\small (2011)}}${\small =}$\pi${\small (1603)} & $\ \pi_{1}${\small (1600)}%
& {\small u(3)}$\overline{d{\small (6)}}${\small ,u(3)}$\overline
{u{\small (3)}}${\small ,d(6)}$\overline{d{\small (6)}}$\\\hline
{\small q}$_{N}^{\ast}${\small (931)}$\overline{q_{\Delta}^{\ast
}{\small (2011)}}${\small =}$T${\small (1603)} & $\chi${\small (1600)}%
$_{I=2}^{\#}$ & \ \ \ \ \ \ \ \ \ \ \ \ \ \ \ ?\\\hline
{\small q}$_{N}^{\ast}${\small (931)}$\overline{q_{N}^{\ast}{\small (2191)}}%
${\small =}$\pi${\small (1664)} & $\bullet\overline{\pi}${\small (1672)} &
{\small u(3)}$\overline{d{\small (6)}}${\small ,u(3)}$\overline{u{\small (3)}%
}${\small ,d(6)}$\overline{d{\small (6)}}$\\\hline
{\small q}$_{S}^{\ast}${\small (1111)}$\overline{{\small q}_{\Sigma}^{\ast
}{\small (1921)}}${\small =}$\pi${\small (1861)} & $\bullet\overline{\pi}%
${\small (1775)} & {\small u(3)}$\overline{d{\small (6)}}${\small ,u(3)}%
$\overline{u{\small (3)}}${\small ,d(6)}$\overline{d{\small (6)}}$\\\hline
{\small q}$_{N}^{\ast}${\small (931)}$\overline{{\small q}_{\Delta}^{\ast
}{\small (2371)}}${\small =}$\pi${\small (1924)} & $\ \chi${\small (2000)} &
{\small u(3)}$\overline{d{\small (6)}}${\small ,u(3)}$\overline{u{\small (3)}%
}${\small ,d(6)}$\overline{d{\small (6)}}$\\\hline
{\small q}$_{N}^{\ast}${\small (931)}$\overline{q_{N}^{\ast}{\small (2641)}}%
${\small =}$\pi${\small (2065)} & $\bullet a_{4}${\small (2040)} &
{\small u(3)}$\overline{d{\small (6)}}${\small ,u(3)}$\overline{u{\small (3)}%
}${\small ,d(6)}$\overline{d{\small (6)}}$\\\hline
{\small q}$_{N}^{\ast}${\small (931)}$\overline{q_{N}^{\ast}{\small (2731)}}%
${\small =}$\pi${\small (2146)} & $\ \overline{\pi}${\small (2125)} &
{\small u(3)}$\overline{d{\small (6)}}${\small ,u(3)}$\overline{u{\small (3)}%
}${\small ,d(6)}$\overline{d{\small (6)}}$\\\hline
{\small q}$_{N}^{\ast}${\small (931)}$\overline{q_{\Delta}^{\ast
}{\small (2731)}}${\small =}$\pi${\small (2246)} & $\ {\small \rho}%
_{3}{\small (2250)}$ & {\small u(3)}$\overline{d{\small (6)}}${\small ,u(3)}%
$\overline{u{\small (3)}}${\small ,d(6)}$\overline{d{\small (6)}}$\\\hline
{\small q}$_{S}^{\ast}${\small (1111)}$\overline{q_{\Sigma}^{\ast
}{\small (2551)}}${\small =}$\pi${\small (2434)} & $\ \overline{\pi}%
${\small (2400)} & {\small u(3)}$\overline{d{\small (6)}}${\small ,u(3)}%
$\overline{u{\small (3)}}${\small ,d(6)}$\overline{d{\small (6)}}$\\\hline
{\small q}$_{N}^{\ast}${\small (931)}$\overline{q_{\Delta}^{\ast
}{\small (3091)}}${\small =}$\pi${\small (2567)} & \ \ \ \ \ \ ? &
{\small u(3)}$\overline{d{\small (6)}}${\small ,u(3)}$\overline{u{\small (3)}%
}${\small ,d(6)}$\overline{d{\small (6)}}$\\\hline
{\small q}$_{S}^{\ast}${\small (1111)}$\overline{q_{\Sigma}^{\ast
}{\small (2731)}}${\small =}$\pi${\small (2598)} & \ \ \ \ \ \ ? &
{\small u(3)}$\overline{d{\small (6)}}${\small ,u(3)}$\overline{u{\small (3)}%
}${\small ,d(6)}$\overline{d{\small (6)}}$\\\hline
\ \ \ \ \ \ \ \ \ \ \ \ \ \ \ \ ... & \ \ \ \ \ \ ... &
\ \ \ \ \ \ \ \ \ \ \ \ \ \ \ \ ...\\\hline
\end{tabular}
$

\ \ \ \ \#$\chi${\small (1600)}$_{I=2}$ is not established

\bigskip\newpage

\ \ \ \ \ \ \ \ \ \ \ \ \ \ \ \ \ \ \ \ \ \ \ \ \ \ \ \ \ \ \ \ \ \ \ \ \ \ \ \ \ \ \ 

\ \ \ \ \ Table III. Strange Mesons ({\small S = }$\pm1${\small , C = b = 0, I
= 1/2, Q = 1, 0})%

\begin{tabular}
[c]{|l|l|l|}\hline
\ \ \ The BCC Quark Model & Experiment & The Quark Model\\\hline
q$_{i}^{\ast}$(m$_{k}$)$\overline{q_{j}^{\ast}(m_{l})}$ = K(m) & \ K(m) &
\ \ \ \ \ q$_{i}^{\ast}$(m$_{k}$)$\overline{q_{j}^{\ast}(m_{l})}$\\\hline
q$_{N}^{\ast}$(931)$\overline{q_{S}^{\ast}(1111)}$=K(494) & $\bullet$K(494) &
$\overline{\text{{\small s(123)}}}${\small u(3), }$\overline
{\text{{\small s(123)}}}${\small d(6)}\\\hline
q$_{N}^{\ast}$(931)$\overline{q_{S}^{\ast}(1391)}$=K(899) & $\bullet$K(892) &
$\overline{\text{{\small s(123)}}}${\small u(3), }$\overline
{\text{{\small s(123)}}}${\small d(6)}\\\hline
q$_{N}^{\ast}$(931)$\overline{q_{\Sigma}^{\ast}(1651)}$=K(1310) & $\bullet
$K(1270) & $\overline{\text{{\small s(123)}}}${\small u(3), }$\overline
{\text{{\small s(123)}}}${\small d(6)}\\\hline
q$_{N}^{\ast}$(931)$\overline{q_{S}^{\ast}(2011)}$=K(1463) & $\bullet
\overline{\text{K}}$(1426) & $\overline{\text{{\small s(123)}}}${\small u(3),
}$\overline{\text{{\small s(123)}}}${\small d(6)}\\\hline
q$_{N}^{\ast}$(931)$\overline{q_{\Sigma}^{\ast}(1921)}$=K(1556) & \ K(1580) &
$\overline{\text{{\small s(123)}}}${\small u(3), }$\overline
{\text{{\small s(123)}}}${\small d(6)}\\\hline
q$_{N}^{\ast}$(931)$\overline{q_{\Sigma}^{\ast}(2011)}$=K(1638) & $\bullet
$K(1653) & $\overline{\text{{\small s(123)}}}${\small u(3), }$\overline
{\text{{\small s(123)}}}${\small d(6)}\\\hline
q$_{N}^{\ast}$(2191)$\overline{q_{S}^{\ast}(1111)}$=K(1769) & $\bullet
\overline{\text{K}}$(1775) & $\overline{\text{{\small s(123)}}}${\small u(3),
}$\overline{\text{{\small s(123)}}}${\small d(6)}\\\hline
q$_{N}^{\ast}$(931)$\overline{q_{S}^{\ast}(2551)}$=K(1804) & $\bullet
\overline{\text{K}}$(1825) & $\overline{\text{{\small s(123)}}}${\small u(3),
}$\overline{\text{{\small s(123)}}}${\small d(6)}\\\hline
q$_{N}^{\ast}$(931)$\overline{q_{\Sigma}^{\ast}(2371)}$=K(1966) &
$\ \overline{\text{K}}$(1965) & $\overline{\text{{\small s(123)}}}%
${\small u(3), }$\overline{\text{{\small s(123)}}}${\small d(6)}\\\hline
q$_{N}^{\ast}$(931)$\overline{q_{S}^{\ast}(2641)}$=K(2036) & $\bullet
$K(2045) & $\overline{\text{{\small s(123)}}}${\small u(3), }$\overline
{\text{{\small s(123)}}}${\small d(6)}\\\hline
q$_{N}^{\ast}$(931)$\overline{q_{\Sigma}^{\ast}(2551)}$=K(2129) &
\ \ \ \ \ \ \ ? & $\overline{\text{{\small s(123)}}}${\small u(3), }%
$\overline{\text{{\small s(123)}}}${\small d(6)}\\\hline
q$_{N}^{\ast}$(931)$\overline{q_{\Sigma}^{\ast}(2641)}$=K(2211) & \ K(2250) &
$\overline{\text{{\small s(123)}}}${\small u(3), }$\overline
{\text{{\small s(123)}}}${\small d(6)}\\\hline
q$_{N}^{\ast}$(931)$\overline{q_{\Sigma}^{\ast}(2731)}$=K(2293) &
$\ \overline{\text{K}}$(2350) & $\overline{\text{{\small s(123)}}}%
${\small u(3), }$\overline{\text{{\small s(123)}}}${\small d(6)}\\\hline
q$_{N}^{\ast}$(931)$\overline{q_{\Sigma}^{\ast}(3091)}$=K(2621) & \ K(2500) &
$\overline{\text{{\small s(123)}}}${\small u(3), }$\overline
{\text{{\small s(123)}}}${\small d(6)}\\\hline
q$_{N}^{\ast}$(931)$\overline{q_{S}^{\ast}(4271)}$=K(3597) & \ \ \ \ \ \ \ ? &
$\ \ \ \ \ \ \ \ \ \ \ \ \ ?$\\\hline
q$_{N}^{\ast}$(931)$\overline{q_{S}^{\ast}(10031)}$=K(9429) &
\ \ \ \ \ \ \ ? & \ \ \ \ \ \ \ \ \ \ \ \ \ ?\\\hline
\ \ \ \ \ \ \ \ \ \ \ \ \ \ \ \ \ ... & \ \ \ \ \ \ ... &
\ \ \ \ \ \ \ \ \ \ \ \ \ ...\\\hline
\end{tabular}

\ \ \ \ \ \ \ \ \ \ \ \ \ \ \ \ \ \ \ \ \ \ \ \ \ \ \ \ \ \ \ \ \ \ \ \ \ \ \ \ \ \ \ 

\bigskip\newpage

\ \ \ \ \ \ \ \ \ \ \ \ \ \ \ \ \ \ \ \ \ \ \ \ \ \ \ \ \ \ \ \ \ \ \ \ \ \ \ \ \ \ \ \ \ \ \ \ \ \ \ \ \ \ \ \ \ \ \ \ \ \ \ \ \ \ \ \ \ \ \ \ \ \ \ \ \ \ \ \ \ \ \ \ \ \ \ \ \ \ \ \ \ \ \ \ \ \ \ \ \ \ \ \ \ \ \ \ \ \ \ \ \ \ \ \ \ \ 

\ \ \ \ \ \ \ \ \ \ \ \ \ \ Table IV Charmed and Charmed Strange Mesons%

\begin{tabular}
[c]{|l|l|l|}\hline
\ \ The BCC Quark Model & \ Experiment & \ The Quark Model\\\hline
\ \ \ \ q$_{i}^{\ast}$(m$_{k}$)$\overline{q_{j}^{\ast}(m_{l})}$ = D(m) &
Meson(m) & \ \ \ \ \ q$_{i}^{\ast}$(m$_{k}$)$\overline{q_{j}^{\ast}(m_{l})}%
$\\\hline
Charmed Mesons (C = $\pm1)$ &  & \ \ \ \ \ \ \ \ \ \ \ c$\overline{u}%
,$c$\overline{d}$\\\hline
{\small q}$_{C}^{\ast}${\small (2271)}$\overline{q_{N}^{\ast}({\small 931})}%
${\small = }$D(1866)$ & $\bullet D(1869)$ & c{\small (1250)}$\overline
{{\small u(3)}}$, c{\small (1250)}$\overline{{\small d(6)}}$\\\hline
{\small q}$_{C}${\small (2441)}$\overline{q_{N}^{\ast}({\small 931})}%
${\small = }$D(2029)$ & $\bullet D(2008)$ & c{\small (1250)}$\overline
{{\small u(3)}}$, c{\small (1250)}$\overline{{\small d(6)}}$\\\hline
{\small q}$_{C}${\small (2531)}$\overline{q_{N}^{\ast}({\small 931})}%
${\small = }$D(2115)$ & \ \ \ \ \ \ \ ? & c{\small (1250)}$\overline
{{\small u(3)}}$, c{\small (1250)}$\overline{{\small d(6)}}$\\\hline
{\small q}$_{C}^{\ast}${\small (2271)}$\overline{{\small q}_{N}^{\ast
}{\small (1471)}}${\small =}$D(2349)$ & $\bullet D(2420)$ & c{\small (1250)}%
$\overline{{\small u(3)}}$, c{\small (1250)}$\overline{{\small d(6)}}$\\\hline
{\small q}$_{C}${\small (2969)}$\overline{q_{N}^{\ast}({\small 931})}%
${\small = }$D(2526)$ & $\bullet\overline{D}(2507)$ & c{\small (1250)}%
$\overline{{\small u(3)}}$, c{\small (1250)}$\overline{{\small d(6)}}$\\\hline
{\small q}$_{C}^{\ast}${\small (2271)}$\overline{{\small q}_{N}^{\ast
}{\small (1921)}}${\small =}$D(2750)$ & \ \ \ \ \ \ \ ? & c{\small (1250)}%
$\overline{{\small u(3)}}$, c{\small (1250)}$\overline{{\small d(6)}}$\\\hline
{\small q}$_{C}^{\ast}${\small (2271)}$\overline{{\small q}_{N}^{\ast
}{\small (2191)}}${\small =}$D(2991)$ & \ \ \ \ \ \ \ ? & c{\small (1250)}%
$\overline{{\small u(3)}}$, c{\small (1250)}$\overline{{\small d(6)}}$\\\hline
{\small q}$_{C}^{\ast}${\small (6591)}$\overline{q_{N}^{\ast}({\small 931})}%
${\small = }$D(5996)$ & \ \ \ \ \ \ \ ? &
\ \ \ \ \ \ \ \ \ \ \ \ \ \ \ ?\\\hline
{\small q}$_{C}${\small (13791)}$\overline{q_{N}^{\ast}({\small 931})}%
${\small =}$D(13274)$ & \ \ \ \ \ \ \ ? &
\ \ \ \ \ \ \ \ \ \ \ \ \ \ \ ?\\\hline
&  & \\\hline
Charmed, Strange (C=S=$\pm$1) &  & \ \ \ \ \ \ \ \ \ \ \ \ \ \ c$\overline{s}%
$\\\hline
{\small q}$_{C}^{\ast}${\small (2271)}$\overline{q_{S}^{\ast}({\small 1111})}%
${\small = D}$_{S}(2021)$ & $\bullet$D$_{S}(1969)$ &
\ \ \ \ \ c(1250)$\overline{s(123)}$\\\hline
{\small q}$_{C}^{\ast}${\small (2271})$\overline{q_{S}^{\ast}{\small (1391)}}%
${\small = D}$_{S}${\small (2126)} & $\bullet$D$_{S}^{\ast}(2112)$ &
\ \ \ \ \ c(1250)$\overline{s(123)}$\\\hline
{\small q}$_{C}^{\ast}${\small (2961)}$\overline{q_{S}^{\ast}({\small 1111})}%
${\small = D}$_{S}(2531)$ & $\bullet\overline{\text{D}_{S}}(2555)^{\pm}$ &
\ \ \ \ \ c(1250)$\overline{s(123)}$\\\hline
{\small q}$_{C}^{\ast}${\small (2271)}$\overline{q_{S}^{\ast}({\small 2551})}%
${\small = D}$_{S}${\small (3332)} & \ \ \ \ \ \ \ \ ? &
\ \ \ \ \ c(1250)$\overline{s(123)}$\\\hline
{\small q}$_{C}^{\ast}${\small (6591)}$\overline{q_{S}^{\ast}({\small 1111})}%
${\small = D}$_{S}(6151)$ & \ \ \ \ \ \ \ \ ? &
\ \ \ \ \ \ \ \ \ \ \ \ \ \ ?\\\hline
{\small q}$_{C}^{\ast}${\small (6591)}$\overline{q_{S}^{\ast}({\small 2551})}%
${\small = D}$_{S}(7215)$ & \ \ \ \ \ \ \ \ ? &
\ \ \ \ \ \ \ \ \ \ \ \ \ \ ?\\\hline
{\small q}$_{C}${\small (13791)}$\overline{q_{S}^{\ast}({\small 1111})}%
${\small =D}$_{S}(13432)$ & \ \ \ \ \ \ \ \ ? &
\ \ \ \ \ \ \ \ \ \ \ \ \ \ ?\\\hline
{\small q}$_{C}${\small (13791)}$\overline{q_{S}^{\ast}({\small 2551})}%
${\small =D}$_{S}(14604)$ & \ \ \ \ \ \ \ \ ? &
\ \ \ \ \ \ \ \ \ \ \ \ \ \ ?\\\hline
\ \ \ \ \ \ \ \ \ \ \ \ \ \ \ \ \ \ \ \ ... & \ \ \ \ \ \ \ ... &
\ \ \ \ \ \ \ \ \ \ \ \ \ \ ...\\\hline
\end{tabular}
\ \ \ \ \ \ \ \ \ \ \ \ \ \ \ \ \ \ \ \ \ \ \ \ \ \ \ \ \ \ \ \ \ \ \ \ \ \ \ \ \ \ \ \ \ \ \ \ \ \ \ \ \ \ \ \ \ \ \ \ \ \ \ \ \ \ \ \ \ \ \ \ \ \ \ \ \ \ \ \ \ \ \ \ \ \ \ \ \ 

\ \ \ \ \ \ \ \ \ \ \ \ \ \ \ \ \ \ \ \ \ \ \ \ \ \ \ \ \ \ \ \ \ \ \ \ \ \ \ \ \ \ \ \ \ \ \ \ \ \ \ \ \ \ \ \ \ \ \ \ \ \ \ \ \ \ \ \ \ \ \ \ \ \ \ \ \ \ 

\ \ \newpage

\ \ \ \ \ \ \ \ \ \ \ \ \ \ \ \ \ \ \ \ \ \ \ \ \ \ \ \ \ \ \ \ \ \ \ \ \ \ \ \ \ \ \ \ \ \ \ \ \ \ \ \ \ 

Table V. \ Bottom, Bottom Strange, and Bottom Charmed Mesons%

\begin{tabular}
[c]{|l|l|l|}\hline
\ \ \ \ \ \ \ The BCC Quark Model & \ Exper. & The Quark Model\\\hline
\ \ \ \ q$_{i}^{\ast}$(m$_{k}$)$\overline{q_{j}^{\ast}(m_{l})}=B(m)$ &
\ M(m) & \ \ \ \ $\ \ \ \ \ \ $\\\hline
\ \ \ Bottom Mesons ({\small b=}$\pm1)$ &  & $\ \ \ \ \ \ \ \ \ \ \overline
{b}$u, $\overline{b}$d\\\hline
$\overline{q_{b}^{\ast}(5531)}${\small q}$_{N}^{\ast}${\small (931)=B(5164)} &
$\bullet${\small B(5279)} & $\overline{\text{{\small b(4200)}}}${\small u(3),
}$\overline{\text{{\small b(4200)}}}${\small d(6)}\\\hline
$\overline{q_{b}^{\ast}(5531)}q_{N}^{\ast}${\small (1201)=B(5655)} &
\ {\small B(5732)} & $\overline{\text{{\small b(4200)}}}${\small u(3),
}$\overline{\text{{\small b(4200)}}}${\small d(6)}\\\hline
$\overline{q_{b}^{\ast}(5531)}q_{N}^{\ast}${\small (1471)=B(5896)} &
\ \ \ \ \ \ ? & $\overline{\text{{\small b(4200)}}}${\small u(3), }%
$\overline{\text{{\small b(4200)}}}${\small d(6)}\\\hline
$\overline{q_{b}^{\ast}(5531)}${\small q}$_{N}^{\ast}${\small (1831)=B(6217)}%
& \ \ \ \ \ \ ? & $\overline{\text{{\small b(4200)}}}${\small u(3),
}$\overline{\text{{\small b(4200)}}}${\small d(6)}\\\hline
$\overline{q_{b}^{\ast}(9951)}${\small q}$_{N}^{\ast}${\small (931)=B(9504)} &
\ \ \ \ \ \ ? & \ \ \ \ \ \ \ \ \ \ \ \ \ \ ?\\\hline
$\overline{q_{b}^{\ast}(15811)}${\small q}$_{N}^{\ast}${\small (931)=B(15385)}%
& \ \ \ \ \ \ ? & \ \ \ \ \ \ \ \ \ \ \ \ \ \ ?\\\hline
&  & \\\hline
Bottom, Strange ({\small b =}$\pm${\small \ 1,S=}$\mp1$) &  &
$\ \ \ \ \ \ \ \ \ \ \ \ \overline{b}$s\\\hline
$\overline{q_{b}^{\ast}(5531)}${\small q}$_{S}^{\ast}${\small (1111)=B}%
$_{S}(5319)$ & $\bullet${\small B}$_{S}${\small (5370)} &
$\ \ \ \ \ \ \ \overline{\text{{\small b(4200)}}}${\small s(123)}\\\hline
$\overline{q_{b}^{\ast}(5531)}q_{S}^{\ast}${\small (1391)=B}$_{S}(5674)$ &
\ {\small B}$_{S}${\small (5850)} & $\ \ \ \ \ \ \ \overline
{\text{{\small b(4200)}}}${\small s(123)}\\\hline
$\overline{q_{b}^{\ast}(5531)}${\small q}$_{S}^{\ast}${\small (2551)=B}%
$_{S}(6639)$ & {\small \ \ \ \ \ \ \ ?} & $\ \ \ \ \ \ \ \overline
{\text{{\small b(4200)}}}${\small s(123)}\\\hline
$\overline{q_{b}^{\ast}(9951)}${\small q}$_{S}^{\ast}${\small (1111)=B}%
$_{S}(9659)$ & {\small \ \ \ \ \ \ \ ?} & $\ \ \ \ \ \ \ \ \ \ \ \ \ \ ?$%
\\\hline
$\overline{q_{b}^{\ast}(15811)}${\small q}$_{S}^{\ast}${\small (1111)=B}%
$_{S}(15540)$ & {\small \ \ \ \ \ \ \ ?} & $\ \ \ \ \ \ \ \ \ \ \ \ \ \ ?$%
\\\hline
&  & \\\hline
Bottom, Charmed ({\small b =}C{\small =}$\pm$1) &  &
$\ \ \ \ \ \ \ \ \ \ \ \ \ \overline{b}$c\\\hline
$\overline{q_{b}^{\ast}(5531)}${\small q}$_{C}^{\ast}${\small (2271)=B}%
$_{C}(6691)$ & $\bullet${\small B}$_{C}${\small (6400)} &
$\ \ \ \ \ \ \overline{\text{{\small b(4200)}}}${\small c(1250)}\\\hline
$\overline{q_{b}^{\ast}(5531)}q_{C}^{\ast}${\small (6591)=B}$_{C}(10822)$ &
\ \ \ \ \ \ ? & \ \ \ \ \ \ \ \ \ \ \ \ \ \ ?\\\hline
$\overline{q_{b}^{\ast}(9951)}${\small q}$_{C}^{\ast}${\small (2271)=B}%
$_{C}(11031)$ & \ \ \ \ \ \ ? & $\ \ \ \ \ \ \ \ \ \ \ \ \ \ ?$\\\hline
\ \ \ \ \ \ \ \ \ \ \ \ \ \ \ \ \ \ \ \ \ ... & \ \ \ \ \ ... &
\ \ \ \ \ \ \ \ \ \ \ \ ...\\\hline
\end{tabular}
\ \ \ \ \ \ \ \ \ \ \ \ \ \ \ \ \ \ 

\ \ \ \ \ \ \ \ \ \ \ \ \ \ \ \ \ \ \ \ \ \ \ \ \ \ \ \ \ \ \ \ \ \ \ \ \ \ \ \qquad
\ \ \ \ \ \ \ \ \ \ \ \ \ \ \ \ \ \ \ \ \ \ \ \ \ \ \ \ \ \ \ \ \ \ \ \ \ \ \ \ \ \ \ \ \ \ \ \ \ \ \ \ \ \ \ \ \ \ \ \ \ 

\ \ \ \ \ \ \ \ \ \ \ \ \ \ \ \ \ \ \ \ \ \ \newpage

\ \ \ \ \ \ \ \ \ \ \ \ \ \ \ \ \ \ \ \ \ \ \ \ \ \ \ \ \ \ \ \ \ \ \ \ \ \ \ \ \ \ \ \ \ \ \ \ \ \ \ \ \ \ \ \ \ \ \ \ \ \ \ \ \ \ \ \ \ \ \ \ \ \ \ \ \ \ \ \ \ \ \ \ \ \ \ \ \ \ \ \ \ \ \ \ \ \ \ \ \ \ \ \ \ \ \ \ \ \ \ \ \ \ \ \ \ \ \ \ \ \ \ \ \ \ \ \ \ \ \ \ \ \ \ \ \ \ \ \ \ \ \ \ \ \ \ \ \ \ \ \ \ \ \ \ \ \ \ \ \ \ \ \ \ \ \ \ \ \ \ \ \ \ \ \ \ \ \ \ \ \ \ \ \ \ \ \ \ \ \ \ \ \ \ \ \ \ \ \ \ \ \ \ \ \ \ \ \ \ \ \ \ \ \ \ \ \ \ \ \ \ \ \ \ \ \ \ \ \ \ \ \ \ \ \ \ \ \ \ \ \ \ \ \ \ \ \ \ \ \ \ \ \ \ \ \ \ \ \ \ \ \ \ \ \ \ \ \ \ \ \ \ \ \ \ \ \ \ \ \ \ \ \ \ \ \ \ \ \ \ \qquad
\ \ \ \ \ \ \ \ \ \ \ \ \ \ \ \ \ \ \ \ \ \ \ \ \ \ \ \ \ \ \ \ \ \ \ \ \ \ \ \ \ \ \ \ \ \ \ \ \ \ \ \ \ \ \ \ \ \ \ \ \ \ \ \ \ \ \ \ \ \ \ \ \ \ \ \ \ \ \ \ \ \ \ \ \ \ \ \ \ \ \ \ \ \ \ \ \ \ \ \ \ \ \ \ \ \ \ \ \ \ \ \ \ \ \ \ \ \ \ \ \ \ \ \ \ \ \ \ \ \ \ \ \ \ \ \ \ \ \ \ \ \ \ \ \ \ \ \ \ \ \ \ \ \ \ \ \ \ \ \ \ \ \ \ \ \ \ \ \ \ \ \ \ \ \ \ \ \ \ \ \ \ \ \ \ \ \ 

\ \ Table VI. The Slightly Heavy\ Mesons (S = C = b =I = Q =0)%

\begin{tabular}
[c]{|l|l|l|}\hline
\ The BCC Quark Model & Experiment & Quark Model\\\hline
\ \ \ \ \ \ q$_{i}^{\ast}$(m$_{k}$)$\overline{q_{j}^{\ast}(m_{l})}=M(m)$ &
Meson(m), $\Gamma$ & \ q$_{i}^{\ast}$(m$_{k}$)$\overline{q_{j}^{\ast}(m_{l})}%
$\\\hline
{\small (48)q}$_{S}^{\ast}${\small (2551)}$\overline{q_{S}^{\ast}(2551)}%
${\small =}$\eta${\small (2902)} & $\bullet\eta_{C}${\small (2980)}$_{13Mev}$%
& {\small c(1250)}$\overline{\text{{\small c(1250)}}}$\\\hline
{\small (120)q}$_{C}^{\ast}${\small (2271)}$\overline{q_{C}^{\ast}(2271)}%
${\small =J/}$\Psi${\small (3044)} & $\bullet${\small J/}$\Psi${\small (3097)}%
$_{87Kev}$ & {\small c(1250)}$\overline{\text{{\small c(1250)}}}$\\\hline
{\small (24)q}$_{\Sigma}^{\ast}${\small (2641)}$\overline{q_{\Sigma}^{\ast
}(2641)}${\small =}$\eta${\small (3394)} & $\bullet\chi${\small (3415)}%
$_{14Mev}$ & {\small c(1250)}$\overline{\text{{\small c(1250)}}}$\\\hline
{\small (24)q}$_{\Sigma}^{\ast}${\small (2731)}$\overline{q_{\Sigma}^{\ast
}(2731)}${\small =}$\eta${\small (3535)} & $\bullet\chi${\small (3510)}%
$_{0.88Mev}$ & {\small c(1250)}$\overline{\text{{\small c(1250)}}}$\\\hline
{\small (40)q}$_{C}^{\ast}${\small (2441)}$\overline{q_{C}^{\ast}(2441)}%
${\small =}$\psi${\small (3568)} & $\bullet\chi${\small (3556)}$_{2.0Mev}$ &
{\small c(1250)}$\overline{\text{{\small c(1250)}}}$\\\hline
{\small (60)q}$_{C}^{\ast}${\small (2271)}$\overline{q_{C}^{\ast}(2531)}%
${\small =}$\psi${\small (3693)} & $\bullet\psi${\small (3686)}$_{277Kev}$ &
{\small c(1250)}$\overline{\text{{\small c(1250)}}}$\\\hline
{\small (40)q}$_{C}^{\ast}${\small (2531)}$\overline{q_{C}^{\ast}(2531)}%
${\small =}$\psi${\small (3740)} & $\bullet\psi${\small (3770)}$_{24Mev}$ &
{\small c(1250)}$\overline{\text{{\small c(1250)}}}$\\\hline
{\small (30)q}$_{C}^{\ast}${\small (2441)}$\overline{q_{C}^{\ast}(2531)}%
${\small =}$\psi${\small (3854)} & $\ \psi${\small (3836) \ \ \ ?} &
{\small c(1250)}$\overline{\text{{\small c(1250)}}}$\\\hline
{\small (54)q}$_{S}^{\ast}${\small (1111)}$\overline{q_{S}^{\ast}(4271)}%
${\small =}$\eta${\small (3952)} & $\bullet\psi${\small (4040)}$_{52Mev}$ &
{\small c(1250)}$\overline{\text{{\small c(1250)}}}$\\\hline
{\small (60)q}$_{C}^{\ast}${\small (2271)}$\overline{q_{C}^{\ast}(2961)}%
${\small =}$\psi${\small (4104)} & $\bullet\psi${\small (4160)}$_{78Mev}$ &
{\small c(1250)}$\overline{\text{{\small c(1250)}}}$\\\hline
{\small (40)q}$_{C}^{\ast}${\small (2961)}$\overline{q_{C}^{\ast}(2961)}%
${\small =}$\psi${\small (4554)} & $\bullet\psi${\small (4415)}$_{43Mev}$ &
{\small c(1250)}$\overline{\text{{\small c(1250)}}}$\\\hline
{\small (36)q}$_{S}^{\ast}${\small (2551)}$\overline{q_{S}^{\ast}(4271)}%
${\small =}$\eta${\small (4944)} & \ \ \ \ \ \ \ \ ? & {\small c(1250)}%
$\overline{\text{{\small c(1250)}}}$\\\hline
{\small (75)}$\overline{q_{C}^{\ast}(6591)}${\small q}$_{C}^{\ast}%
${\small (2271)=}$\psi${\small (7374)} & \ \ \ \ \ \ \ \ ? &
\ \ \ \ \ \ \ \ ?\\\hline
{\small (75)}$\overline{\text{q}_{C}^{\ast}\text{(13791)}}${\small q}%
$_{C}^{\ast}${\small (2271)=}$\psi${\small (14654)} & \ \ \ \ \ \ \ \ ? &
\ \ \ \ \ \ \ \ ?\\\hline
{\small (80)}$\overline{\text{q}_{C}^{\ast}\text{(13791)}}${\small q}$_{C}%
${\small (13791)=}$\psi${\small (25596)} & \ \ \ \ \ \ \ \ ? &
\ \ \ \ \ \ \ \ ?\\\hline
\end{tabular}
\ \ \ \ \ \ \ \ \ \newpage

\newpage

\ \ \ \ \ \ \ \ \ \ \ \ \ \ \ \ \ \ \ \ \ \ \ \ \ \ \ \ \ \ \ \ \ \ \ \ \ \ \ \ \ \ \ \ \ \ \ \ \ \ \ \ \ \ \ \ \ \ \ \ \ \ \ \ \ \ \ \ \ \ \ \ \ \ \ \ \ \ \ \ \ \ 

\ \ \ \ \ \ \ \ \ Table VII. \ The\ Heavy\ Mesons (S = C = b = I = Q = 0)%

\begin{tabular}
[c]{|l|l|l|}\hline
\ The BCC Quark Model & \ \ Experiment & Quark Model\\\hline
\ \ \ q$_{i}^{\ast}$(m$_{k}$)$\overline{q_{j}^{\ast}(m_{l})}$ = M(m) &
\ \ Meson(m), $\Gamma$ & \ q$_{i}^{\ast}$(m$_{k}$)$\overline{q_{j}^{\ast
}(m_{l})}$\\\hline
{\small (72)q}$_{b}^{\ast}${\small (5531)}$\overline{q_{b}^{\ast}(5531)}%
${\small =}$\Upsilon${\small (9489)} & $\bullet\Upsilon$%
{\small (1S)(9460),52kev} & {\small b(5531)}$\overline{{\small b(5531)}}%
$\\\hline
{\small (54)q}$_{S}^{\ast}${\small (1111)}$\overline{q_{S}^{\ast}(10031)}%
${\small =}$\eta${\small (9734)} &
\begin{tabular}
[c]{l}%
$\bullet\chi_{b0}${\small (1P)(9860)}\\
$\bullet\chi_{b1}${\small (1P)(9892)}\\
$\bullet\chi_{b2}${\small (1P)(9913)}%
\end{tabular}
& {\small b(5531)}$\overline{{\small b(5531)}}$\\\hline
{\small (36)q}$_{S}^{\ast}${\small (1391)}$\overline{q_{S}^{\ast}(10031)}%
${\small =}$\eta${\small (9955)} & $\bullet\Upsilon${\small (2S)(10023),44kev}%
& {\small b(5531)}$\overline{{\small b(5531)}}$\\\hline
{\small (27)q}$_{S}${\small (2011)}$\overline{q_{S}^{\ast}(10031)}$%
{\small =}$\eta${\small (10443)} &
\begin{tabular}
[c]{l}%
$\bullet\chi_{b0}${\small (2P)(10232)}\\
$\bullet\chi_{b1}${\small (2P)(10255)}\\
$\bullet\chi_{b2}${\small (2P)(10269)}%
\end{tabular}
& {\small b(5531)}$\overline{{\small b(5531)}}$\\\hline
{\small (80)q}$_{C}^{\ast}${\small (6591)}$\overline{q_{C}^{\ast}(6591)}%
${\small =}$\psi${\small (10792)} & $\bullet\Upsilon$%
{\small (3S)(10355),26kev} & {\small b(5531)}$\overline{{\small b(5531)}}%
$\\\hline
{\small (27)q}$_{S}${\small (2451)}$\overline{q_{S}^{\ast}(10031)}$%
{\small =}$\eta${\small (10791)} & $\bullet\Upsilon${\small (4S)(10580),14Mev}%
& {\small b(5531)}$\overline{{\small b(5531)}}$\\\hline
{\small (36)q}$_{S}^{\ast}${\small (2551)}$\overline{q_{S}^{\ast}(10031)}%
${\small =}$\eta${\small (10870)} & $\bullet\Upsilon${\small (10860),110Mev} &
{\small b(5531)}$\overline{{\small b(5531)}}$\\\hline
{\small (27)q}$_{S}${\small (2641)}$\overline{q_{S}^{\ast}(10031)}$%
{\small =}$\eta${\small (10941)} & \ \ \ \ \ \ \ \ \ \ ? & {\small b(5531)}%
$\overline{{\small b(5531)}}$\\\hline
{\small (27)q}$_{S}${\small (2731)}$\overline{q_{S}^{\ast}(10031)}$%
{\small =}$\eta${\small (11012)} & $\bullet\Upsilon${\small (11020),79Mev} &
{\small b(5531)}$\overline{{\small b(5531)}}$\\\hline
{\small (45)q}$_{b}^{\ast}${\small (5531)}$\overline{q_{b}^{\ast}(9951)}%
${\small =}$\Upsilon${\small (14329)} & \ \ \ \ \ \ \ \ \ \ \ ? &
\ \ \ \ \ \ \ \ \ ?\\\hline
{\small (48)q}$_{b}^{\ast}${\small (9951)}$\overline{q_{b}^{\ast}(9951)}%
${\small =}$\Upsilon${\small (17805)} & \ \ \ \ \ \ \ \ \ \ \ ? &
\ \ \ \ \ \ \ \ \ ?\\\hline
{\small (45)q}$_{b}^{\ast}${\small (5531)}$\overline{q_{b}^{\ast}(15811)}%
${\small =}$\Upsilon${\small (20210)} & \ \ \ \ \ \ \ \ \ \ \ ? &
\ \ \ \ \ \ \ \ \ ?\\\hline
{\small (48)q}$_{b}^{\ast}${\small (15811)}$\overline{q_{b}^{\ast}(15811)}%
${\small =}$\Upsilon${\small (29597)} & \ \ \ \ \ \ \ \ \ \ \ ? &
\ \ \ \ \ \ \ \ \ ?\\\hline
\end{tabular}

\ \ \ \ \ \ \ \ \ \ \ \ \ \ \ \ \ \ \ 

In summary, the BCC Quark Model explains the experimental intrinsic quantum
numbers and masses of all mesons. Virtually all experimentally confirmed
mesons are included in the BCC Quark Model. However, we do not give out the
angular momenta and parities of the mesons. They are needed to consider the
wave functions of the quarks, the point groups (the point group O$_{h}$, the
point group P, the point group N,...) of the body center cubic quark lattice.
Those are out the scope of this paper.

\section{ Evidence for Some New Quarks \ \ \ \ \ \ \ \ \ \ \ \ \ \ \ \ \ \ \ }

According to the confinement theory \cite{CONFINEMENT} of the Quark Model and
the accompanying excitation concept of the BCC Quark Model \cite{NetXu
(section II)}, we cannot see any individual quark. We can only infer the
existence of quarks from the existence of baryons and mesons. Thus, if we find
the mesons which were made of certain quarks, it means that we find the
quarks. For example, from meson J/$\Psi$(3097) = [q$_{C}^{\ast}$%
(2271)$\overline{q_{C}^{\ast}(2271)}$], we discovered the quark q$_{C}^{\ast}%
$(2271). Similarly, the experimental meson spectrum \cite{Particle(2000)}\ has
already provided some evidence of the new quarks q$_{S}^{\ast}$(1391),
q$_{S}^{\ast}$(2551), and q$_{C}^{\ast}$(6591)...

\subsection{Four Brother Quarks $q_{S}^{\ast}(1111),$ $q_{S}^{\ast}(2551),$
$q_{b}^{\ast}(5531),$ and $q_{b}^{\ast}(9951)$}

From Fig. 5 (c) (see Appendix I) of the BCC Quark Model \cite{NetXu (Quark)},
we see the following four ``brother'' quarks: at E$_{N}=1/2,\ \overrightarrow
{n}=(1,1,0),$ $q_{S}^{\ast}(1111)$; at E$_{N}=9/2,$ $\overrightarrow
{n}=(2,2,0),$ q$_{S}^{\ast}(2551)$; at E$_{N}=25/2,$ $\overrightarrow
{n}=(3,3,0),$ $q_{b}^{\ast}(5531);$ and at E$_{N}=49/2,$ $\overrightarrow
{n}=(4,4,0),$ $q_{b}^{\ast}(9951)$. They are born on the single energy bands
of the $\Sigma$-axis and at the same symmetry point $N$. The four ``brothers''
have the same isospin (I = 0), the same electric charge (Q = -1/3), and the
same generalized strange number (S$_{G}$ = S + C + b = -1).

A1. The baryons made by the brothers have I =0, Q = 0, S$_{G}$= -1, with long
life times \cite{BCC MODEL}:
\begin{equation}%
\begin{tabular}
[c]{|l|}\hline
q$_{S}^{\ast}$(1111)q$_{u}^{\prime}(3)$q$_{d}^{\prime}(6)$=$\Lambda
(1120)\leftrightarrow\Lambda$(1116), $\tau$ = (2.632$\pm0.020)\times10^{-10}s$
;\\\hline
q$_{S}^{\ast}$(2551)q$_{u}^{\prime}(3)$q$_{d}^{\prime}(6)$= $\Lambda
(2560)\leftrightarrow\Omega_{C}^{0}$(2704), $\tau$ = (0.64$\pm0.020)\times
10^{-11}s$;\\\hline
q$_{b}^{\ast}$(5531)q$_{u}^{\prime}(3)$q$_{d}^{\prime}(6)$=$\Lambda_{b}%
$(5540)$\leftrightarrow\Lambda_{b}$(5641), $\tau$ = (1.14$\pm0.08)\times
10^{-12}s;$\\\hline
q$_{b}^{\ast}$(9951)q$_{u}^{\prime}(3)$q$_{d}^{\prime}(6)$= $\Lambda_{b}%
$(9960)$\leftrightarrow?$, $\ ?.$\\\hline
\end{tabular}
\end{equation}

A2. The mesons that are composed of four brothers and their own antiquarks
are:
\begin{equation}%
\begin{tabular}
[c]{|l|}\hline
q$_{S}^{\ast}$(1111)$\overline{q_{S}^{\ast}({\small 1111})}$=$\eta
(549)\leftrightarrow\lbrack\eta(549),\Gamma=(1.18\pm0.11)kev$]\\\hline
q$_{S}^{\ast}$({\small 2551})$\overline{q_{S}^{\ast}({\small 2551})}$= $\eta
($2902)$\leftrightarrow$[ $\eta_{C}($2969),$\Gamma=(13.2$ Mev]\\\hline
q$_{b}^{\ast}$(5531)$\overline{q_{b}^{\ast}({\small 5531})}$=$\Upsilon
(1S)(9489)\leftrightarrow\lbrack\Upsilon(1S)(9460)$, $\Gamma=$ (52.5$\pm1.8)$
kev],\\\hline
q$_{b}^{\ast}$(9951)$\overline{q_{b}^{\ast}({\small 9951})}$=$\Upsilon
(17805)\leftrightarrow?,?.$\\\hline
\end{tabular}
\end{equation}

A3. The mesons that are composed of quark q$_{N}^{\ast}$(931) and four brother
antiquarks are:%

\begin{equation}%
\begin{tabular}
[c]{|l|}\hline
q$_{N}^{\ast}$(931)$\overline{q_{S}^{\ast}(1111)}${\small = }%
K(494)$\leftrightarrow\lbrack K(494),$ $\Gamma=$ (1.2386 $\pm$ 0.0024)
$\times$ 10$^{-8}$s]\\\hline
q$_{N}^{\ast}$(931)$\overline{q_{S}^{\ast}(2551)}$=K(1804)$\leftrightarrow
\lbrack$K$_{2}$(1820), $\Gamma$ = 276 $\pm$ 35 Mev]\\\hline
q$_{N}^{\ast}$(931)$\overline{q_{b}^{\ast}(5531)}${\small =}
B(5164)$\leftrightarrow$ [B(5279), $\Gamma$ = (1.62 $\pm$ 0.06) $\times$
10$^{-12}$s]\\\hline
q$_{N}^{\ast}$(931)$\overline{q_{b}^{\ast}(9951)}${\small =}
B(9504)$\leftrightarrow$ \ (?. ?)\\\hline
\end{tabular}
\end{equation}

A4. The mesons that are composed of quark q$_{S}^{\ast}$(1111) and four
brother antiquarks are:%

\begin{equation}%
\begin{tabular}
[c]{|l|}\hline
q$_{S}^{\ast}$(1111)$\overline{q_{S}^{\ast}(1111)}$= $\eta(547)\leftrightarrow
\lbrack\eta(549),$$\Gamma$= (1.18 $\pm$0.11 kev)]\\\hline
q$_{S}^{\ast}$(1111)$\overline{q_{S}^{\ast}(2551)}$=$\eta$%
(1960)$\leftrightarrow\lbrack f_{2}(1950),$$\Gamma$= 208 Mev]\\\hline
q$_{S}^{\ast}$(1111)$\overline{q_{b}^{\ast}(5531)}$= B$_{S}$%
(5319)$\leftrightarrow$[B$_{S}$(5279), $\Gamma$= (1.61 $\pm$0.10) $\times
$10$^{-12}$s]\\\hline
q$_{S}^{\ast}$(1111)$\overline{q_{b}^{\ast}(9951)}$= B$_{S}$%
(9659)$\leftrightarrow$\ (?. ?)\\\hline
\end{tabular}
\end{equation}

There is much evidence of the first three brother quarks, q$_{S}^{\ast}%
$(1111), q$_{S}^{\ast}$(2551), and q$_{b}^{\ast}$(5531).

\subsection{The Quark q$_{S}^{\ast}$(1391)}

There are three brother quarks, q$_{S}^{\ast}$(1391)$,$ q$_{S}^{\ast}%
$(4271)$,$ and q$_{S}^{\ast}$(10031). They are born on the single energy bands
of the $\Delta$-axis at the same symmetry point ($\Gamma)$ from $\Delta
S=-1$(see Fig. 5 \ b of Appendix I$)$. Since\ the quarks q$_{S}^{\ast}$(4271)
and q$_{S}^{\ast}$(10031) have higher energies, we will see the evidence later
(the q$_{C}\overline{q_{C}^{\ast}}$ Mesons and the q$_{b}\overline{q_{b}%
^{\ast}}$ Mesons). We only study the quark q$_{S}^{\ast}$(1391) now.

B1. The baryons of the quarks%

\begin{equation}
q_{S}^{\ast}(1391)q_{u}^{\prime}(3)q_{d}^{\prime}(6)=\langle%
\begin{tabular}
[c]{l}%
$\Lambda(1400)\leftrightarrow\Lambda(1405)$\\
$\Sigma(1400)\leftrightarrow\Sigma(1385)$%
\end{tabular}
\end{equation}
\ \ \ 

B2. The meson\ that is composed of the quark q$_{S}^{\ast}$(1391) and its own
antiquarks \ \ \ \ \ \ \ \
\begin{equation}
q_{S}^{\ast}(1391)\overline{q_{S}^{\ast}(1391)}=\eta(952)\leftrightarrow
\eta^{\prime}(958)
\end{equation}

B3. The meson that is composed of $\overline{q_{S}^{\ast}(1391)}$ and the
ground quark $q_{N}^{\ast}(931)$ \qquad%
\begin{equation}
q_{N}^{\ast}(931)\overline{q_{S}^{\ast}(1391)}=K(899)\leftrightarrow K(892)
\end{equation}
\ \ 

B4. The meson that is composed of $\overline{q_{S}^{\ast}(1391)}$ and the
ground quark $q_{C}^{\ast}(2271)$
\begin{equation}
q_{C}^{\ast}(2271)\overline{q_{S}^{\ast}(1391)}=D_{S}(2126)\leftrightarrow
D_{S}^{\ast}(2112)
\end{equation}

B5. The meson that is composed of $\overline{q_{S}^{\ast}(1391)}$ and the
ground quark $q_{b}^{\ast}(5531)$\qquad%
\begin{equation}
q_{b}^{\ast}(5531)\overline{q_{S}^{\ast}(1391)}=B_{S}(5674)\leftrightarrow
B_{sj}^{\ast}(5850)
\end{equation}

There is enough evidence to show that the quark q$_{S}^{\ast}$(1391) really exists.\ \ \ \ \ \ \ \ \ \ \ \ \ 

\ \ \ \ \ \ \ \ \ \ \ \ \ \ \ \ \ \ \ \ \ \ \ \ \ \ \ \ \ \ \ \ \ \ \ 

\subsection{The q$_{C}\overline{q_{C}^{\ast}}$ Mesons\ $\rightarrow$%
q$_{S}^{\ast}$(2551)}

According to the Quark Model, the 10 experimental mesons [$\eta_{C}$(2980)
$\Gamma$=13 Mev, J/$\Psi$(3097) $\Gamma$=87 kev, $\chi_{co}$(3415) $\Gamma$=15
Merv, $\chi_{c1}$(3511) $\Gamma$= 0.88 Mev, $\chi_{c2}$(3556) $\Gamma$= 2.0
Mev, $\psi$(3686) $\Gamma$= 277 kev, $\psi$(3770) $\Gamma$= 24 Mev, $\psi
$(4040) $\Gamma$= 52 Mev, $\psi$(4160) $\Gamma$=78 Mev, and $\psi$(4415)
$\Gamma$=43 Mev] are all \ c(1250)$\overline{c(1250)}$ with different angular
momenta and parities ( sea Table VI).

However, the ground state $\eta_{C}(2980)$ ($\Gamma=$13 Mev=1300 kev)\ of
\ c$\overline{c}$ is not the most long lived meson. The most long lived meson
of \ c(1250)$\overline{c(1250)}$ is $J/\Psi(3097)$ ($\Gamma=87$ kev$).$ This
means that $\eta_{C}(2980)$ is not the ground state, while $J/\Psi(3097)$ is
the ground state. What is $\eta_{C}(2980)$ ? The BCC\ Quark Model shows that
$\eta_{C}(2980)$ is q$_{S}^{\ast}$(2551)$\overline{q_{S}^{\ast}(2551)}$. Using
the quark q$_{S}^{\ast}$(2551), we can explain the puzzle, i. e. why
$J/\Psi(3097)$ has a longer lifetime ($\Gamma=87$ kev$)$ than $\eta_{C}($2969)
($\Gamma=$13 Mev=1300 kev). According to the BCC Quark Model, $\eta_{C}($2969)
is composed of q$_{S}^{\ast}$(2551) and $\overline{q_{S}^{\ast}(2551)},$\ but
$J/\Psi(3097)$ is made of q$_{C}^{\ast}$(2271) and $\overline{q_{C}^{\ast
}(2271)}$. They are not the same kind of mesons. The quark pair q$_{C}^{\ast}%
$(2271)$\overline{q_{C}^{\ast}(2271)}$ ( $J/\Psi(3097)$) has a better chance
(O(q$_{C}^{\ast}$(2271)$\overline{q_{C}^{\ast}(2271)}$)= 120 units) than the
pair quark q$_{S}^{\ast}$(2551)$\overline{q_{S}^{\ast}(2551)}$ ($\eta
_{C}(2980)$) has (O(q$_{S}^{\ast}$(2551)$\overline{q_{S}^{\ast}(2551)}$) = 48
units).\ Thus, $J/\Psi(3097)$ has a longer lifetime than $\eta_{C}($2969) has.\ \ \ \ \ \ \ \ \ \ \ \ 

The experimental mesons show that the quark q$_{S}^{\ast}(2551)$ really
exists. At the same time, the quark q$_{S}^{\ast}$(4271) may exist too. It is
inside the meson {\small [(54)}$\bullet$q$_{S}^{\ast}$(1111)$\overline
{q_{S}^{\ast}(4271)}$ {\small = }$\chi$(3952)$\rightarrow\psi$(4040)] and the
meson{\small \ [(54)}$\bullet$q$_{S}^{\ast}$(1391)$\overline{q_{S}^{\ast
}(4271)}$ {\small = }$\chi$(4105)$\rightarrow\psi(4160)$] (sea Table VI).

\subsection{The q$_{b}\overline{q_{b}^{\ast}}$ Mesons\ $\rightarrow$
q$_{C}^{\ast}$(6591)}

According to the Quark Model, there are 12 experimental mesons that are
composed of q$_{b}^{\ast}$(5531)$\overline{q_{b}^{\ast}(5531)}$: $\Upsilon
$(1S)(9460) ($\Gamma$=53kev), $\chi_{b0}$(1P) (9860), $\chi_{b1}$(1P) (9893),
$\chi_{b2}$(1P)(9913), $\Upsilon$(2S)(10023) ($\Gamma$=44kev), $\chi_{b0}%
$(2P)(10232), $\chi_{b1}$(2P)(10255), $\chi_{b2}$(2P)(10269), $\Upsilon
$(3S)(10355) ($\Gamma$=26kev), $\Upsilon$(4S)(10580) ($\Gamma$=14Mev),
$\Upsilon$(10860) ($\Gamma$=110Mev), and $\Upsilon$(11020) ($\Gamma$=79Mev).
The ground state is q$_{b}^{\ast}$(5531)$\overline{q_{b}^{\ast}(5531)}$
=$\Upsilon$(1S)(9460) ($\Gamma$=52kev).

First, it looks like there are too many mesons corresponding to the same quark
pair (q$_{b}^{\ast}(5531)\overline{q_{b}^{\ast}(5531)}$). Second, there exists
a puzzle--why do the excitation $\Upsilon$(3S)(10355) ($\Gamma$= 26 kev) have
longer lifetimes than the ground state $\Upsilon$(1S)(9460) ($\Gamma$ = 52
kev)? According to the BCC Quark Model, they are not all the excitations of
q$_{b}^{\ast}$(5531)$\overline{q_{b}^{\ast}(5531)}$. The meson with the
longest lifetime ($\tau\sim$1/$\Gamma$) is not the meson $\Upsilon$(1S)(9460)
($\Gamma$= 52 kev) [q$_{b}^{\ast}$(5531)$\overline{q_{b}^{\ast}(5531)}$]. It
is the meson $\Upsilon(3S)(10355)$ ($\Gamma$ = 26 kev). It cannot be an
excited state of the q$_{b}^{\ast}$(5531)$\overline{q_{b}^{\ast}(5531)}$ (the
explanation of the Quark Model). There is not any quark pair that can explain
it in the Quark Model. However, the BCC Quark Model can explain it using the
quark pair (80)q$_{C}^{\ast}$(6591)$\overline{q_{C}^{\ast}(6591)}$ =
$\Upsilon$(10792) [I = 0, Q = 0, S= C = b = 0]{\small . }The meson $\Upsilon
$(10792)has the longest lifetime ($\tau\sim$1/$\Gamma$=1/26 kev)). The small
error of the mass may come from its angular momentum. These experimental
results show that there will be a new quark q$_{C}^{\ast}$(6591) in addition
to the five quarks (u, d, s, c, and b) of the Quark Model. There may be
another new quark, q$_{S}^{\ast}$(10031), also. It will be inside the meson
$\eta$(9734) {\small [}(36)$\bullet$q$_{S}^{\ast}$(1111)$\overline{q_{S}%
^{\ast}(10031)}$ = $\eta$(9734)], the meson $\eta$(9955) {\small [}%
(36)$\bullet$q$_{S}^{\ast}$(1391)$\overline{q_{S}^{\ast}(10031)}$ = $\eta
$(9955)], the meson $\eta$(10446) [(27)$\bullet$q$_{S}$(2011)$\overline
{q_{S}^{\ast}(10031)}$ = $\eta$(10446)], the meson $\eta$(10791)
[(27)$\bullet$q$_{S}$(2451)$\overline{\text{q}_{S}^{\ast}(10031)}$ = $\eta
$(10446)], and the meson $\eta$(10870) [(36)$\bullet$q$_{S}^{\ast}%
$(2551)$\overline{q_{S}^{\ast}(10031)}$ = $\eta$(10870)] (sea Table VII).

Summarizing the section, the experimental meson spectrum provides some
evidence that shows the new quarks q$_{S}^{\ast}$(1391), q$_{S}^{\ast}$(2551),
and q$_{C}^{\ast}$(6591) really exist in nature.\ The new quarks q$_{S}^{\ast
}$(4271) and q$_{S}^{\ast}$(10031) may exist also. Thus, there are three
``brother'' quark families: (1) the three brothers $q_{S}^{\ast}(1111),$
$q_{S}^{\ast}(2551),$ and $q_{b}^{\ast}(5531);$ (2) the three brothers
q$_{S}^{\ast}$(1391)$,$ q$_{S}^{\ast}$(4271)$,$ and q$_{S}^{\ast}%
$(10031)$;(3)$ the two brothers q$_{C}^{\ast}$(2271), q$_{C}^{\ast}$(6591).
Therefore, we have previously shown that the experimental meson spectrum
supports the BCC Quark Model.\ \ \ \ \ \ \ \ \ \ \ \ \ 

\section{Predictions and Discussion}

\subsection{Some New Mesons}

According to the BCC Quark Model (see Appendix II), a series of possible new
mesons with high energies exist. However, when energy goes higher and higher,
on one hand, the theoretical mesons will become denser and denser; while on
the other hand, the experimental full widths of these mesons will become wider
and wider. This case makes these new mesons extremely difficult to seperate.
Therefore, currently it is very difficult\ to discover the higher energy
mesons predicted by the BCC Quark Model. We believe that many new mesons will
be discovered in the future with the development of more sensitive
experimental techniques. The following new mesons predicted by the model seem
to have a better chance of being discovered in the near future:

A1. \ The slightly higher mass mesons%
\begin{equation}%
\begin{tabular}
[c]{|l|l|}\hline
(O(I)) \ q$_{i}^{\ast}$(m$_{k}$)$\overline{q_{j}^{\ast}(m_{l})}=M(m);$ &
\ Quantum numbers\ \ \ \\\hline
(60)$\bullet$q$_{N}^{\ast}$(931)$\overline{q_{S}^{\ast}(4271)}$=K(3597); &
{\small S =1, b = 0, C = 0, I =}$\frac{1}{2}${\small , Q = 1, 0}\\\hline
(38)$\bullet$q$_{C}^{\ast}$(2271)$\overline{q_{N}^{\ast}(1921)}$=D(2750); &
{\small S = b = 0, C = 1, I = }$\frac{1}{2}${\small , Q = 1, 0}\\\hline
(38)$\bullet$q$_{C}^{\ast}$(2271)$\overline{q_{N}^{\ast}(2191)}$=D(2991); &
{\small S = b = 0, C = 1, I =}$\frac{1}{2}${\small , Q = 1, 0}\\\hline
(36)$\bullet$q$_{C}^{\ast}$(2271)$\overline{q_{S}^{\ast}(2011)}$=D$_{S}%
$(2690) & {\small S = C = 1, b = 0, I = 0, Q = 1}\\\hline
(42)$\bullet$q$_{C}^{\ast}$(2271)$\overline{q_{S}^{\ast}(2551)}$=D$_{S}%
$(3332); & {\small S = C = 1, b = 0, I = 0, Q = 1}\\\hline
(26)$\bullet\overline{q_{b}^{\ast}(5531)}$q$_{N}^{\ast}$(1471)=B(5896);\  &
{\small S= C = 0, b = 1, I = }$\frac{1}{2}${\small , Q = 1, 0}\\\hline
(26)$\bullet\overline{q_{b}^{\ast}(5531)}$q$_{N}^{\ast}$(1831)=B(6217); &
{\small S= C = 0, b = 1, I = $\frac{1}{2}$, Q = 1, 0\ }\\\hline
(24)$\bullet$q$_{b}^{\ast}$(5531)$\overline{q_{S}^{\ast}(2011)}$=B$_{S}%
$(6238); & {\small S= -b = 1, C = 0, I = }$0${\small , Q = 0}\ \\\hline
(30)$\bullet\overline{{\small q}_{b}^{\ast}{\small (5531)}}$q$_{S}^{\ast}%
$(2551)=B$_{S}$(6629); & {\small S= -b = 1, C = 0, I = }$0${\small , Q =
0}\\\hline
\end{tabular}
\label{Litter High Meson}%
\end{equation}

A2. The high mass mesons%
\begin{equation}%
\begin{tabular}
[c]{|l|l|}\hline
O(I) q$_{i}^{\ast}$(m$_{k}$)$\overline{q_{j}^{\ast}(m_{l})}=M(m);$ & Quantum
numbers\\\hline
{\small (68)}$\bullet\overline{q_{N}^{\ast}(931)}$q$_{C}^{\ast}$(6591)
{\small = D(5996);} & {\small S = b = 0, C = 1, I =}$\frac{1}{2}${\small , Q
=0,-1}\\\hline
{\small (32)}$\bullet$q$_{C}^{\ast}$(6591)$\overline{q_{S}^{\ast}(1111)}$
{\small = D}$_{S}${\small (6151);} & {\small S = C = 1, b = 0, I = 0, Q =
1}\\\hline
{\small (36)}$\bullet$q$_{N}^{\ast}$(931)$\overline{q_{b}^{\ast}(9951)}%
${\small \ = B(9504);} & {\small S = C = 0, b = 1, I =}$\frac{1}{2}${\small ,
Q = 1, 0}\\\hline
{\small (24)}$\bullet$q$_{b}^{\ast}$(9951)$\overline{q_{S}^{\ast}(1111)\text{
}}${\small = B}$_{S}${\small (9659);} & {\small S = -b = 1, C = -0, I = }%
$0${\small , Q = 0}\\\hline
{\small (48]}$\bullet$q$_{S}^{\ast}$(4271)$\overline{q_{S}^{\ast}(4271)}$
{\small = }$\eta${\small (5926);} & {\small S = 0, C = b = 0, I = 0, Q =
\ 0.}\\\hline
{\small (32)}$\bullet$q$_{b}^{\ast}$(5531)$\overline{q_{C}^{\ast}(6591)}%
${\small =B}$_{C}${\small (10822);} & {\small S = 0, C = b = -1 I = 1, Q =
-1}\\\hline
\end{tabular}
\label{High Mass Mesons}%
\end{equation}

A3. The super mass mesons\ \ \ \ \ \ \ \ \ \ \ \ \ \ \ \ \ \ \
\begin{equation}%
\begin{tabular}
[c]{|l|l|}\hline
O(I) \ \ \ \ \ \ q$_{i}^{\ast}$(m$_{k}$)$\overline{q_{j}^{\ast}(m_{l})}$ =
$\eta$(m); & \ Quantum numbers\\\hline
(48)$\bullet$q$_{S}^{\ast}$(10031)$\overline{q_{S}^{\ast}(10031)}$ = $\chi
$(17837); & S = C = b = I = Q = 0\\\hline
(80)$\bullet$q$_{C}^{\ast}$(13791)$\overline{q_{C}^{\ast}(13791)}$= $\psi
$(25596); & S = C = b = I = Q = 0\\\hline
(48)$\bullet$q$_{b}^{\ast}$(9951)$\overline{q_{b}^{\ast}(9951)}$ = $\Upsilon
$(17805); & S = C = b = I = Q = 0\\\hline
(48)$\bullet$q$_{b}^{\ast}$(15811)$\overline{q_{b}^{\ast}(15811)}$ =
$\Upsilon$(29597); & S = C = b = I = Q = 0\\\hline
\end{tabular}
\label{Super Mass Mesons}%
\end{equation}

A4. The high isospin mesons

The BCC Quark Model predicts many high isospin mesons (I = 3/2, I = 2, and I =
5/2, and I = 3). They cannot be predicted by the Quark Model. They have not
been discovered by experiment (except a meson $\chi(1600)$ with I = 2) yet. We
list the high isospin mesons predicted by the BCC Quark Model in Appendix III.\ 

The mesons T(m) with I =2 (from q$_{N}^{\ast}$(931)$\overline{q_{\Delta}%
^{\ast}(m)}$) have O(q$_{i}^{\ast}\overline{q_{j}^{\ast}}$)= 37 (%
$>$%
24). Thus they should be discovered by today's experiments. However, there are
five members (I$_{z}$=2, Q=2; I$_{z}$=1, Q=1; I$_{z}$=0, Q=0; I$_{z}$=-1,
Q=-1; I$_{z}$=-2, Q=-2) in the mesons. The member with Q = 2 will be composed
by q$_{N}^{\ast}$(931)$^{2/3}$(Q= 2/3) and $\overline{q_{\Delta}^{\ast}}%
$(Q=4/3). The two quarks with the same kind of electric charges will repel
each other with 4 times (2/3$\times$4/3=8/9) the strength that the mesons with
Q = 1 (1/3$\times2/3=2/9)$ exhibit. Therefore, the members are very difficult
to find. If we could not find the members with Q = $\pm2$, we would find the
other three members with Q = 1, 0, -1. They may be observed as the $\pi$
mesons with S =C = b =0, I =1, Q = 1, 0, -1.

Although the mesons with I = 2 are very difficult to find, our great
experimental physicists have already discovered one--$\chi(1600)$ [I$^{G}%
$(J$^{PC}$) = 2$^{+}$(2$^{++}$)] \cite{Meson (I=2)}. According to the BCC
Quark Model, there is a series of mesons with I = 2: \ %

\begin{equation}%
\begin{tabular}
[c]{|l|l|l|}\hline
O(T)q(m)$\overline{q(m)}$=Meson(m) & {\small Experiment[I}$^{G}$%
{\small (J}$^{PC}${\small )]} & May observed\\\hline
{\small 37q}$_{N}^{\ast}${\small (931)}$\overline{q_{\Delta}^{\ast}(1291)}%
${\small =T(960)} & ? & $\pi${\small (960), I=1,Q=1,0,-1}\\\hline
{\small 37q}$_{N}^{\ast}${\small (931)}$\overline{q_{\Delta}^{\ast}(1651)}%
${\small =T(1282)} & ? & $\pi${\small (1282), I=1,Q=1,0,-1}\\\hline
{\small 37q}$_{N}^{\ast}${\small (931)}$\overline{q_{\Delta}^{\ast}(2011)}%
${\small =T(1603)} & $\chi${\small (1600)}$^{\ast}${\small [2}$^{+}%
${\small (2}$^{++}${\small )]} & $\pi${\small (1603), I=1,Q=1,0,-1}\\\hline
{\small 37q}$_{N}^{\ast}${\small (931)}$\overline{q_{\Delta}^{\ast}(2371)}%
${\small =T(1924)} & ? & $\pi${\small (1924), I=1,Q=1,0,-1}\\\hline
{\small 37q}$_{N}^{\ast}${\small (931)}$\overline{q_{\Delta}^{\ast}(2731)}%
${\small =T(2246)} & ? & $\pi${\small (2246), I=1,Q=1,0,-1}\\\hline
\end{tabular}
\label{Meson  with I=2}%
\end{equation}

\qquad\ \ \ \ \ \ {\small * The meson }$\chi${\small (1600)}$\ ${\small has
not been established. It still needs to be confirmed.}$\ $

\ \ \ \ \ \ \ \ \ \ \ \ \ \ \ \ \ \ \ \ \ \ \ \ \ \ \ \ \ \ \ \ \ \ \ \ \ \ 

\subsection{ Experimental Verification of The BCC Quark Model}

From the preceding predictions, we can find the key experiments that determine
whether the BCC Quark Model is a good modification of the Quark Model.

B1. The Quark Model, from 6 flavored quarks [u$_{2/3}(3$ Mev), d$_{-1/3}$(6
Mev), s$_{-1/3}(123$ Mev), c$_{2/3}(1.25$ Gev), b$_{-1/3}(4.2$ Gev), and
t$_{2/3}(174$ Gev)] and the formula meson = q$\overline{q},$ cannot give any
meson with isospin I = 2. However, the BCC Quark Model predicts many mesons
with I = 2 [see (\ref{Meson with I=2})]. If we can discover any meson with I =
2, we have shown that the BCC Quark Model is a good modification of the Quark Model.

B2. The Quark Model cannot give any meson with 100000
$>$%
M
$>$%
15000 Mev. However, the BCC Quark Model predicts many mesons with 100000%
$>$%
M
$>$%
15000 Mev, such as:
\begin{equation}%
\begin{tabular}
[c]{|l|}\hline
O(I=0) \ \ q(m)$\overline{q(m)}$ = Meson(M) \ (Quantum Numbers)\\\hline
(80)$\bullet$q$_{C}^{\ast}$(13791)$\overline{q_{C}^{\ast}(13791)}$ = $\psi
$(25596) (S = C = b =0, I = 0; Q = 0),\\\hline
(48)$\bullet$q$_{b}^{\ast}$(9951)$\overline{q_{b}^{\ast}(9951)}$ = $\Upsilon
$(17806) (S = C = b =0, I=0;Q=0),\\\hline
(48)$\bullet$q$_{b}^{\ast}$(15811)$\overline{q_{b}^{\ast}(15811)}$ =
$\Upsilon$(29597) (S = C = b =0, I=0;Q=0),\\\hline
(48)$\bullet$q$_{S}^{\ast}$(10031)$\overline{q_{S}^{\ast}(10031)}=\eta($17837)
(S = C = b =0, I = 0, Q = 0).\\\hline
\end{tabular}
\label{Meson With L-Mass}%
\end{equation}

If we can discover the mesons with 100000%
$>$%
M
$>$%
15000 Mev, we will show that the BCC Quark Model is a good modification of the
Quark Model. Thus, we propose to search for the above mesons. The discovery of
any one of the above mesons will provide strong support for the BCC Quark Model.

\subsection{Discussions}

C1. The BCC Quark Model deals mainly with the low energy state properties of
the baryons and the mesons. In the low energy cases, we shall consider the
periodic field of the body center cubic quark lattice. However, in the high
energy scattering cases of the baryons and mesons, because the strong
interactions (color) of the quarks are short range and saturable (a baryon
that is composed of three different colored quarks is a colorless system), we
can only consider the three quark system (the primitive cell approximation of
the BCC Quark Model \cite{NetXu (Section III)}){\small .} In this
approximation, we consider the excited quark (q$^{\ast}$) and the primitive
cell \textbf{(}$u\prime+d\prime$) only, omitting the quark lattice. Thus,
there are only three quarks in the system of a baryon: one excited quark
(q$^{\ast}$)\ and two accompanying excited quarks ($u\prime$\ and $d\prime$).
Similarly, there are only two quarks (q$_{i}^{\ast}\overline{\text{q}%
_{j}^{\ast}}$) in a meson. In other words, in high energy scattering cases,
the Quark Model is an excellent approximation of the BCC Quark Model. We do
not need to consider the whole BCC quark lattice.

C2. Similar to the discovery of new stars in the sky with the improvements of
the telescope, many new mesons and new baryons will be discovered with the
birth of new techniques.

C3. There are always some limitations for any physical theory. If the theory
is applied outside the limitations, it will not be useful. The Quantum Field
Theory is no exception. It is powerful in dealing with the point particles'
(the point model's) scattering problems; however, for the bound states, which
are constructed by many particles, it is not so powerful.\ Thus, for a long
time we have been looking forward to the day when a new theory will be born.
Not only will it deal with the scattering problems, but also it will solve the
bound state problems.\ The BCC Quark Model might play the role of providing a
hint along the path to a new theory, like the Bohr atom model did to quantum
mechanics. The new theory may be completely different from the quantum field
theory in mathematical form, as quantum mechanics is so different from classic
mechanics in the mathematical form.\ \ \ \ \ \ \ \ \ \ \ \ \ \ \ \ 

\section{Conclusions\ \ \ }

1. Using the phenomenological formula of the binding energies of the mesons,
from the quark spectrum (\ref{Quark-Spectrum}) and (\ref{Quark Quantum
Number}) of the BCC Quark Model, we deduce a meson spectrum that agrees well
with experimental results. These mass spectrum of mesons has not been obtained
by any other model, including the Quark Model.

2. The BCC Quark Model does not need the mixture of three quark-antiquark
pairs (u$\overline{u}$, d$\overline{d}$, and s$\overline{s})$ to explain the
mesons ($\eta,\varpi,\phi$, h, and f) with I = 0 and S=C=b=0 as the Quark
Model does. The BCC Quark Model really has enough quarks to construct the full
meson spectrum, according to the principle that a meson is made of a quark and
an antiquark. Thus, the BCC Quark Model saves this principle of the Quark Model.

3. The quarks q$_{S}^{\ast}$(1391), q$_{S}^{\ast}$(2551), and q$_{C}^{\ast}%
$(6591), that are predicted by the BCC Quark Model, might have been discovered
by experiments already.

4. The Quark Model assumes that there are five independent elementary quarks
(u, d, s, c, and b)--they are the five ground states of the quark spectrum of
the BCC\ Quark Model [see (\ref{Quark-Spectrum})]. The Quark Model uses only
these five quarks to explain the baryon spectrum and the meson spectrum.
Therefore, the Quark Model is an approximation of the five ground states of
the BCC Quark Model.

5. After the BCC Quark Model successfully deduces the baryon
spectrum{\small \ }\cite{NetXu (Baryons)}, it also successfully deduces the
meson spectrum. These results show that the vacuum material \cite{Wilczek}
really has the body center cubic symmetries (the point groups and the space group).

6. In this paper, we use only the energy bands and symmetries of the BCC quark
lattice (half of the results of the free particle approximation--0 order
approximation). We need to consider the symmetry wave functions (the other
half of the results of the free particle approximation) of the quarks to find
the angular momenta and parities of the quarks, the mesons, and the baryons
(see our next paper: Xin Yu and Jiao-Lin Xu, ``The Symmetry Quark Wave
Functions of the BCC Quark Model''). We will do higher order approximation later.

\begin{center}
\bigskip\textbf{Acknowledgment}
\end{center}

We sincerely thank Professor Robert L. Anderson for his valuable advice. We
also acknowledge\textbf{\ }our indebtedness to Professor D. P. Landau for his
help. One of the two authors, J. L. Xu, especially thanks Professor T. D. Lee,
for his \textit{particle physics} class in Beijing and his CUSPEA program that
gave him an opportunity to earn his Ph.D. He thanks Professor Y. S. Wu, H. Y.
Guo, and S. Chen \cite{XUarticle} very much for many very useful discussions.
We thank our friend Z. Y. Wu very much for his help in preparing this paper.
We especially thank our friend Fugao Wang for helping us post this paper on
the web.

Fig. 1. \ The first Brillouin zone of the body center cubic lattice. The
symmetry points and axes are indicated. The center of the first Brillouin zone
is at the point $\Gamma$ with 48 symmetry operations, the general ground quark
q$_{N}^{\ast}(931)$ will bear at this point. The $\Delta$-axis (the axis
$\Gamma$-H) is a $4$ fold rotation axis with 8 symmetry operations and the
strange number S = 0. The q$_{\Delta}^{\ast}$ quark families (q$_{\Delta
}^{5/3},$q$_{\Delta}^{2/3},$q$_{\Delta}^{-1/3},$q$_{\Delta}^{-4/3}$) will
appear on the axis. The $\Lambda$-axes (the axis $\Gamma$-P) and the F-axis
(the axis P-H) are $3$ fold rotation axes with 6 symmetry operations and the
strange number S =\ -1; the q$_{\Sigma}^{\ast}$ quark families (q$_{\Sigma
}^{2/3},$q$_{\Sigma}^{-1/3},$q$_{\Sigma}^{-4/3}$) and the q$_{\Lambda}^{\ast}$
quarks will appear on these axes. The $\Sigma$-axis (the axis $\Gamma$-N) and
the G-axis (the axis M-N) are $2$ fold rotation axes with 4 symmetry
operations and the strange number S = -2; the q$_{\Xi}^{\ast}$ quark families
(q$_{\Xi}^{-1/3},$q$_{\Xi}^{-4/3}$) will appear on these axes. The D-axis (the
axis P-N) is parallel to the $\Delta$-axis, S = 0, and the axis is a $2$ fold
rotation axis; the quark q$_{N}^{\ast}$ families (q$_{N}^{2/3}$, q$_{N}%
^{-1/3}$) will be on the axis.

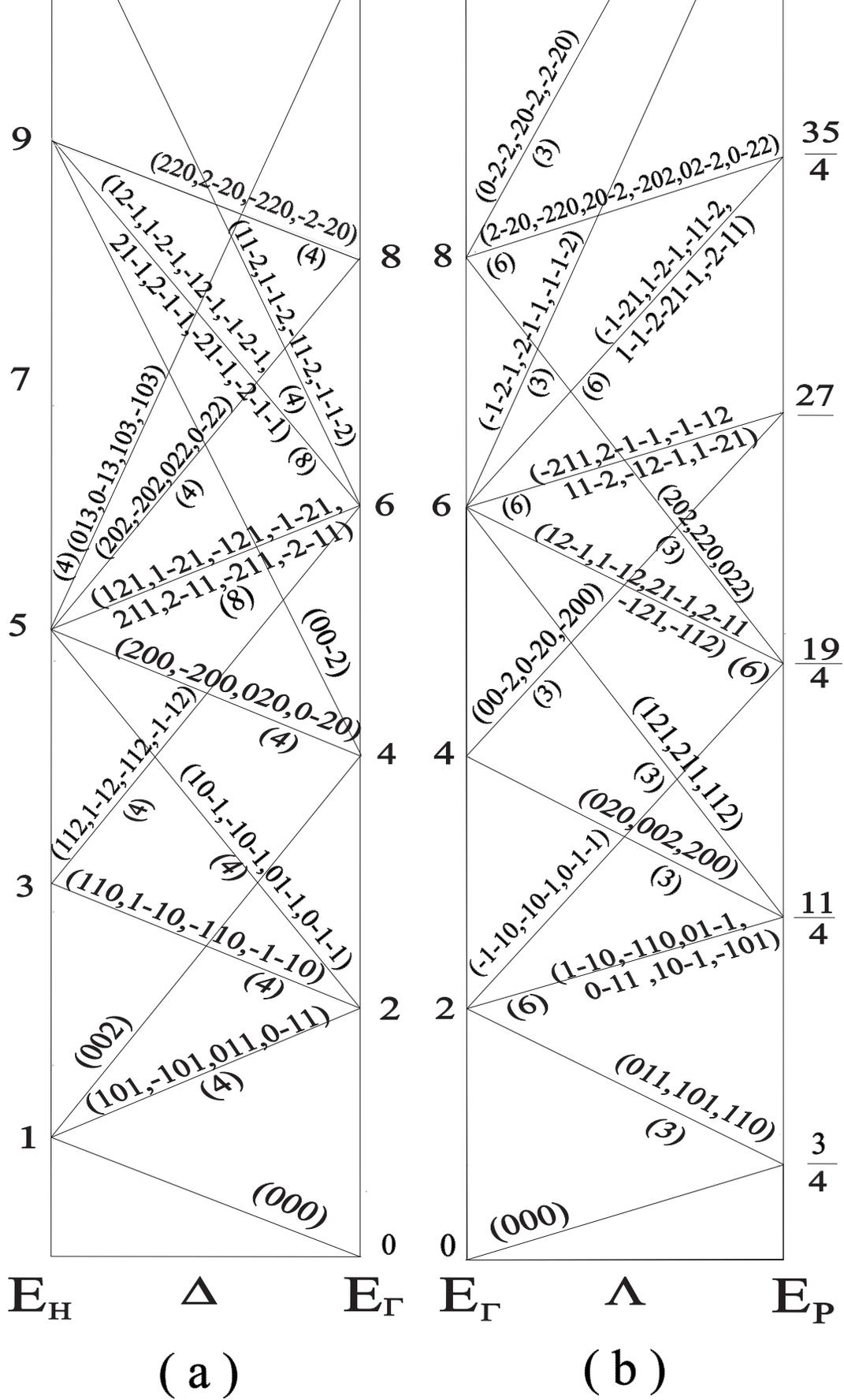
Fig. 2. \ \ (a) The energy bands on the $\Delta$-axis (the axis $\Gamma-$H).
The numbers above the lines are the values of $\vec{n}$ = ($n_{1}$, $n_{2}$,
$n_{3}$). The numbers under the lines are the fold numbers of the degeneracy.
E$_{\Gamma}$ is the value of E($\vec{k}$, $\vec{n}$) at the end point
$\Gamma,$ while E$_{H}$ is the value of E($\vec{k}$, $\vec{n}$) at the other
end point H. \ \ (b) The energy bands on the $\Lambda$-axis (the axis $\Gamma
$-P). E$_{\Gamma}$ is the value of E($\vec{k}$, $\vec{n}$) at the end point
$\Gamma,$ while E$_{P}$ is the value of E($\vec{k}$, $\vec{n}$) at the other
end point P. The numbers above the lines are the values of $\vec{n}$ =
($n_{1}$, $n_{2}$, $n_{3}$). The numbers under the lines are the fold numbers
of the degeneracy.

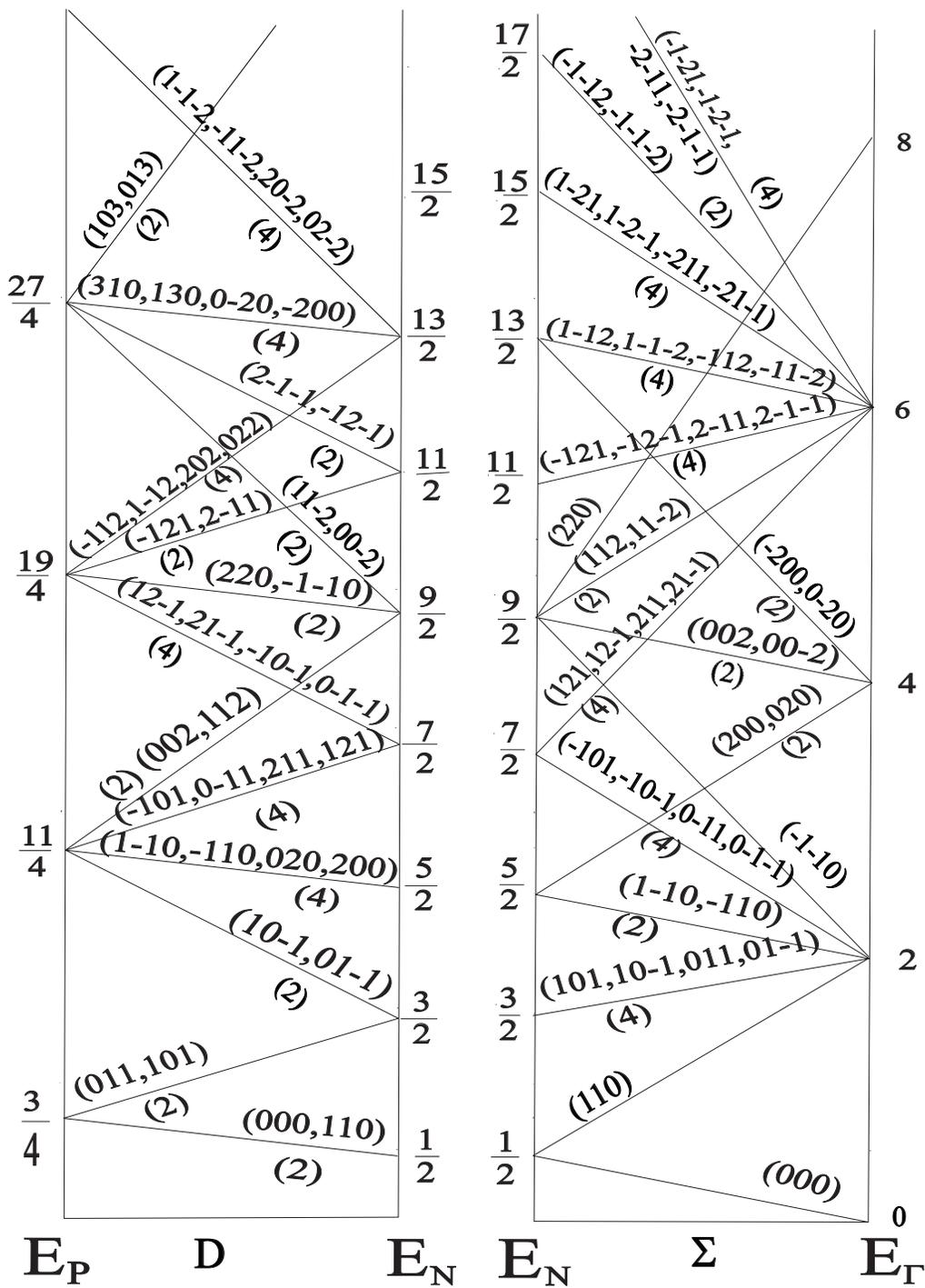
Fig. 3. \ \ (a) The energy bands on the $\Sigma$-axis (the axis $\Gamma$-N).
The numbers above the lines are the values of $\vec{n}$ = ($n_{1}$, $n_{2}$,
$n_{3}$). The numbers under the lines are the fold numbers of the degeneracy.
E$_{\Gamma}$ is the value of E($\vec{k}$, $\vec{n}$) at the end point
$\Gamma,$ while E$_{N}$ is the value of E($\vec{k}$, $\vec{n}$) at the other
end point N. \ (b) The energy bands on the D-axis (the axis P-N). E$_{P}$ is
the value of E($\vec{k}$, $\vec{n}$) at the end point P$,$ while E$_{N}$ is
the value of E($\vec{k}$, $\vec{n}$) at the other end point N. The numbers
above the lines are the values of $\vec{n}$ = ($n_{1}$, $n_{2}$, $n_{3}$). The
numbers under the lines are the fold numbers of the degeneracy.

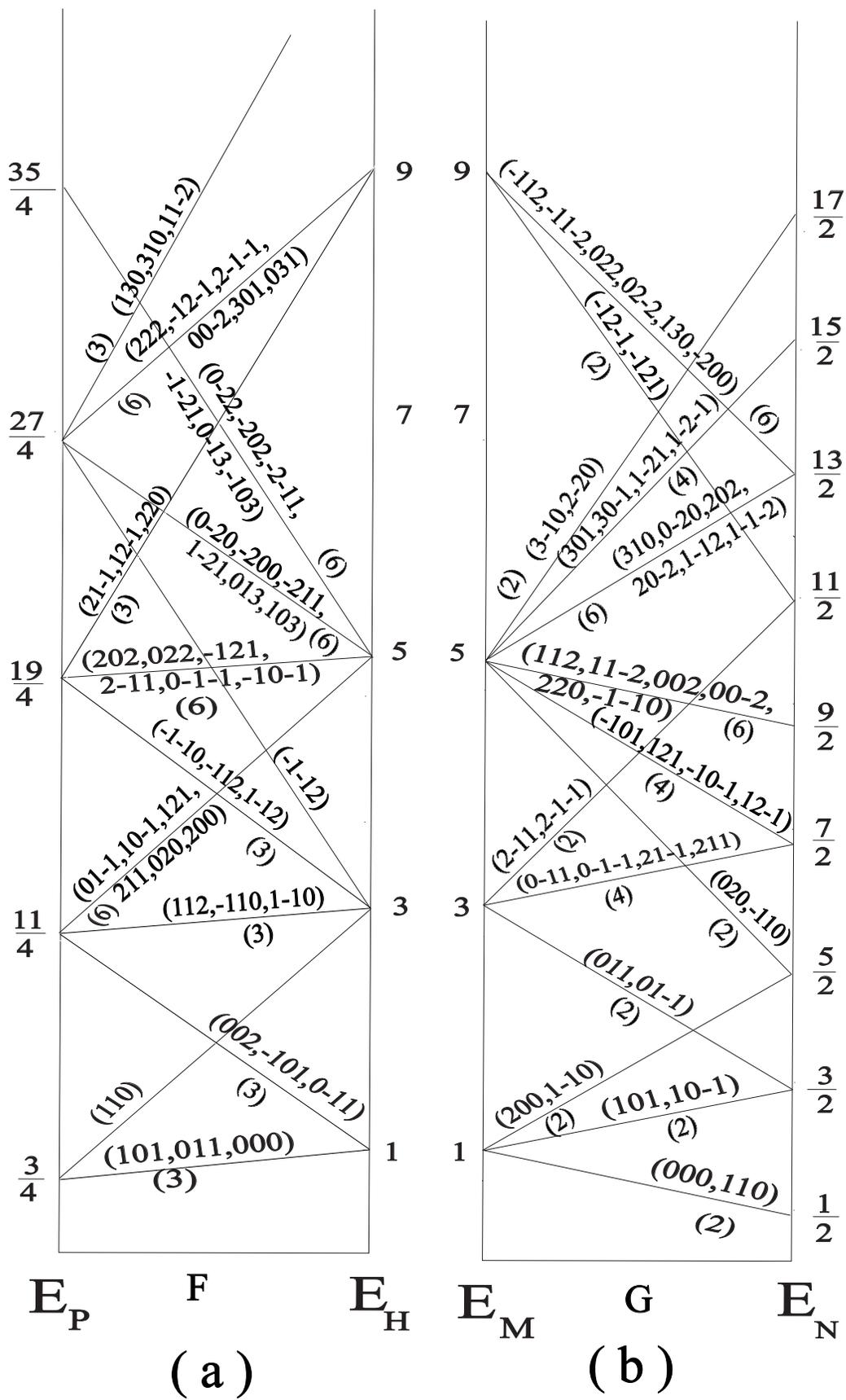
Fig. 4. \ \ (a) The energy bands on the $F$-axis (the axis P-H). The numbers
above the lines are the values of $\vec{n}$ = ($n_{1}$, $n_{2}$, $n_{3}$). The
numbers under the lines are the fold numbers of the degeneracy. E$_{P}$ is the
value of E($\vec{k}$, $\vec{n}$) at the end point $P,$ while E$_{H}$ is the
value of E($\vec{k}$, $\vec{n}$) at the other end point H.\ \ (b) The energy
bands on the G-axis (the axis M-N).\ E$_{M}$ is the value of E($\vec{k}$,
$\vec{n}$) at the end point M$,$ while E$_{N}$ is the value of E($\vec{k}$,
$\vec{n}$) at the other end point N. The numbers above the lines are the
values of $\vec{n}$ = ($n_{1}$, $n_{2}$, $n_{3}$). The numbers under the lines
are the fold numbers of the degeneracy.\ \ 

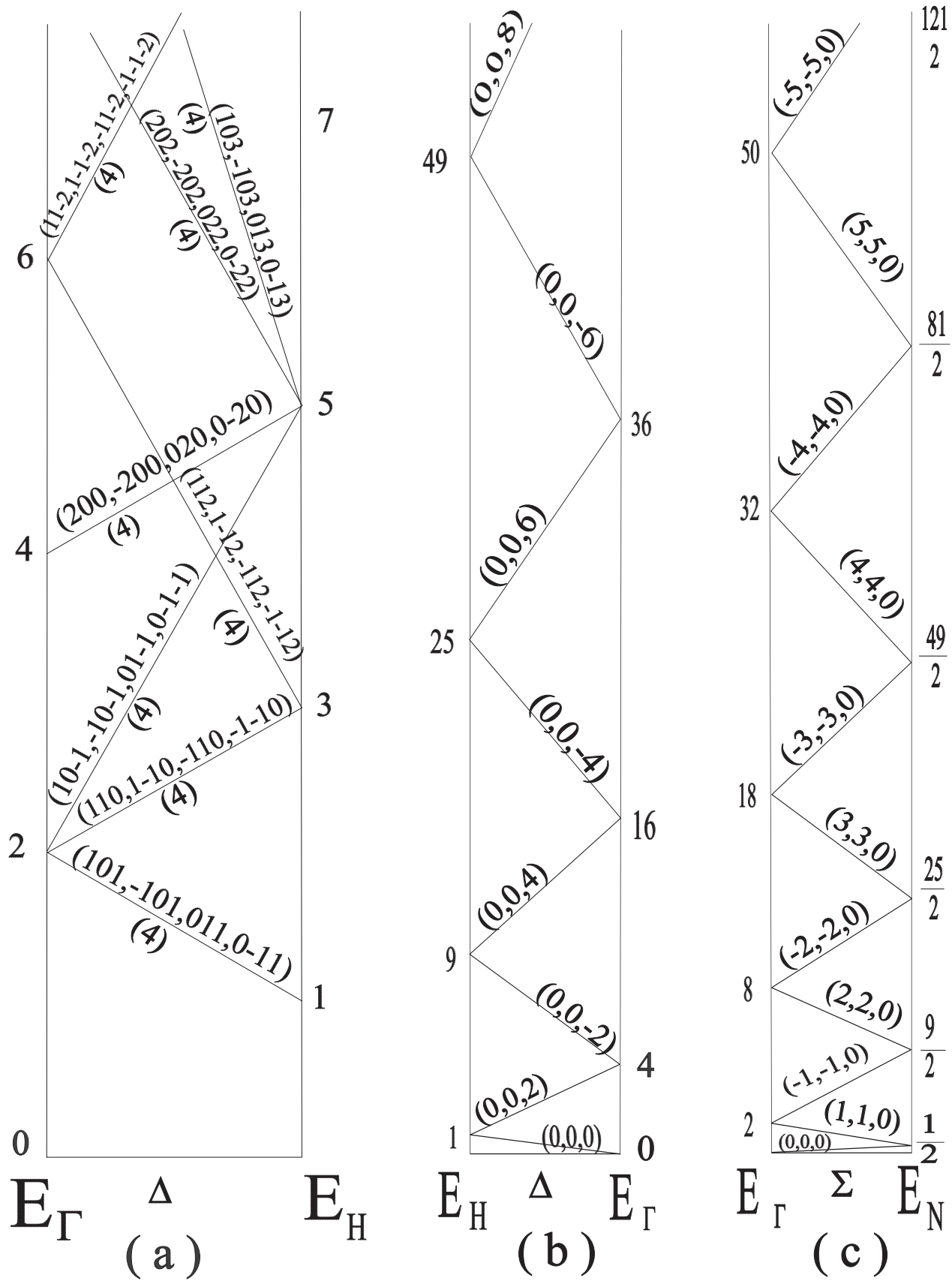
Fig. 5. \ \ (a) The $4$ fold degenerate energy bands (selected from Fig. 2(a))
on the $\Delta$ -axis (the axis $\Gamma$-H). The numbers above the lines are
the values of $\vec{n}$ (n$_{1}$, n$_{2}$, n$_{3}$). The numbers under the
lines are the numbers of the degeneracy of the energy bands. \ \ (b) The
single energy bands (selected from Fig. 2(a)) on the $\Delta$- axis (the axis
$\Gamma$-H). The numbers above the lines are the values of $\vec{n}$ (n$_{1}$,
n$_{2}$, n$_{3}$). \ \ (c) The single energy band (selected from Fig. 3(a)) on
the $\Sigma$-axis (the axis $\Gamma$-N). The numbers above the lines are the
values of $\vec{n}$ (n$_{1}$, n$_{2}$, n$_{3}$).

\newpage\qquad\ \ \qquad\qquad\qquad\qquad\qquad{\LARGE Appendix II \ }

\ \ \ \ \ \ \ {\Large The Comparing of the Theory Meson Spectrum }

\ \ \ \ \ \ \ \ \ \ \ \ \ \ \ \ \ \ \ \ \ \ \ {\Large with the Experimental Results}

\bigskip

\ \ \ \ \ \ \ \ \ \ \ \ \ \ \ \ \ \ \ \ \ \ \ \ \ \ \ \ Table I. \ Light
Unflavored Mesons

$%
\begin{tabular}
[c]{|l|l|l|}\hline
\ \ \ \ \ \ The BCC Quark Model & {\small Experiment} & {\small The Quark
Model}\\\hline
\ \ {\small [d}$_{q}${\small +d}$_{\overline{q}}${\small +O(I)]q}$_{i}^{\ast}%
${\small (m}$_{k}${\small )}$\overline{q_{j}^{\ast}(m_{l})}${\small =M(m)} &
{\small M{\tiny eason}(m)\ I} & \\\hline%
\begin{tabular}
[c]{l}%
\textbf{1+1+192}$\bullet$\textbf{q}$_{N}^{\ast}$\textbf{(931)}$\overline
{q_{N}^{\ast}(931)}$\textbf{=}$\pi$\textbf{(139)1}%
\end{tabular}
&
\begin{tabular}
[c]{l}%
$\bullet\pi${\small (139)1}%
\end{tabular}
& {\small u}$_{3}\overline{d_{6}}${\small ,u}$_{3}\overline{u_{3}}$%
{\small ,d}$_{6}\overline{d_{6}}$\\\hline%
\begin{tabular}
[c]{l}%
\textbf{1+1+96}$\bullet$\textbf{q}$_{S}^{\ast}$\textbf{(1111)}$\overline
{q_{S}^{\ast}(1111)}$\textbf{=}$\eta$\textbf{(549)0}%
\end{tabular}
{\small \ } &
\begin{tabular}
[c]{l}%
$\bullet\eta${\small (547)0}%
\end{tabular}
& {\small u}$_{3}\overline{u_{3}}${\small ,d}$_{6}\overline{d_{6}}$%
{\small ,s}$_{{\tiny 123}}\overline{{\small s}_{{\tiny 123}}}$\\\hline%
\begin{tabular}
[c]{l}%
\textbf{1+1+42}$\bullet$\textbf{q}$_{N}^{\ast}$\textbf{(931)}$\overline
{q_{N}^{\ast}(1201)}$\textbf{=}$\pi$\textbf{(780)1}%
\end{tabular}
&
\begin{tabular}
[c]{l}%
$\bullet\rho${\small (770)1}%
\end{tabular}
{\small \ \ \ } & {\small u}$_{3}\overline{d_{6}}${\small ,u}$_{3}%
\overline{u_{3}}${\small ,d}$_{6}\overline{d_{6}}$\\\hline%
\begin{tabular}
[c]{l}%
\textbf{1+1+84q}$_{N}^{\ast}$\textbf{(931)}$\overline{q_{N}^{\ast}(1201)}%
$\textbf{=}$\eta$\textbf{(780)0}%
\end{tabular}
&
\begin{tabular}
[c]{l}%
$\bullet\omega${\small (782)\ 0}%
\end{tabular}
& {\small u}$_{3}\overline{u_{3}}${\small ,d}$_{6}\overline{d_{6}}$%
{\small ,s}$_{{\tiny 123}}\overline{{\small s}_{{\tiny 123}}}$\\\hline%
\begin{tabular}
[c]{l}%
\textbf{1+1+32}$\bullet$\textbf{q}$_{N}^{\ast}$\textbf{(1201)}$\overline
{q_{N}^{\ast}(1201)}$\textbf{=}$\varpi$\textbf{(813)0}%
\end{tabular}
&
\begin{tabular}
[c]{l}%
{\small f}$_{0}${\small ({\tiny 400-1200})0}%
\end{tabular}
& {\small u}$_{3}\overline{u_{3}}${\small ,d}$_{6}\overline{d_{6}}$%
{\small ,s}$_{{\tiny 123}}\overline{{\small s}_{{\tiny 123}}}$\\\hline%
\begin{tabular}
[c]{l}%
\textbf{1+1+48}$\bullet$\textbf{q}$_{S}^{\ast}$\textbf{(1391)}$\overline
{q_{S}^{\ast}{\small (}\text{1391)}}$\textbf{=}$\eta$\textbf{(952)0}%
\end{tabular}
&
\begin{tabular}
[c]{l}%
$\bullet\eta^{\prime}${\small (958)\ 0}%
\end{tabular}
& {\small u}$_{3}\overline{u_{3}}${\small ,d}$_{6}\overline{d_{6}}$%
{\small ,s}$_{{\tiny 123}}\overline{{\small s}_{{\tiny 123}}}$\\\hline%
\begin{tabular}
[c]{l}%
{\small 1+1+24}$\bullet${\small q}$_{\Sigma}^{\ast}${\small (1201)}%
$\overline{q_{\Sigma}^{\ast}(1201)}${\small =}$\eta${\small (962)0}\\
\textbf{1+1+32}$\bullet$\textbf{q}$_{\Delta}^{\ast}$\textbf{(1291)}%
$\overline{q_{\Delta}^{\ast}(1291)}$\textbf{=}$\eta$\textbf{(967)0}%
\end{tabular}
&
\begin{tabular}
[c]{l}%
$\bullet${\small f}$_{0}${\small (980)0}%
\end{tabular}
& {\small u}$_{3}\overline{u_{3}}${\small ,d}$_{6}\overline{d_{6}}$%
{\small ,s}$_{{\tiny 123}}\overline{{\small s}_{{\tiny 123}}}$\\\hline%
\begin{tabular}
[c]{l}%
\textbf{1+1+56}$\bullet$\textbf{q}$_{N}^{\ast}$\textbf{(931)}$\overline
{q_{\Delta}^{\ast}(1291)}$\textbf{=}$\pi$\textbf{(960)1}\\
{\small 1+1+24}$\bullet$q$_{\Delta}^{\ast}${\small (1291)}$\overline
{q_{\Delta}^{\ast}(1291)}${\small =}$\pi${\small (967)1}\\
{\small 1+1+32}$\bullet${\small q}$_{\Sigma}^{\ast}${\small (1201)}%
$\overline{q_{\Sigma}^{\ast}(1201)}${\small =}$\pi${\small (979)1\ }\\
{\small 1+1+42}$\bullet${\small q}$_{N}^{\ast}${\small (931)}$\overline
{{\small q}_{N}^{\ast}{\small (}\text{{\small 1471}}{\small )}}${\small =}%
$\pi${\small (1021)1}%
\end{tabular}
&
\begin{tabular}
[c]{l}%
$\bullet${\small a}$_{0}${\small (980)1}%
\end{tabular}
& {\small u}$_{3}\overline{d_{6}}${\small ,u}$_{3}\overline{u_{3}}$%
{\small ,d}$_{6}\overline{d_{6}}$\\\hline%
\begin{tabular}
[c]{l}%
\textbf{1+1+84}$\bullet$\textbf{q}$_{N}^{\ast}$\textbf{(931)}$\overline
{{\small q}_{N}^{\ast}{\small (1471)}}$\textbf{=}$\eta$\textbf{(1021)0}%
\end{tabular}
&
\begin{tabular}
[c]{l}%
$\bullet\phi${\small (1020)0}%
\end{tabular}
& {\small u}$_{3}\overline{u_{3}}${\small ,d}$_{6}\overline{d_{6}}$%
{\small ,s}$_{{\tiny 123}}\overline{{\small s}_{{\tiny 123}}}$\\\hline%
\begin{tabular}
[c]{l}%
\textbf{1+1+54}$\bullet$\textbf{q}$_{S}^{\ast}$\textbf{(1111)}$\overline
{\text{q}_{S}^{\ast}\text{({\small 1391})}}$\textbf{=}$\eta$\textbf{(1204)0}%
\end{tabular}
{\small \ } &
\begin{tabular}
[c]{l}%
$\bullet${\small h}$_{1}${\small (1170)0}%
\end{tabular}
& {\small u}$_{3}\overline{u_{3}}${\small ,d}$_{6}\overline{d_{6}}$%
{\small ,s}$_{{\tiny 123}}\overline{{\small s}_{{\tiny 123}}}$\\\hline%
\begin{tabular}
[c]{l}%
{\small 1+1+30}$\bullet${\small q}$_{S}^{\ast}${\small (1111)}$\overline
{{\small q}_{\Sigma}^{\ast}{\small (1201)}}${\small =}$\pi${\small (1206)1}\\
\textbf{1+2+56}$\bullet$\textbf{q}$_{N}^{\ast}$\textbf{(931)}$\overline
{{\small q}_{\Delta}^{\ast}{\small (1651)}}$\textbf{=}$\pi$\textbf{(1282)1}%
\end{tabular}
&
\begin{tabular}
[c]{l}%
$\bullet${\small b}$_{1}${\small (1235)1}\\
$\bullet${\small a}$_{1}${\small (1260)1}\\
$\bullet\overline{\pi}$\textbf{(1248)1}%
\end{tabular}
& {\small u}$_{3}\overline{d_{6}}${\small ,u}$_{3}\overline{u_{3}}$%
{\small ,d}$_{6}\overline{d_{6}}$\\\hline%
\begin{tabular}
[c]{l}%
\textbf{1+1+32}$\bullet$\textbf{q}$_{N}^{\ast}$\textbf{(1471)}$\overline
{\text{q}_{N}^{\ast}\text{({\small 1471)}}}$\textbf{=}$\eta$\textbf{(1269)0}%
\end{tabular}
&
\begin{tabular}
[c]{l}%
$\bullet${\small f}$_{2}${\small (1270)0}\\
$\bullet${\small f}$_{1}${\small (1285)0}\\
$\bullet\eta${\small (1295)0}\\
$\bullet\overline{\eta}$\textbf{(1283)0}%
\end{tabular}
& {\small u}$_{3}\overline{u_{3}}${\small ,d}$_{6}\overline{d_{6}}$%
{\small ,s}$_{{\tiny 123}}\overline{{\small s}_{{\tiny 123}}}$\\\hline%
\begin{tabular}
[c]{l}%
\textbf{1}$\times$\textbf{2}$\times$\textbf{42}$\bullet$\textbf{q}$_{N}^{\ast
}$\textbf{(931)}$\overline{\text{q}_{N}^{\ast}\text{{\small (1831)}}}%
$\textbf{=}$\pi$\textbf{(1342)1}%
\end{tabular}
&
\begin{tabular}
[c]{l}%
$\bullet\pi${\small (1300)1}\\
$\bullet a_{2}${\small (1320)1}\\
$\pi_{1}${\small (1400)1}\ \\
$\bullet\overline{\pi}$\textbf{(1340)1}%
\end{tabular}
& {\small u}$_{3}\overline{d_{6}}${\small ,u}$_{3}\overline{u_{3}}$%
{\small ,d}$_{6}\overline{d_{6}}$\\\hline
\end{tabular}%
\begin{tabular}
[c]{l}
\end{tabular}
$

\bigskip\
\begin{tabular}
[c]{|l|l|l|}\hline%
\begin{tabular}
[c]{l}%
\textbf{1+2+84}$\bullet$\textbf{q}$_{N}^{\ast}$\textbf{(931)}$\overline
{\text{{\small q}}_{N}^{\ast}\text{{\small (1831)}}}$\textbf{=}$\eta
$\textbf{(1342)0}%
\end{tabular}
&
\begin{tabular}
[c]{l}%
$\bullet$f$_{0}${\small (1370)0}\\
h$_{1}{\small (1380)0}$\\
$\bullet\overline{\eta}$\textbf{(1375)0}%
\end{tabular}
& {\small u}$_{3}\overline{u_{3}}${\small ,d}$_{6}\overline{d_{6}}$%
{\small ,s}$_{{\tiny 123}}\overline{{\small s}_{{\tiny 123}}}$\\\hline%
\begin{tabular}
[c]{l}%
\textbf{1+4+84}$\bullet$\textbf{q}$_{N}^{\ast}$\textbf{(931)}$\overline
{{\small q}_{N}^{\ast}\text{{\small (1921)}}}$\textbf{=}$\eta$\textbf{(1423)0}%
\\
{\small 1+3+24}$\bullet${\small q}$_{\Delta}^{\ast}${\small (1291)}%
$\overline{{\small q}_{\Delta}^{\ast}\text{{\small (1651)}}}${\small =}$\eta
${\small (1473)0}%
\end{tabular}
&
\begin{tabular}
[c]{l}%
$\bullet$f$_{1}${\small (1420)0}\\
$\bullet\omega${\small (1420)0}\\
f$_{2}${\small (1430)0}\\
$\bullet\eta{\small (1440)0}$\\
$\bullet\overline{\eta}$\textbf{(1428)0}%
\end{tabular}
& {\small u}$_{3}\overline{u_{3}}${\small ,d}$_{6}\overline{d_{6}}$%
{\small ,s}$_{{\tiny 123}}\overline{{\small s}_{{\tiny 123}}}$\\\hline%
\begin{tabular}
[c]{l}%
\textbf{1+4+42}$\bullet$\textbf{q}$_{N}^{\ast}$\textbf{(931)}$\overline
{\text{{\small q}}_{N}^{\ast}\text{{\small (1921)}}}$\textbf{=}$\pi
$\textbf{(1423)1}\\
{\small 2+2+24}$\bullet$q$_{\Delta}^{\ast}${\small (1651)}$\overline
{\text{{\small q}}_{\Delta}^{\ast}\text{{\small (1651)}}}${\small =}$\pi
${\small (1565)1}%
\end{tabular}
&
\begin{tabular}
[c]{l}%
$\bullet a_{0}${\small (1450)1}\\
$\bullet\rho${\small (1450)1}\\
$\bullet\overline{\pi}$\textbf{(1450)1}%
\end{tabular}
& {\small u}$_{3}\overline{d_{6}}${\small ,u}$_{3}\overline{u_{3}}$%
{\small ,d}$_{6}\overline{d_{6}}$\\\hline%
\begin{tabular}
[c]{l}%
\textbf{2}$+$\textbf{2}$+$\textbf{32}$\bullet$\textbf{q}$_{\Delta}^{\ast}%
$\textbf{(1651)}$\overline{{\small q}_{\Delta}^{\ast}\text{{\small (1651)}}}%
$\textbf{=}$\eta$\textbf{(1565)0}%
\end{tabular}
&
\begin{tabular}
[c]{l}%
$\bullet$f$_{0}${\small (1500)0}\\
f$_{1}${\small (1510)0}\\
$\bullet$f$_{2}^{^{\prime}}${\small (1525)0}\\
\ f$_{2}${\small (1565)0}\\
$\bullet\overline{\eta}$\textbf{(1525)0}%
\end{tabular}
& {\small u}$_{3}\overline{u_{3}}${\small ,d}$_{6}\overline{d_{6}}$%
{\small ,s}$_{{\tiny 123}}\overline{{\small s}_{{\tiny 123}}}$\\\hline%
\begin{tabular}
[c]{l}%
\textbf{1+1+56}$\bullet$\textbf{q}$_{N}^{\ast}$\textbf{(931)}$\overline
{{\small q}_{\Delta}^{\ast}\text{{\small (2011)}}}=\pi$\textbf{(1603)1}\\
{\small 1+3+30}$\bullet${\small q}$_{S}^{\ast}${\small (1111)}$\overline
{{\small q}_{\Sigma}^{\ast}\text{{\small (1651)}}}${\small =}$\pi
${\small (1616)1}%
\end{tabular}
& $\
\begin{tabular}
[c]{l}%
$\pi_{1}$\textbf{(1600)0}%
\end{tabular}
$ & {\small u}$_{3}\overline{d_{6}}${\small ,u}$_{3}\overline{u_{3}}%
${\small ,d}$_{6}\overline{d_{6}}$\\\hline%
\begin{tabular}
[c]{l}%
\textbf{1+1+37}$\bullet$\textbf{q}$_{N}^{\ast}$\textbf{(931)}$\overline
{q_{\Delta}^{\ast}(2011)}$\textbf{=T(1603)2}%
\end{tabular}
&
\begin{tabular}
[c]{l}%
$\chi$\textbf{(1600)2}%
\end{tabular}
$\ $ & \ \ \ \ \ \ \ \ \ \ {\small ?}\\\hline%
\begin{tabular}
[c]{l}%
\textbf{1+2+84}$\bullet$\textbf{q}$_{N}^{\ast}$\textbf{(931)}$\overline
{\text{{\small q}}_{N}^{\ast}\text{{\small (2191)}}}$\textbf{=}$\eta
$\textbf{(1664)0}\\
{\small 1+1+30}$\bullet$q$_{S}^{\ast}${\small (1111)}$\overline
{\text{{\small q}}_{S}^{\ast}\text{{\small (2011)}}}$=$\eta${\small (1669)0}%
\end{tabular}
&
\begin{tabular}
[c]{l}%
f$_{2}${\small (1640)0}\\
$\eta_{2}${\small (1645)0}\\
$\bullet\omega{\small (1650)0}$\\
$\chi${\small (1650)0}\\
$\bullet\omega_{3}${\small (1670)0}\\
$\bullet\phi_{2}${\small (1680)0}\\
$\bullet\overline{\eta}$\textbf{(1656)0}%
\end{tabular}
& {\small u}$_{3}\overline{u_{3}}${\small ,d}$_{6}\overline{d_{6}}$%
{\small ,s}$_{{\tiny 123}}\overline{{\small s}_{{\tiny 123}}}$\\\hline%
\begin{tabular}
[c]{l}%
\textbf{1+2+42}$\bullet$\textbf{q}$_{N}^{\ast}($\textbf{931)}$\overline
{{\small q}_{N}^{\ast}{\small (2191)}}$\textbf{=}$\pi$\textbf{(1664)1}%
\end{tabular}
&
\begin{tabular}
[c]{l}%
$a_{1}${\small (1640)1}\\
a$_{2}${\small (1660)1}\\
$\bullet\pi_{2}${\small (1670)1}\\
$\bullet\rho_{3}${\small (1690)1}\\
$\bullet\rho${\small (1700)1}\\
$\bullet\overline{\pi}$\textbf{(1672)1}%
\end{tabular}
& {\small u}$_{3}\overline{d_{6}}${\small ,u}$_{3}\overline{u_{3}}$%
{\small ,d}$_{6}\overline{d_{6}}$\\\hline
\end{tabular}%
$\backslash$%

\bigskip%
\begin{tabular}
[c]{|l|l|l|}\hline%
\begin{tabular}
[c]{l}%
{\small 3}$+${\small 3}$+${\small 24}$\bullet${\small q}$_{\Sigma}^{\ast}%
${\small (1651)}$\overline{q_{\Sigma}^{\ast}(1651)}${\small =}$\eta
${\small (1758)0}\\
\textbf{1}$+$\textbf{1}$+$\textbf{45}$\bullet$\textbf{q}$_{S}^{\ast}%
$\textbf{(1111)}$\overline{q_{S}^{\ast}(2011)}$\textbf{=}$\eta$%
\textbf{(1768)0}%
\end{tabular}
&
\begin{tabular}
[c]{l}%
$\bullet${\small f}$_{0}${\small (1710)0}\\
$\eta${\small (1760)0}\\
f$_{2}${\small (1810)0}\\
$\bullet\overline{\eta}$\textbf{(1760)0}%
\end{tabular}
& {\small u}$_{3}\overline{u_{3}}${\small ,d}$_{6}\overline{d_{6}}$%
{\small ,s}$_{{\tiny 123}}\overline{{\small s}_{{\tiny 123}}}$\\\hline
\
\begin{tabular}
[c]{l}%
\textbf{1+5}$+$\textbf{30}$\bullet$\textbf{q}$_{S}^{\ast}$\textbf{(1111)}%
$\overline{\text{{\small q}}_{\Sigma}^{\ast}\text{{\small (1921)}}}$%
\textbf{=}$\pi$\textbf{(1861)1}%
\end{tabular}
&
\begin{tabular}
[c]{l}%
a$_{2}${\small (1750)1}\\
$\chi${\small (1775)1}\\
$\bullet\pi${\small (1800)1}\\
$\bullet\overline{\pi}$\textbf{(1775)1}%
\end{tabular}
& {\small u}$_{3}\overline{d_{6}}${\small ,u}$_{3}\overline{u_{3}}$%
{\small ,d}$_{6}\overline{d_{6}}$\\\hline%
\begin{tabular}
[c]{l}%
\textbf{2}$+$\textbf{2}$+$\textbf{32}$\bullet$\textbf{q}$_{N}^{\ast}%
$\textbf{(1831)}$\overline{q_{N}^{\ast}(1831)}$\textbf{=}$\eta$%
\textbf{(1852)0}%
\end{tabular}
&
\begin{tabular}
[c]{l}%
$\bullet\phi_{3}${\small (1850)0}\\
$\eta_{2}${\small (1870)0}\\
$\bullet\overline{\eta}$\textbf{(1860)0}%
\end{tabular}
& {\small u}$_{3}\overline{u_{3}}${\small ,d}$_{6}\overline{d_{6}}$%
{\small ,s}$_{{\tiny 123}}\overline{{\small s}_{{\tiny 123}}}$\\\hline%
\begin{tabular}
[c]{l}%
{\small 1+1+54}$\bullet${\small q}$_{S}^{\ast}${\small (1111)}$\overline
{q_{S}^{\ast}(2551)}${\small =}$\eta${\small (1960)0}\\
\textbf{1}$+$\textbf{2}$+$\textbf{84}$\bullet$\textbf{q}$_{N}^{\ast}%
$\textbf{(931)}$\overline{q_{N}^{\ast}(2551)}$\textbf{=}$\eta$\textbf{(1985)0}%
\\
{\small 4}$+${\small 4}$+${\small 32}$\bullet${\small q}$_{N}^{\ast}%
(${\small 1921)}$\overline{q_{N}^{\ast}(1921)}${\small =}$\eta$%
{\small (1993)0}%
\end{tabular}
&
\begin{tabular}
[c]{l}%
$\chi${\small (1910)0}\\
f$_{2}${\small (1950)0}\\
$\overline{\mathbf{\eta}}$\textbf{(1930)0}%
\end{tabular}
& {\small u}$_{3}\overline{u_{3}}${\small ,d}$_{6}\overline{d_{6}}$%
{\small ,s}$_{{\tiny 123}}\overline{{\small s}_{{\tiny 123}}}$\\\hline%
\begin{tabular}
[c]{l}%
\textbf{1}$+$\textbf{1}$+$\textbf{56}$\bullet$\textbf{q}$_{N}^{\ast}%
($\textbf{931)}$\overline{q_{\Delta}^{\ast}(\text{2371})}$\textbf{=}$\pi
$\textbf{(1924)1}\\
{\small 1+2}$+${\small 30}$\bullet${\small q}$_{S}^{\ast}${\small (1111)}%
$\overline{q_{\Sigma}^{\ast}(2011)}${\small =}$\pi${\small (1943)1}\\
{\small 1+2+42}$\bullet${\small q}$_{N}^{\ast}${\small (931)}$\overline
{q_{N}^{\ast}(2551)}${\small =}$\pi${\small (1985)1}%
\end{tabular}
&
\begin{tabular}
[c]{l}%
$\chi$\textbf{(2000)1}%
\end{tabular}
& {\small u}$_{3}\overline{d_{6}}${\small ,u}$_{3}\overline{u_{3}}$%
{\small ,d}$_{6}\overline{d_{6}}$\\\hline%
\begin{tabular}
[c]{l}%
\textbf{1+3+42}$\bullet$\textbf{q}$_{N}^{\ast}$\textbf{(931)}$\overline
{q_{N}^{\ast}(2641)}$\textbf{=}$\pi$\textbf{(2065)1}%
\end{tabular}
&
\begin{tabular}
[c]{l}%
$\bullet${\small a}$_{4}${\small (2040) 1}%
\end{tabular}
& {\small u}$_{3}\overline{d_{6}}${\small ,u}$_{3}\overline{u_{3}}$%
{\small ,d}$_{6}\overline{d_{6}}$\\\hline
$%
\begin{tabular}
[c]{l}%
\textbf{1+3+84}$\bullet$\textbf{q}$_{N}^{\ast}$\textbf{(931)}$\overline
{q_{N}^{\ast}(2641)}$\textbf{=}$\eta$\textbf{(2065)0}\\
{\small 1+1+24}$\bullet${\small q}$_{\Delta}^{\ast}${\small (1291)}%
$\overline{q_{\Delta}^{\ast}(2371)}${\small =}$\eta${\small (2086)0}%
\end{tabular}
$ &
\begin{tabular}
[c]{l}%
$\bullet${\small f}$_{2}${\small (2010)0}\\
{\small f}$_{0}${\small (2020) 0}\\
$\bullet${\small f}$_{4}${\small (2050)0}\\
{\small f}$_{0}${\small (2060)0}\\
$\bullet\overline{\eta}$\textbf{(2035)0}%
\end{tabular}
& {\small u}$_{3}\overline{u_{3}}${\small ,d}$_{6}\overline{d_{6}}$%
{\small ,s}$_{{\tiny 123}}\overline{{\small s}_{{\tiny 123}}}$\\\hline%
\begin{tabular}
[c]{l}%
{\small 1+1+32}$\bullet$q$_{\Delta}^{\ast}${\small (2011)}$\overline
{q_{\Delta}^{\ast}(2011)}${\small =}$\eta${\small (2132)0}\\
\textbf{1+2+84}$\bullet$\textbf{q}$_{N}^{\ast}$\textbf{(931)}$\overline
{q_{N}^{\ast}(2731)}$\textbf{=}$\eta$\textbf{(2146)0}\\
{\small 1+1+45}$\bullet$q$_{S}^{\ast}$(1111)$\overline{\text{q}_{S}^{\ast
}\text{(2451)}}${\small =}$\eta${\small (2168)0}\\
{\small 5+5+24}$\bullet${\small q}$_{\Sigma}^{\ast}${\small (1921)}%
$\overline{q_{\Sigma}^{\ast}(1921)}${\small =}$\eta${\small (2220)0}%
\end{tabular}
&
\begin{tabular}
[c]{l}%
{\small f}$_{2}${\small (2150) 0}\\
{\small f}$_{0}${\small (2200) 0}\\
{\small f}$_{j}${\small (2220) 0}\\
$\eta${\small (2225) 0}\\
$\overline{\eta}$\textbf{(2199)0}%
\end{tabular}
& {\small u}$_{3}\overline{u_{3}}${\small ,d}$_{6}\overline{d_{6}}$%
{\small ,s}$_{{\tiny 123}}\overline{{\small s}_{{\tiny 123}}}$\\\hline%
\begin{tabular}
[c]{l}%
{\small 1+1+24}$\bullet$q$_{\Delta}^{\ast}${\small (2011)}$\overline
{q_{\Delta}^{\ast}(2011)}${\small =}$\pi${\small (2132)1}\\
\textbf{1+2+42}$\bullet$\textbf{q}$_{N}^{\ast}$\textbf{(931)}$\overline
{q_{N}^{\ast}(2731)}$\textbf{=}$\pi$\textbf{(2146)1}%
\end{tabular}
&
\begin{tabular}
[c]{l}%
$\pi_{2}${\small (2100) 1}\\
$\rho${\small (2150) 1}\\
$\overline{\mathbf{\pi}}$\textbf{(2125)1}%
\end{tabular}
& {\small u}$_{3}\overline{d_{6}}${\small ,u}$_{3}\overline{u_{3}}$%
{\small ,d}$_{6}\overline{d_{6}}$\\\hline
\end{tabular}

\ \ \ \ \ \ \ \ \ \ \ \ \ \ \ \ \ \ \ \ \ \ \ \ \ \ \ \ \ \ \ \ \ \ \ 

\ \ \ \ \ \ \ \ \ \ \ \ \ \ \ \ \ \ \ \ \ \ \ \ \ \ %

\begin{tabular}
[c]{|l|l|l|}\hline%
\begin{tabular}
[c]{l}%
\textbf{1}$+$\textbf{4}$+$\textbf{56}$\bullet$\textbf{q}$_{N}^{\ast}%
$\textbf{(931)}$\overline{q_{\Delta}^{\ast}(2731)}$\textbf{=}$\pi
$\textbf{(2246)1}\\
{\small 1+1+30}$\bullet$q$_{S}^{\ast}$(1111)$\overline{q_{\Sigma}^{\ast
}(2371)}${\small =}$\pi${\small (2271)1}%
\end{tabular}
&
\begin{tabular}
[c]{l}%
$\rho_{3}$\textbf{(2250)1}%
\end{tabular}
& {\small u}$_{3}\overline{d_{6}}${\small ,u}$_{3}\overline{u_{3}}$%
{\small ,d}$_{6}\overline{d_{6}}$\\\hline%
\begin{tabular}
[c]{l}%
{\small 1+1+24}$\bullet${\small q}$_{S}^{\ast}${\small (2011)}$\overline
{q_{S}^{\ast}(2011)}${\small =}$\eta${\small (2316)0}\\
{\small 2+2+24}$\bullet${\small q}$_{\Sigma}^{\ast}${\small (2011)}%
$\overline{q_{\Sigma}^{\ast}(2011)}${\small =}$\eta${\small (2371)0}\\
\textbf{1+3+45}$\bullet$\textbf{q}$_{S}^{\ast}$\textbf{(1111)}$\overline
{q_{S}^{\ast}(2641)}$\textbf{=}$\eta$\textbf{(2341)0}%
\end{tabular}
&
\begin{tabular}
[c]{l}%
$\bullet${\small f}$_{2}${\small (2300)0}\\
{\small f}$_{4}${\small (2300)0}\\
$\bullet${\small f}$_{2}${\small (2340)0}\\
$\bullet\overline{\eta}$\textbf{(2313)0}%
\end{tabular}
& {\small u}$_{3}\overline{u_{3}}${\small ,d}$_{6}\overline{d_{6}}$%
{\small ,s}$_{{\tiny 123}}\overline{{\small s}_{{\tiny 123}}}$\\\hline%
\begin{tabular}
[c]{l}%
\\
\textbf{1+3+30}$\bullet$\textbf{q}$_{S}^{\ast}$\textbf{(1111)}$\overline
{q_{\Sigma}^{\ast}(2551)}$\textbf{=}$\pi$\textbf{(2434)1}%
\end{tabular}
&
\begin{tabular}
[c]{l}%
$\rho_{5}${\small (2350)1}\\
{\small a}$_{6}${\small (2450)1}\\
$\overline{\pi}$\textbf{(2400)1}%
\end{tabular}
& {\small u}$_{3}\overline{d_{6}}${\small ,u}$_{3}\overline{u_{3}}$%
{\small ,d}$_{6}\overline{d_{6}}$\\\hline
$%
\begin{tabular}
[c]{l}%
{\small 1+3+24}$\bullet${\small q}$_{\Delta}^{\ast}${\small (1291)}%
$\overline{q_{\Delta}^{\ast}(1293)}${\small =}$\eta${\small (2392)0}\\
{\small 2+2+32}$\bullet$q$_{N}^{\ast}$(2191)$\overline{q_{N}^{\ast}(2191)}%
${\small =}$\eta${\small (2405)0}\\
\textbf{1+1+45}$\bullet$\textbf{q}$_{S}^{\ast}$\textbf{(1111)}$\overline
{q_{S}^{\ast}(2731)}$\textbf{=}$\eta$\textbf{(2423)0}%
\end{tabular}
$ &
\begin{tabular}
[c]{l}%
\textbf{f}$_{6}$\textbf{(2510) 0}%
\end{tabular}
& {\small u}$_{3}\overline{u_{3}}${\small ,d}$_{6}\overline{d_{6}}$%
{\small ,s}$_{{\tiny 123}}\overline{{\small s}_{{\tiny 123}}}$\\\hline%
\begin{tabular}
[c]{l}%
{\small 1+2+30q}$_{S}^{\ast}${\small (1111)}$\overline{q_{\Sigma}^{\ast
}(2641)}${\small =}$\pi${\small (2516)1}\\
\textbf{1+3+56}$\bullet$\textbf{q}$_{N}^{\ast}$\textbf{(931)}$\overline
{q_{\Delta}^{\ast}(3091)}$\textbf{=}$\pi$\textbf{(2567)1}%
\end{tabular}
& \ \ \ \ \ \ \ \ \ {\small ?} & {\small u}$_{3}\overline{d_{6}}$%
{\small ,u}$_{3}\overline{u_{3}}${\small ,d}$_{6}\overline{d_{6}}$\\\hline%
\begin{tabular}
[c]{l}%
\textbf{1+3+30}$\bullet$\textbf{q}$_{S}^{\ast}$\textbf{(1111)}$\overline
{q_{\Sigma}^{\ast}(2731)}$\textbf{=}$\pi$\textbf{(2598)1}\\
{\small 1+1+24}$\bullet$q$_{\Delta}^{\ast}${\small (2371)}$\overline
{q_{\Delta}^{\ast}(2371)}${\small =}$\pi${\small (2670)1}%
\end{tabular}
& \ \ \ \ \ \ \ \ \ {\small ?} & {\small u}$_{3}\overline{d_{6}}$%
{\small ,u}$_{3}\overline{u_{3}}${\small ,d}$_{6}\overline{d_{6}}$\\\hline%
\begin{tabular}
[c]{l}%
{\small 1+3+24}$\bullet$q$_{\Delta}^{\ast}${\small (1291)}$\overline
{q_{\Delta}^{\ast}(3091)}${\small =}$\eta${\small (2699)0}\\
\textbf{1+1+32}$\bullet$\textbf{q}$_{\Delta}^{\ast}$\textbf{(2371)}%
$\overline{q_{\Delta}^{\ast}(2371)}$\textbf{=}$\eta$\textbf{(2670)0}%
\end{tabular}
& \ \ \ \ \ \ \ \ \ {\small ?} & {\small u}$_{3}\overline{u_{3}}$%
{\small ,d}$_{6}\overline{d_{6}}${\small ,s}$_{{\tiny 123}}\overline
{{\small s}_{{\tiny 123}}}$\\\hline%
\begin{tabular}
[c]{l}%
\textbf{2+2+32}$\bullet$\textbf{q}$_{N}^{\ast}$\textbf{(2551)}$\overline
{q_{N}^{\ast}(2551)}$\textbf{=}$\eta$\textbf{(2928)0}%
\end{tabular}
& \ \ \ \ \ \ \ \ \ {\small ?} & {\small u}$_{3}\overline{u_{3}}$%
{\small ,d}$_{6}\overline{d_{6}}${\small ,s}$_{{\tiny 123}}\overline
{{\small s}_{{\tiny 123}}}$\\\hline%
\begin{tabular}
[c]{l}%
\textbf{1+1+24}$\bullet$\textbf{q}$_{\Sigma}^{\ast}$\textbf{(2371)}%
$\overline{q_{\Sigma}^{\ast}(2371)}$\textbf{=}$\eta$\textbf{(2963)0}%
\end{tabular}
& \ \ \ \ \ \ \ \ \ {\small ?} & {\small u}$_{3}\overline{u_{3}}$%
{\small ,d}$_{6}\overline{d_{6}}${\small ,s}$_{{\tiny 123}}\overline
{{\small s}_{{\tiny 123}}}$\\\hline%
\begin{tabular}
[c]{l}%
{\small 1+1+24}$\bullet${\small q}$_{S}^{\ast}${\small (2451)}$\overline
{q_{S}^{\ast}(2451)}${\small =}$\eta${\small (3037)0}\\
\textbf{3+3+32}$\bullet$\textbf{q}$_{N}^{\ast}$\textbf{(2641)}$\overline
{q_{N}^{\ast}(2641)}$\textbf{=}$\eta$\textbf{(3054)0}%
\end{tabular}
& \ \ \ \ \ \ \ \ \ {\small ?} & {\small u}$_{3}\overline{u_{3}}$%
{\small ,d}$_{6}\overline{d_{6}}${\small ,s}$_{{\tiny 123}}\overline
{{\small s}_{{\tiny 123}}}$\\\hline%
\begin{tabular}
[c]{l}%
\textbf{4+4+24}$\bullet$\textbf{q}$_{\Delta}^{\ast}$\textbf{(2731)}%
$\overline{q_{\Delta}^{\ast}(2731)}$\textbf{=}$\pi$\textbf{(3178)1}%
\end{tabular}
& \ \ \ \ \ \ \ \ \ {\small ?} & {\small u}$_{3}\overline{d_{6}}$%
{\small ,u}$_{3}\overline{u_{3}}${\small ,d}$_{6}\overline{d_{6}}$\\\hline%
\begin{tabular}
[c]{l}%
\textbf{4+4+32}$\bullet$\textbf{q}$_{\Delta}^{\ast}$\textbf{(2731)}%
$\overline{q_{\Delta}^{\ast}(2731)}$\textbf{=}$\eta$\textbf{(3178)0}\\
{\small 2+2+32}$\bullet q_{N}^{\ast}${\small (2731)}$\overline{q_{N}^{\ast
}(2731)}${\small =}$\eta${\small (3179)0}\\
{\small 3+3+24}$\bullet${\small q}$_{\Sigma}^{\ast}${\small (2551)}%
$\overline{q_{\Sigma}^{\ast}(2551)}${\small =}$\eta${\small (3252)0}\\
{\small 3+3+24}$\bullet${\small q}$_{S}^{\ast}${\small (2641)}$\overline
{q_{S}^{\ast}(2641)}${\small =}$\eta${\small (3339)0}%
\end{tabular}
&
\begin{tabular}
[c]{l}%
$\chi${\small (3250) 0}%
\end{tabular}
& {\small u}$_{3}\overline{u_{3}}${\small ,d}$_{6}\overline{d_{6}}$%
{\small ,s}$_{{\tiny 123}}\overline{{\small s}_{{\tiny 123}}}$\\\hline%
\begin{tabular}
[c]{l}%
\textbf{3+3+24}$\bullet$\textbf{q}$_{\Sigma}^{\ast}$\textbf{(2731)}%
$\overline{q_{\Sigma}^{\ast}(2731)}$\textbf{=}$\eta$\textbf{(3535)0}%
\end{tabular}
& \ \ \ \ \ \ \ \ \ {\small ?} & {\small u}$_{3}\overline{u_{3}}$%
{\small ,d}$_{6}\overline{d_{6}}${\small ,s}$_{{\tiny 123}}\overline
{{\small s}_{{\tiny 123}}}$\\\hline%
\begin{tabular}
[c]{l}%
\textbf{1+1+48}$\bullet$\textbf{q}$_{S}^{\ast}$\textbf{(4271)}$\overline
{q_{S}^{\ast}(4271)}$\textbf{=}$\eta$\textbf{(5926)0}%
\end{tabular}
& \ \ \ \ \ \ {\small \ \ \ \ ?} & \ \ \ \ \ \ \ \ \ \ ?\\\hline%
\begin{tabular}
[c]{l}%
\textbf{1+1+48}$\bullet\overline{q_{S}^{\ast}(10031)}$\textbf{q}$_{S}^{\ast}%
$\textbf{(10031)=}$\eta$\textbf{(17837)}%
\end{tabular}
& \ \ \ \ \ \ \ \ \ {\small ?} & \ \ \ \ \ \ \ \ \ \ ?\\\hline
\end{tabular}

\newpage

\qquad\ \ \ \ \ \ \ \ \ \ \ \ \ \ \ Table II. \ \ Strange Mesons (S = $\pm$1,C
= b = 0)

$%
\begin{tabular}
[c]{|l|l|l|}\hline
\ \ \ \ \ \ \ \ \ The BCC Quark Model & Experiment & {\small The Quark
Model}\\\hline
\ \ {\small [d}$_{q}${\small +d}$_{\overline{q}}${\small +O(I)]q}$_{i}^{\ast}%
${\small (m}$_{k}${\small )}$\overline{q_{j}^{\ast}(m_{l})}${\small =K(m)} &
\ \ \ K(m) & $\ \ \ \ \ \overline{s}$u, $\overline{s}$d\\\hline%
\begin{tabular}
[c]{l}%
\textbf{1+1+(72)}$\bullet$\textbf{q}$_{N}^{\ast}$\textbf{(931)}$\overline
{q_{S}^{\ast}(1111)}$\textbf{=K(494)}%
\end{tabular}
&
\begin{tabular}
[c]{l}%
$\bullet$\textbf{K(494)}%
\end{tabular}
& $\overline{\text{{\small S}}_{\text{{\small 123}}}}${\small u}$_{{\small 3}%
}${\small , }$\overline{\text{{\small S}}_{\text{{\small 123}}}}$%
{\small d}$_{{\small 6}}$\\\hline%
\begin{tabular}
[c]{l}%
{\small 1+1+(32)q}$_{N}^{\ast}${\small (1201)}$\overline{q_{S}^{\ast}(1111)}%
${\small =K(885)}\\
\textbf{1+1+(60)}$\bullet$\textbf{q}$_{N}^{\ast}$\textbf{(931)}$\overline
{q_{S}^{\ast}(1391)}$\textbf{=K(899)}\\
{\small 1+1+(54)}$\bullet${\small q}$_{N}^{\ast}${\small (931)}$\overline
{q_{\Sigma}^{\ast}(1201)}${\small =K(901)}%
\end{tabular}
&
\begin{tabular}
[c]{l}%
$\bullet$\textbf{K}$^{\ast}$\textbf{(892)}%
\end{tabular}
& $\overline{\text{{\small S}}_{\text{{\small 123}}}}${\small u}$_{{\small 3}%
}${\small , }$\overline{\text{{\small S}}_{\text{{\small 123}}}}$%
{\small d}$_{{\small 6}}$\\\hline%
\begin{tabular}
[c]{l}%
{\small 1+1+(32)q}$_{N}^{\ast}${\small (1471)}$\overline{q_{S}^{\ast}(1111)}%
${\small =K(1126)}\\
\textbf{1+3+(54)}$\bullet$\textbf{q}$_{N}^{\ast}$\textbf{(931)}$\overline
{q_{\Sigma}^{\ast}(1651)}$\textbf{=K(1310)}%
\end{tabular}
&
\begin{tabular}
[c]{l}%
$\bullet$\textbf{K}$_{1}$\textbf{(1270)}%
\end{tabular}
& $\overline{\text{{\small S}}_{\text{{\small 123}}}}${\small u}$_{{\small 3}%
}${\small , }$\overline{\text{{\small S}}_{\text{{\small 123}}}}$%
{\small d}$_{{\small 6}}$\\\hline%
\begin{tabular}
[c]{l}%
{\small 2+1+(32)}q$_{N}^{\ast}$(1831)$\overline{q_{S}^{\ast}(1111)}%
${\small =K(1447)}\\
\textbf{1+1+(54)}$\bullet$\textbf{q}$_{N}^{\ast}$\textbf{(931)}$\overline
{q_{S}^{\ast}(2011)}$\textbf{=K(1463)}%
\end{tabular}
&
\begin{tabular}
[c]{l}%
$\bullet${\small K}$_{1}${\small (1400)}\\
$\bullet${\small K}$^{\ast}${\small (1410)}\\
$\bullet${\small K}$_{0}^{\ast}${\small (1430)}\\
$\bullet${\small K}$_{2}^{\ast}${\small (1430)}\\
\ \ {\small K(1460)}\\
$\bullet\overline{\mathbf{K}}$\textbf{(1426)}%
\end{tabular}
& $\overline{\text{{\small S}}_{\text{{\small 123}}}}${\small u}$_{{\small 3}%
}${\small , }$\overline{\text{{\small S}}_{\text{{\small 123}}}}$%
{\small d}$_{{\small 6}}$\\\hline%
\begin{tabular}
[c]{l}%
{\small 4+1+(32)q}$_{N}^{\ast}${\small (1921)}$\overline{q_{S}^{\ast}(1111)}%
${\small =K(1528)}\\
\textbf{1+5+(54)}$\bullet$\textbf{q}$_{N}^{\ast}$\textbf{(931)}$\overline
{q_{\Sigma}^{\ast}(1921)}$\textbf{=K(1556)}%
\end{tabular}
&
\begin{tabular}
[c]{l}%
\ \textbf{\ K}$_{2}$\textbf{(1580)}%
\end{tabular}
& $\overline{\text{{\small S}}_{\text{{\small 123}}}}${\small u}$_{{\small 3}%
}${\small , }$\overline{\text{{\small S}}_{\text{{\small 123}}}}$%
{\small d}$_{{\small 6}}$\\\hline%
\begin{tabular}
[c]{l}%
\textbf{1+2+(54)}$\bullet$\textbf{q}$_{N}^{\ast}$\textbf{(931)}$\overline
{q_{\Sigma}^{\ast}(2011)}$\textbf{=K(1638)}%
\end{tabular}
&
\begin{tabular}
[c]{l}%
\ \ {\small K(1630)}\\
\ \ {\small K(1650)}\\
$\bullet${\small K}$^{\ast}${\small (1680)}\\
$\bullet\overline{\mathbf{K}}$\textbf{(1653)}%
\end{tabular}
& $\overline{\text{{\small S}}_{\text{{\small 123}}}}${\small u}$_{{\small 3}%
}${\small , }$\overline{\text{{\small S}}_{\text{{\small 123}}}}$%
{\small d}$_{{\small 6}}$\\\hline%
\begin{tabular}
[c]{l}%
\textbf{2+1+(32)}$\bullet$\textbf{q}$_{N}^{\ast}$\textbf{(2191)}%
$\overline{q_{S}^{\ast}(1111)}$\textbf{=K(1769)}%
\end{tabular}
&
\begin{tabular}
[c]{l}%
$\bullet${\small K}$_{2}${\small (1770)}\\
$\bullet${\small K(1780)}\\
$\bullet\overline{\mathbf{K}}$\textbf{(1775)}%
\end{tabular}
& $\overline{\text{{\small S}}_{\text{{\small 123}}}}${\small u}$_{{\small 3}%
}${\small , }$\overline{\text{{\small S}}_{\text{{\small 123}}}}$%
{\small d}$_{{\small 6}}$\\\hline%
\begin{tabular}
[c]{l}%
\textbf{1+1+(60)}$\bullet$\textbf{q}$_{N}^{\ast}$\textbf{(931)}$\overline
{q_{S}^{\ast}(2551)}$\textbf{=K(1804)}\\
{\small 1+1+(54)}$\bullet${\small q}$_{N}^{\ast}${\small (931)}$\overline
{q_{S}^{\ast}(2451)}${\small =K(1863)}%
\end{tabular}
&
\begin{tabular}
[c]{l}%
\ \ {\small K(1830)}\\
$\bullet\mathbf{K}_{2}$\textbf{(1820)}\\
$\bullet\overline{K}$(1825)
\end{tabular}
& $\overline{\text{{\small S}}_{\text{{\small 1125}}}}${\small u}%
$_{{\small 3}}${\small , }$\overline{\text{{\small S}}_{\text{{\small 1125}}}%
}${\small d}$_{{\small 6}}$\\\hline%
\begin{tabular}
[c]{l}%
\textbf{1+1+(54)}$\bullet$\textbf{q}$_{N}^{\ast}$\textbf{(931)}$\overline
{q_{\Sigma}^{\ast}(2371)}$\textbf{=K(1966)}%
\end{tabular}
&
\begin{tabular}
[c]{l}%
\ \ {\small K}$_{0}^{\ast}${\small (1950)}\\
\ \ {\small K}$_{2}^{\ast}${\small (1980)}\\
$\overline{\mathbf{K}}$\textbf{(1965)}%
\end{tabular}
& $\overline{\text{{\small S}}_{\text{{\small 123}}}}${\small u}$_{{\small 3}%
}${\small , }$\overline{\text{{\small S}}_{\text{{\small 123}}}}$%
{\small d}$_{{\small 6}}$\\\hline
\end{tabular}
$

\ \ \ \ \ \ \ \ \ \ \ \ \ \ \ \ \ \ \ \ \ \ \ \ %

\begin{tabular}
[c]{|l|l|l|}\hline%
\begin{tabular}
[c]{l}%
\textbf{1+3+(54)}$\bullet$\textbf{q}$_{N}^{\ast}$\textbf{(931)}$\overline
{q_{S}^{\ast}(2641)}$\textbf{=K(2036)}\\
{\small 2+1+(32)q}$_{N}^{\ast}${\small (2551)}$\overline{q_{S}^{\ast}(1111)}%
${\small =K(2090)}%
\end{tabular}
&
\begin{tabular}
[c]{l}%
$\bullet$\textbf{K(2045}%
\end{tabular}
& $\overline{\text{{\small S}}_{\text{{\small 123}}}}${\small u}$_{{\small 3}%
}${\small , }$\overline{\text{{\small S}}_{\text{{\small 123}}}}$%
{\small d}$_{{\small 6}}$\\\hline%
\begin{tabular}
[c]{l}%
\textbf{1+3+(54)}$\bullet$\textbf{q}$_{N}^{\ast}$\textbf{(931)}$\overline
{q_{\Sigma}^{\ast}(2551)}$\textbf{=K(2129)}\\
{\small 1+1+(54)}$\bullet${\small q}$_{N}^{\ast}${\small (931)}$\overline
{q_{S}^{\ast}(2731)}${\small =K(2118)}\\
{\small 3+1+(32)}$\bullet${\small q}$_{N}^{\ast}${\small (2641)}%
$\overline{q_{S}^{\ast}(1111)}${\small =K(2170)}%
\end{tabular}
& \ \ \ \ \ \ \ {\small ?} & $\overline{\text{{\small S}}_{\text{{\small 123}%
}}}${\small u}$_{{\small 3}}${\small , }$\overline{\text{{\small S}%
}_{\text{{\small 123}}}}${\small d}$_{{\small 6}}$\\\hline%
\begin{tabular}
[c]{l}%
\textbf{1+2+(54)}$\bullet$\textbf{q}$_{N}^{\ast}$\textbf{(931)}$\overline
{q_{\Sigma}^{\ast}(2641)}$\textbf{=K(2211)}\\
{\small 2+1+(32)}$\cdot${\small q}$_{N}^{\ast}${\small (2731)}$\overline
{q_{S}^{\ast}(1111)}${\small =K(2251)}%
\end{tabular}
& {\small
\begin{tabular}
[c]{l}%
\textbf{K}$_{2}$\textbf{(2250)}%
\end{tabular}
\ } & $\overline{\text{{\small S}}_{\text{{\small 123}}}}${\small u}%
$_{{\small 3}}${\small , }$\overline{\text{{\small S}}_{\text{{\small 123}}}}%
${\small d}$_{{\small 6}}$\\\hline%
\begin{tabular}
[c]{l}%
\textbf{1+3+(54)}$\bullet$\textbf{q}$_{N}^{\ast}$\textbf{(931)}$\overline
{q_{\Sigma}^{\ast}(2731)}$\textbf{=K(2293)}%
\end{tabular}
&
\begin{tabular}
[c]{l}%
{\small K}$_{3}${\small (2320)}\\
{\small K}$_{5}^{\ast}${\small (2380)}\\
$\overline{\mathbf{K}}$\textbf{(2350)}%
\end{tabular}
& $\overline{\text{{\small S}}_{\text{{\small 123}}}}${\small u}$_{{\small 3}%
}${\small , }$\overline{\text{{\small S}}_{\text{{\small 123}}}}$%
{\small d}$_{{\small 6}}$\\\hline%
\begin{tabular}
[c]{l}%
\textbf{1+5+(54)}$\bullet$\textbf{q}$_{N}^{\ast}$\textbf{(931)}$\overline
{q_{\Sigma}^{\ast}(3091)}$\textbf{=K(2621)}%
\end{tabular}
& {\small
\begin{tabular}
[c]{l}%
\textbf{K}$_{4}$\textbf{(2500)}%
\end{tabular}
\ \ } & $\overline{\text{{\small S}}_{\text{{\small 123}}}}${\small u}%
$_{{\small 3}}${\small , }$\overline{\text{{\small S}}_{\text{{\small 123}}}}%
${\small d}$_{{\small 6}}$\\\hline
{\small ...} & \ \ \ \ \ \ \ \ ? & $\overline{\text{{\small S}}%
_{\text{{\small 123}}}}${\small u}$_{{\small 3}}${\small , }$\overline
{\text{{\small S}}_{\text{{\small 123}}}}${\small d}$_{{\small 6}}$\\\hline%
\begin{tabular}
[c]{l}%
\textbf{1+1+(60)}$\bullet$\textbf{q}$_{N}^{\ast}$\textbf{(931)}$\overline
{q_{S}^{\ast}(4271)}$\textbf{=K(3597)}%
\end{tabular}
& \ \ \ \ \ \ \ \ ? & \ \ \ \ \ \ \ \ \ \ ?\\\hline%
\begin{tabular}
[c]{l}%
\textbf{1+1+(60)}$\bullet$\textbf{q}$_{N}^{\ast}$\textbf{(931)}$\overline
{q_{S}^{\ast}(10031)}$\textbf{=K(9429)}%
\end{tabular}
& \ \ \ \ \ \ \ \ ? & \ \ \ \ \ \ \ \ \ \ ?\\\hline
\ \ \ \ \ \ \ \ \ \ \ \ \ \ \ \ \ \ \ \ \ \ \ \ \ \ \ \ ... &
\ \ \ \ \ \ ... & \ \ \ \ \ \ \ \ ...\\\hline
\end{tabular}

\ \ \ \ \ \ \ \ \ \ \ \ \ \ \ \ \ \ \ \ \ \ \ \ \ \ \ \ \ \ \ \ \ \ \ \ \newpage

\qquad\ \ \ \ \ \ \ \ \ \ \ \ \ \ \ \ \ \ \ \ \ Table III. \ Charmed Mesons (C
= $\pm1)$%

\begin{tabular}
[c]{|l|l|l|}\hline
\ \ \ \ \ \ The BCC Quark Model & {\small Experiment} & {\small The Quark
Model}\\\hline
\ \ \ {\small (O(I)) q}$_{i}^{\ast}${\small (m}$_{k}${\small )}$\overline
{q_{j}^{\ast}(m_{l})}${\small =M(m)} & $\ \ \ \ ${\small M(m)} &
{\small \ \ \ \ \ \ \ \ c}$\overline{d}${\small , c}$\overline{u}$\\\hline%
\begin{tabular}
[c]{l}%
\textbf{(78)}$\bullet$\textbf{q}$_{C}^{\ast}$\textbf{(2271)}$\overline
{q_{N}^{\ast}{\small (931)}}$\textbf{=D(1866)}%
\end{tabular}
&
\begin{tabular}
[c]{l}%
$\bullet${\small D(1869)}%
\end{tabular}
& {\small c}$_{{\small 1250}}\overline{d_{6}}${\small , c}$_{{\small 1250}%
}\overline{u_{3}}$\\\hline%
\begin{tabular}
[c]{l}%
\textbf{(58)}$\bullet$\textbf{q}$_{C}$\textbf{(2441)}$\overline{q_{N}^{\ast
}{\small (931)}}$\textbf{=D(2029)}%
\end{tabular}
&
\begin{tabular}
[c]{l}%
$\bullet${\small D(2007)}$^{0}$\\
$\bullet${\small D(2010)}$^{\pm}$%
\end{tabular}
& {\small c}$_{{\small 1250}}\overline{d_{6}}${\small , c}$_{{\small 1250}%
}\overline{u_{3}}$\\\hline%
\begin{tabular}
[c]{l}%
{\small (38)}$\bullet${\small q}$_{C}^{\ast}${\small (2271)}$\overline
{q_{N}^{\ast}(1201)}${\small =D(2107)}\\
\textbf{(58)}$\bullet$\textbf{q}$_{C}^{\ast}$\textbf{(2531)}$\overline
{q_{N}^{\ast}{\small (931)}}$\textbf{=D(2115)}%
\end{tabular}
& \ \ \ \ \ \ \ \ \ {\small ?} & {\small c}$_{{\small 1250}}\overline{d_{6}}%
${\small , c}$_{{\small 1250}}\overline{u_{3}}$\\\hline%
\begin{tabular}
[c]{l}%
\textbf{(38)}$\bullet$\textbf{q}$_{C}^{\ast}$\textbf{(2271)}$\overline
{q_{N}^{\ast}\mathbf{(}1471\mathbf{)}}$\textbf{=D(2348)}%
\end{tabular}
&
\begin{tabular}
[c]{l}%
{\small \ D}$_{1}${\small (2420)}$^{\pm}$\\
$\bullet${\small D}$_{1}${\small (2420)}$^{0}$%
\end{tabular}
& {\small c}$_{{\small 1250}}\overline{d_{6}}${\small , c}$_{{\small 1250}%
}\overline{u_{3}}$\\\hline%
\begin{tabular}
[c]{l}%
\textbf{(58)}$\bullet$\textbf{q}$_{C}^{\ast}$\textbf{(2961)}$\overline
{q_{N}^{\ast}{\small (931)}}$\textbf{=D(2526)}\\
{\small (38)}$\bullet${\small q}$_{C}^{\ast}${\small (2271)}$\overline
{q_{N}^{\ast}(1831)}${\small =D(2670)}%
\end{tabular}
&
\begin{tabular}
[c]{l}%
$\bullet${\small D}$_{2}^{\ast}${\small (2460)}$^{\pm}$\\
$\bullet${\small D}$_{2}^{\ast}${\small (2460)}$^{0}$\\
{\small \ D}$^{\ast}${\small (2640)}$^{\pm}$\\
$\bullet\overline{\mathbf{D}}${\small (2507)}%
\end{tabular}
& {\small c}$_{{\small 1250}}\overline{d_{6}}${\small , c}$_{{\small 1250}%
}\overline{u_{3}}$\\\hline%
\begin{tabular}
[c]{l}%
\textbf{(38)}$\bullet$\textbf{q}$_{C}^{\ast}$\textbf{(2271)}$\overline
{q_{N}^{\ast}(1921)}$\textbf{=D(2750)}%
\end{tabular}
& \ \ \ \ \ \ \ \ \ \ \ ? & {\small c}$_{{\small 1250}}\overline{d_{6}}%
${\small , c}$_{{\small 1250}}\overline{u_{3}}$\\\hline%
\begin{tabular}
[c]{l}%
\textbf{(38)}$\bullet$\textbf{q}$_{C}^{\ast}$\textbf{(2271)}$\overline
{q_{N}(2191)}$\textbf{=D(2991)}%
\end{tabular}
& \ \ \ \ \ \ \ \ \ \ \ ? & {\small c}$_{{\small 1250}}\overline{d_{6}}%
${\small , c}$_{{\small 1250}}\overline{u_{3}}$\\\hline%
\begin{tabular}
[c]{l}%
\textbf{(68)}$\bullet$\textbf{q}$_{C}^{\ast}$\textbf{(6591)}$\overline
{q_{N}^{\ast}{\small (931)}}$\textbf{=D(5996)}%
\end{tabular}
\  & \ \ \ \ \ \ \ \ \ \ \ ? & {\small \ \ \ \ \ \ \ \ \ \ \ \ \ ?}\\\hline%
\begin{tabular}
[c]{l}%
\textbf{(68)}$\bullet$\textbf{q}$_{C}^{\ast}$\textbf{(13791)}$\overline
{q_{N}^{\ast}{\small (931)}}$\textbf{=D(13274)}%
\end{tabular}
\  & \ \ \ \ \ \ \ \ \ \ \ ? & {\small \ \ \ \ \ \ \ \ \ \ \ \ \ ?}\\\hline
&  & \\\hline
\ Charmed, Strange Mesons {\small (C= S= }$\pm${\small 1)} & {\small \ } &
{\small \ \ \ \ \ \ \ \ \ \ \ c}$\overline{s}$\\\hline%
\begin{tabular}
[c]{l}%
\textbf{(54)}$\bullet$\textbf{q}$_{C}^{\ast}$\textbf{(2271)}$\overline
{q_{S}^{\ast}(1111)}$\textbf{=D}$_{S}$\textbf{(2021)}%
\end{tabular}
&
\begin{tabular}
[c]{l}%
$\bullet$\textbf{D}$_{S}(1969)$%
\end{tabular}
$\ $ & {\small \ \ c(1250)}$\overline{S(123)}$\\\hline%
\begin{tabular}
[c]{l}%
{\small (34)}$\bullet${\small q}$_{C}${\small (2441)}$\overline{q_{S}^{\ast
}(1111)}${\small =D}$_{S}${\small (2034)}\\
{\small (34)}$\bullet${\small q}$_{C}${\small (2531)}$\overline{q_{S}^{\ast
}(1111)}${\small =D}$_{S}${\small (2120)}\\
\textbf{(42)}$\bullet$\textbf{q}$_{C}^{\ast}$\textbf{(2271)}$\overline{q}_{S}%
$\textbf{(1391)=D}$_{S}$\textbf{(2126)}%
\end{tabular}
&
\begin{tabular}
[c]{l}%
$\bullet$\textbf{D}$_{S}^{\ast}(2112)$%
\end{tabular}
$\ $ & {\small \ c(1250)}$\overline{S(123)}$\\\hline%
\begin{tabular}
[c]{l}%
\textbf{(34)}$\bullet$\textbf{q}$_{C}$\textbf{(2961)}$\overline{q_{S}^{\ast
}(1111)}$\textbf{=D}$_{S}$\textbf{(2531)}%
\end{tabular}
&
\begin{tabular}
[c]{l}%
$\bullet${\small D}$_{S1}(2536)^{\pm}$\\
$\bullet${\small D}$_{Sj}(2573)^{\pm}$\\
$\bullet\overline{\mathbf{D}}${\small (2555)}$^{\pm}$%
\end{tabular}
& {\small c(1250)}$\overline{S(123)}$\\\hline%
\begin{tabular}
[c]{l}%
{\small (36)}$\bullet${\small q}$_{C}^{\ast}${\small (2271)}$\overline
{q_{S}(2011)}${\small =D}$_{S}${\small (2690)}\\
\textbf{(42)}$\bullet$\textbf{q}$_{C}^{\ast}$\textbf{(2271)}$\overline
{q_{S}^{\ast}\mathbf{(2551)}}$\textbf{=D}$_{S}$\textbf{(3332)}%
\end{tabular}
& {\small \ \ \ \ \ \ \ \ \ \ \ \ \ ?} & {\small \ c(1250)}$\overline{S(123)}%
$\\\hline%
\begin{tabular}
[c]{l}%
\textbf{(44)}$\bullet$\textbf{q}$_{C}^{\ast}$\textbf{(6591)}$\overline
{q_{S}^{\ast}\mathbf{(1111)}}$\textbf{=D}$_{S}$\textbf{(6151)}%
\end{tabular}
\  & {\small \ \ \ \ \ \ \ \ \ \ \ \ \ ?} & {\small \ \ \ \ \ \ \ \ \ \ \ \ ?}%
\\\hline%
\begin{tabular}
[c]{l}%
\textbf{(32)}$\bullet$\textbf{q}$_{C}^{\ast}$\textbf{(6591)}$\overline
{q_{S}^{\ast}\mathbf{(2551)}}$\textbf{=D}$_{S}$\textbf{(7215)}%
\end{tabular}
\  & {\small \ \ \ \ \ \ \ \ \ \ \ \ \ ?} & {\small \ \ \ \ \ \ \ \ \ \ \ \ ?}%
\\\hline%
\begin{tabular}
[c]{l}%
\textbf{(44)}$\bullet$\textbf{q}$_{C}^{\ast}$\textbf{(13791)}$\overline
{q_{S}^{\ast}\mathbf{(1111)}}$\textbf{=D}$_{S}$\textbf{(13432)}%
\end{tabular}
\  & {\small \ \ \ \ \ \ \ \ \ \ \ \ \ ?} & {\small \ \ \ \ \ \ \ \ \ \ \ \ ?}%
\\\hline
\end{tabular}

\qquad\newpage

\ \ \ \ \ \ \ \ \ \ \ \ \ \ \ \ \ \ \ \ \ \ \ \ \ Table\ IV.\ \ Bottom Mesons
(b = $\pm$ 1)%

\begin{tabular}
[c]{|l|l|l|}\hline
{\small \ \ \ \ \ The BCC Quark Model} & {\small Experiment} &
\ \ \ \ {\small The Quark Model}\\\hline
{\small (O(I))q}$_{i}^{\ast}${\small (m}$_{k}${\small )}$\overline{q_{j}%
^{\ast}(m_{l})}${\small =M(m)} & {\small \ \ \ M(m)} &
{\small \ \ \ \ \ \ \ \ \ \ \ \ \ }$\overline{b}${\small u, }$\overline{b}%
${\small d}\\\hline
{\small (66)}$\bullet\overline{q_{b}^{\ast}(5531)}${\small q}$_{N}^{\ast}%
${\small (931)=B(5164)} &
\begin{tabular}
[c]{l}%
$\bullet${\small B(5279)}\\
$\bullet${\small B}$^{\ast}${\small (5325)}%
\end{tabular}
\  & $\overline{{\small b(4200)}}${\small u(3), }$\overline{{\small b(4200)}}%
${\small d(6)}\\\hline
{\small (26)}$\bullet\overline{q_{b}^{\ast}(5531)}${\small q}$_{N}^{\ast}%
${\small (1201)=B(5655)} & \ \ {\small B(5732)} & $\overline{{\small b(4200)}%
}${\small u(3), }$\overline{{\small b(4200)}}${\small d(6)}\\\hline
{\small (26)}$\bullet\overline{q_{b}^{\ast}(5531)}${\small q}$_{N}^{\ast}%
${\small (1471)=B(5896)} & {\small \ \ \ \ \ \ \ \ \ \ ?} & $\overline
{{\small b(4200)}}${\small u(3), }$\overline{{\small b(4200)}}${\small d(6)}%
\\\hline
{\small (26)}$\bullet\overline{q_{b}^{\ast}(5531)}${\small q}$_{N}^{\ast}%
${\small (1831)=B(6217)} & {\small \ \ \ \ \ \ \ \ \ \ ?} & $\overline
{{\small b(4200)}}${\small u(3), }$\overline{{\small b(4200)}}${\small d(6)}%
\\\hline
{\small (26)}$\bullet\overline{q_{b}^{\ast}(5531)}${\small q}$_{N}^{\ast}%
${\small (1930)=B(6298)} & {\small \ \ \ \ \ \ \ \ \ \ ?} & $\overline
{{\small b(4200)}}${\small u(3), }$\overline{{\small b(4200)}}${\small d(6)}%
\\\hline
{\small (60)}$\bullet\overline{q_{b}^{\ast}(9951)}${\small q}$_{N}^{\ast}%
${\small (931)=B(9504)} & {\small \ \ \ \ \ \ \ \ \ \ ?} &
{\small \ \ \ \ \ \ \ \ \ \ \ \ \ \ \ \ \ \ ?}\\\hline
{\small (60)}$\bullet\overline{q_{b}^{\ast}(15811)}${\small q}$_{N}^{\ast}%
${\small (931)=B(15385)} & {\small \ \ \ \ \ \ \ \ \ \ ?} &
{\small \ \ \ \ \ \ \ \ \ \ \ \ \ \ \ \ \ \ ?}\\\hline
&  & \\\hline
{\small \ \ \ \ \ \ Botrtom, Strange Mesons} & {\small (b = S = }$\pm
${\small 1)} & \ \ \ \ \ \ \ \ \ \ \ \ \ \ \ b$\overline{s}$\\\hline
{\small (42)}$\bullet\overline{{\small q}_{b}^{\ast}{\small (5531)}}%
${\small q}$_{S}^{\ast}${\small (1111)=B}$_{S}${\small (5319)} & $\bullet
$B$_{S}(5369)$ & {\small \ \ \ \ \ \ \ }$\overline{{\small b(4200)}}%
${\small s(123)}\\\hline
{\small (30)}$\bullet\overline{{\small q}_{b}^{\ast}{\small (5531)}}%
${\small q}$_{S}^{\ast}${\small (1391)=B}$_{S}${\small (5674)} &
\ B$_{SJ}(5850)$ & {\small \ \ \ \ \ \ \ }$\overline{{\small b(4200)}}%
${\small s(123)}\\\hline
{\small (24)}$\bullet\overline{{\small q}_{b}^{\ast}{\small (5531}}%
${\small )q}$_{S}^{\ast}${\small (2011)=B}$_{S}${\small (6238)} &
\ \ \ \ \ \ \ \ ? & {\small \ \ \ \ \ \ \ }$\overline{{\small b(4200)}}%
${\small s(123)}\\\hline
{\small (24)}$\bullet\overline{{\small q}_{b}^{\ast}{\small (5531)}}%
${\small q}$_{S}^{\ast}${\small (2451)=B}$_{S}${\small (6638)} &
\ \ \ \ \ \ \ \ ? & {\small \ \ \ \ \ \ \ }$\overline{{\small b(4200)}}%
${\small s(123)}\\\hline
{\small (30)}$\bullet\overline{{\small q}_{b}^{\ast}{\small (5531)}}$%
q$_{S}^{\ast}$({\small 2551}){\small =B}$_{S}${\small (6629)} &
\ \ \ \ \ \ \ \ ? & {\small \ \ \ \ \ \ \ }$\overline{{\small b(4200)}}%
${\small s(123)}\\\hline
{\small (36)}$\bullet\overline{{\small q}_{b}^{\ast}{\small (9951)}}$%
q$_{S}^{\ast}$({\small 1111}){\small =B}$_{S}${\small (9659)} &
\ \ \ \ \ \ \ \ ? & \ \ \ \ \ \ \ \ \ \ \ \ \ \ \ ?\\\hline
{\small (24)}$\bullet\overline{{\small q}_{b}^{\ast}{\small (9951)}}$%
q$_{S}^{\ast}$({\small 1391}){\small =B}$_{S}${\small (9994)} &
\ \ \ \ \ \ \ \ ? & \ \ \ \ \ \ \ \ \ \ \ \ \ \ \ ?\\\hline
{\small (24)}$\bullet\overline{{\small q}_{b}^{\ast}{\small (9951)}}$%
q$_{S}^{\ast}$({\small 2551}){\small =B}$_{S}${\small (10866)} &
\ \ \ \ \ \ \ \ ? & \ \ \ \ \ \ \ \ \ \ \ \ \ \ \ ?\\\hline
{\small (36)}$\bullet\overline{{\small q}_{b}^{\ast}{\small (15811)}}$%
q$_{S}^{\ast}$({\small 1111}){\small =B}$_{S}${\small (15540)} &
\ \ \ \ \ \ \ \ ? & \ \ \ \ \ \ \ \ \ \ \ \ \ \ \ ?\\\hline
{\small (24)}$\bullet\overline{{\small q}_{b}^{\ast}{\small (15811)}}$%
q$_{S}^{\ast}$({\small 2551}){\small =B}$_{S}${\small (16774)} &
\ \ \ \ \ \ \ \ ? & \ \ \ \ \ \ \ \ \ \ \ \ \ \ \ ?\\\hline
&  & \\\hline
\ \ \ \ \ \ \ \ {\small Bottom, Charmed Mesons} & {\small (b= }$\mp
${\small 1,C =}$\pm${\small 1)} & $\ \ \ \ \ \ \ \ \ \ \ \ \ \ \overline{b}%
$c\\\hline
{\small (48)}$\overline{q_{b}^{\ast}(5531)}${\small q}$_{C}^{\ast}%
${\small (2271)=B}$_{C}${\small (6691)} & \ $\bullet${\small B}$_{C}%
${\small (6400)} & $\ \ \ \ \ \overline{{\small b(4200)}}${\small c(1250)}%
\\\hline
{\small (38)}$\overline{q_{b}^{\ast}(5531)}${\small q}$_{C}^{\ast}%
${\small (6591)=B}$_{C}${\small (10822)} &  & \\\hline
{\small (42)}$\overline{q_{b}^{\ast}(9951)}${\small q}$_{C}^{\ast}%
${\small (2271)=B}$_{C}${\small (11031)} & \ \ \ \ \ \ \ \ ? &
\ \ \ \ \ \ \ \ \ \ \ \ \ \ \ ?\\\hline
{\small (32)}$\overline{q_{b}^{\ast}(9951)}${\small q}$_{C}^{\ast}%
${\small (6591)=B}$_{C}${\small (15012)} & \ \ \ \ \ \ \ \ ? &
\ \ \ \ \ \ \ \ \ \ \ \ \ \ \ ?\\\hline
{\small (42)}$\overline{q_{b}^{\ast}(15811)}${\small q}$_{C}^{\ast}%
${\small (2271)=B}$_{C}${\small (16912)} & \ \ \ \ \ \ \ \ ? &
\ \ \ \ \ \ \ \ \ \ \ \ \ \ \ ?\\\hline
{\small (38)}$\overline{q_{b}^{\ast}(5531)}${\small q}$_{C}^{\ast}%
${\small (13791)=B}$_{C}${\small (18102)} & \ \ \ \ \ \ \ \ ? &
\ \ \ \ \ \ \ \ \ \ \ \ \ \ \ ?\\\hline
\end{tabular}

\newpage

\ \ \ \ \ \ \ \ \ \ \ \ \ \ \ \ \ \ \ \ Table\ V. The Slightly Heavy\ Mesons
(S = C = b = I = 0)%

\begin{tabular}
[c]{|l|l|l|}\hline
\ \ \ \ \ \ \ The BCC Quark Model & Experiment & {\small Quark Model}\\\hline
\ \ \ {\small [d}$_{q}${\small +d}$_{\overline{q}}${\small +O(I)]q}$_{i}%
^{\ast}${\small (m}$_{k}${\small )}$\overline{q_{j}^{\ast}(m_{l})}%
${\small =M(m)} & \ \ \ M(m)$_{\Gamma}$ & \ \ \ c$\overline{c}$\\\hline%
\begin{tabular}
[c]{l}%
\textbf{\lbrack1+1+48]}$\bullet$\textbf{q}$_{S}^{\ast}$\textbf{(2551)}%
$\overline{q_{S}^{\ast}(2551)}$\textbf{=}$\eta$\textbf{(2902)}%
\end{tabular}
& $\bullet\eta_{C}${\small (2980)}$_{13Mev}$ & {\small c}$_{{\small 1250}%
}\overline{c_{1250}}$\\\hline%
\begin{tabular}
[c]{l}%
\textbf{\lbrack{\small 1+1+120}]}$\bullet$\textbf{q}$_{C}^{\ast}%
$\textbf{(2271)}$\overline{\text{q}_{C}^{\ast}\text{(2271)}}$\textbf{=J/}%
$\Psi$\textbf{(3044)}%
\end{tabular}
& $\bullet${\small J/}$\Psi${\small (3097)}$_{87Kev}$ & {\small c}%
$_{{\small 1250}}\overline{c_{1250}}$\\\hline%
\begin{tabular}
[c]{l}%
\textbf{\lbrack2+2+24]}$\bullet$\textbf{q}$_{\Sigma}$\textbf{(2641)}%
$\overline{q_{\Sigma}^{\ast}(2641)}$\textbf{=}$\eta$\textbf{(3394)}\\
{\small \lbrack1+1+24]}$\bullet${\small q}$_{S}${\small (2731)}$\overline
{q_{S}^{\ast}(2731)}${\small =}$\eta${\small (3480)}%
\end{tabular}
& $\chi_{c0}${\small (3415)}$_{15Mev}$ & {\small c}$_{{\small 1250}}%
\overline{c_{1250}}$\\\hline%
\begin{tabular}
[c]{l}%
\textbf{\lbrack3+3+24]}$\bullet$\textbf{q}$_{\Sigma}$\textbf{(2731)}%
$\overline{q_{\Sigma}^{\ast}(2731)}$\textbf{=}$\eta$\textbf{(3535)}%
\end{tabular}
{\small \ \ } & $\bullet\chi_{c1}${\small (3511)}$_{.88Mev}$ & {\small c}%
$_{{\small 1250}}\overline{c_{1250}}$\\\hline%
\begin{tabular}
[c]{l}%
\textbf{\lbrack1+1+40]}$\bullet$\textbf{q}$_{C}^{\ast}$\textbf{(2441)}%
$\overline{q_{C}^{\ast}(2441)}$\textbf{=}$\psi$\textbf{(3568)}%
\end{tabular}
& $\bullet\chi_{c2}${\small (3556)}$_{2.0Mev}$ & {\small c}$_{{\small 1250}%
}\overline{c_{1250}}$\\\hline%
\begin{tabular}
[c]{l}%
{\small \lbrack1+1+60]}$\bullet${\small q}$_{C}^{\ast}${\small (2271)}%
$\overline{q_{C}^{\ast}(2441)}${\small =}$\psi${\small (3607)}\\
\textbf{\lbrack1+1+60]}$\bullet$\textbf{q}$_{C}^{\ast}$\textbf{(2271)}%
$\overline{q_{C}^{\ast}(2531)}$\textbf{=}$\psi$\textbf{(3693)}%
\end{tabular}
& $\bullet\psi${\small (3686)}$_{277Kev}$ & {\small c}$_{{\small 1250}%
}\overline{c_{1250}}$\\\hline%
\begin{tabular}
[c]{l}%
\textbf{\lbrack1+1+40]}$\bullet$\textbf{q}$_{C}^{\ast}$\textbf{(2531)}%
$\overline{q_{C}^{\ast}(2531)}$\textbf{=}$\psi$\textbf{(3740)}%
\end{tabular}
\  & $\bullet\psi${\small (3770)}$_{24Mev}$ & {\small c}$_{{\small 1250}%
}\overline{c_{1250}}$\\\hline%
\begin{tabular}
[c]{l}%
\textbf{\lbrack1+1+30]}$\bullet$\textbf{q}$_{C}^{\ast}$\textbf{(2441)}%
$\overline{q_{C}^{\ast}(2531)}$\textbf{=}$\psi$\textbf{(3854)}%
\end{tabular}
& $\ \psi${\small (3836)} & {\small c}$_{{\small 1250}}\overline{c_{1250}}%
$\\\hline%
\begin{tabular}
[c]{l}%
\textbf{\lbrack1+1+54]}$\bullet$\textbf{q}$_{S}^{\ast}$\textbf{(1111)}%
$\overline{q_{S}^{\ast}(4271)}$\textbf{=}$\eta$\textbf{(3952)}%
\end{tabular}
{\small \ \ } & $\bullet\psi${\small (4040)}$_{52Mev}$ & {\small c}%
$_{{\small 1250}}\overline{c_{1250}}$\\\hline%
\begin{tabular}
[c]{l}%
\textbf{\lbrack1+1+60]}$\bullet$\textbf{q}$_{C}^{\ast}$\textbf{(2271)}%
$\overline{q_{C}^{\ast}(2961)}$\textbf{=}$\psi$\textbf{(4104)}\\
{\small \lbrack1+1+36]}$\bullet${\small q}$_{S}^{\ast}${\small (1391)}%
$\overline{q_{S}^{\ast}(4271)}${\small =}$\eta${\small (4105)}\\
{\small \lbrack1+1+30]}$\bullet${\small q}$_{C}^{\ast}${\small (2441)}%
$\overline{q_{C}^{\ast}(2961)}${\small =}$\psi${\small (4264)}%
\end{tabular}
& $\bullet\psi${\small (4160)}$_{78Mev}$ & {\small c}$_{{\small 1250}%
}\overline{c_{1250}}$\\\hline%
\begin{tabular}
[c]{l}%
{\small \lbrack1+1+30]}$\bullet${\small q}$_{C}^{\ast}${\small (2531)}%
$\overline{q_{C}^{\ast}(2961)}${\small =}$\psi${\small (4349)}\\
\textbf{\lbrack1+1+40]}$\bullet$\textbf{q}$_{C}^{\ast}$\textbf{(2961)}%
$\overline{q_{C}^{\ast}(2961)}$\textbf{=}$\psi$\textbf{(4554)}\\
{\small \lbrack1+1+27]}$\bullet${\small q}$_{S}${\small (2011)}$\overline
{q_{S}^{\ast}(4271)}${\small =}$\eta${\small (4553)}%
\end{tabular}
& $\bullet\psi${\small (4415)}$_{43Mev}$ & {\small c}$_{{\small 1250}%
}\overline{c_{1250}}$\\\hline%
\begin{tabular}
[c]{l}%
\textbf{\lbrack1+1+36]}$\bullet$\textbf{q}$_{S}^{\ast}$\textbf{(2551)}%
$\overline{q_{S}^{\ast}(4271)}$\textbf{=}$\eta$\textbf{(4944)}%
\end{tabular}
& \ \ \ \ \ \ \ \ \ \ ? & {\small c}$_{{\small 1250}}\overline{c_{1250}}%
$\\\hline%
\begin{tabular}
[c]{l}%
\textbf{\lbrack1+1+75]}$\overline{q_{C}^{\ast}(6591)}$\textbf{q}$_{C}^{\ast}%
$\textbf{(2271)=}$\psi$\textbf{(7374)}%
\end{tabular}
& \ \ \ \ \ \ \ \ \ \ ? & \ \ \ \ \ \ \ ?\\\hline
\textbf{%
\begin{tabular}
[c]{l}%
\textbf{\lbrack1+1+75]}$\overline{q_{C}^{\ast}(13791)}$\textbf{q}$_{C}^{\ast}%
$\textbf{(2271)=}$\psi$\textbf{(14654)}%
\end{tabular}
} & \ \ \ \ \ \ \ \ \ \ ? & \ \ \ \ \ \ \ ?\\\hline
\textbf{%
\begin{tabular}
[c]{l}%
\textbf{\lbrack1+1+80]}$\overline{q_{C}^{\ast}(13791)}$\textbf{q}$_{C}^{\ast}%
$\textbf{(13791)=}$\psi$\textbf{(25596)}%
\end{tabular}
} & \ \ \ \ \ \ \ \ \ \ ? & \ \ \ \ \ \ \ ?\\\hline
\ \ \ \ \ \ \ \ \ \ \ \ \ \ \ \ \ \ \ \ \ \ \ \ \ \ \ \ \ \ \ ... &
\ \ \ \ \ \ \ \ \ \ ... & \ \ \ \ \ \ \ ...\\\hline
\end{tabular}

\bigskip

\ \ \ \ \ \ \ \ \ \ \ \ \ \ \ \ \ \ \ \ \ \ \ \ \ \ \ \ \ \ \ \ \ \ \ \ \ \ \ \ \ \ \ \ \ \ \ \ \ \ \ \ \ 

\begin{center}
\end{center}

\bigskip\newpage

\begin{center}
\end{center}

\ \ \ \ \ \ \ \ \ \ \ \ \ \ \ \ \ \ \ \ \ Table\ VI. The Heavy\ Mesons (S = C
= b = I = 0)%

\begin{tabular}
[c]{|l|l|l|}\hline
\ \ \ The BCC Quark Model & \ \ Experiment $\ $ & {\small The Quark
Model}\\\hline
\ \ {\small (O(I))q}$_{i}^{\ast}${\small (m}$_{k}${\small )}$\overline
{q_{j}^{\ast}(m_{l})}${\small =M(m)} & \ \ \ \ \ \ M(m) $\Gamma$ &
\ \ \ \ \ \ \ \ b$\overline{b}$\\\hline
{\small (72)}$\bullet$q$_{b}^{\ast}$(5531)$\overline{q_{b}^{\ast
}({\small 5531})}${\small =}$\Upsilon(9489)$ &
\begin{tabular}
[c]{l}%
{\small $\Upsilon$}${\small (1S)(9460)}${\small 53kev}%
\end{tabular}
& b{\small (4200)}$\overline{b\text{{\small (4200)}}}$\\\hline
{\small (54)}$\bullet$q$_{S}^{\ast}$(1111)$\overline{q_{S}^{\ast
}{\small (10031)}}${\small = }$\eta(${\small 9734)} &
\begin{tabular}
[c]{l}%
${\small \chi}_{b0}{\small (1P)(9860)}$\\
${\small \chi}_{b1}{\small (1P)(9893)}$\\
${\small \chi}_{b2}{\small (1P)(9913)}$%
\end{tabular}
& b{\small (4200)}$\overline{b\text{{\small (4200)}}}$\\\hline
{\small (36)}$\bullet$q$_{S}^{\ast}$(1391)$\overline{q_{S}^{\ast
}{\small (10031)}}${\small = }$\eta(${\small 9955)} &
\begin{tabular}
[c]{l}%
{\small $\Upsilon$}${\small (2S)(10023)}${\small 44kev}%
\end{tabular}
& b{\small (4200)}$\overline{b\text{{\small (4200)}}}$\\\hline
{\small (27)}$\bullet${\small q}$_{S}${\small (2011)}$\overline{q_{S}^{\ast
}{\small (10031)}}${\small =}$\eta${\small (10443)} &
\begin{tabular}
[c]{l}%
${\small \chi}_{b0}{\small (2P)(10232)}$\\
${\small \chi}_{b1}{\small (2P)(10255)}$\\
${\small \chi}_{b2}{\small (2P)(10269)}$%
\end{tabular}
& b{\small (4200)}$\overline{b\text{{\small (4200)}}}$\\\hline
{\small (80)}$\bullet${\small q}$_{C}^{\ast}${\small (6591)q}$_{C}^{\ast}%
${\small (6591)=}$\psi${\small (10792)} &
\begin{tabular}
[c]{l}%
{\small $\Upsilon$}${\small (3S)(10355)}${\small 26kev}%
\end{tabular}
& b{\small (4200)}$\overline{b\text{{\small (4200)}}}$\\\hline
{\small (27)}$\bullet${\small q}$_{S}${\small (2451)}$\overline{q_{S}^{\ast
}{\small (10031)}}${\small =}$\eta${\small (10791)} &
\begin{tabular}
[c]{l}%
{\small $\Upsilon$}${\small (4S)(10580)}${\small 14Mev}%
\end{tabular}
& b{\small (4200)}$\overline{b\text{{\small (4200)}}}$\\\hline
{\small (36)}$\bullet${\small q}$_{S}^{\ast}${\small (2551)}$\overline
{q_{S}^{\ast}{\small (10031)}}${\small =}$\eta${\small (10870)} &
\begin{tabular}
[c]{l}%
{\small $\Upsilon$}${\small (10860)}${\small 110 Mev}%
\end{tabular}
& b{\small (4200)}$\overline{b\text{{\small (4200)}}}$\\\hline
{\small (27)}$\bullet${\small q}$_{S}${\small (2641)}$\overline{q_{S}^{\ast
}{\small (10031)}}${\small =}$\eta${\small (10941)} &
\ \ \ \ \ \ \ \ \ \ \ \ \ \ ? & b{\small (4200)}$\overline
{b\text{{\small (4200)}}}$\\\hline
{\small (27)}$\bullet${\small q}$_{S}${\small (2731)}$\overline{q_{S}^{\ast
}{\small (10031)}}${\small =}$\eta${\small (11012)} &
\begin{tabular}
[c]{l}%
{\small $\Upsilon$}${\small (11020)}${\small 79 Mev}%
\end{tabular}
& b{\small (4200)}$\overline{b\text{{\small (4200)}}}$\\\hline
{\small (45)}$\bullet${\small q}$_{b}^{\ast}${\small (5531)}$\overline
{q_{b}^{\ast}({\small 9951})}${\small =}$\Upsilon${\small (14329)} &
\ \ \ \ \ \ \ \ \ \ \ \ \ \ ? & \ \ \ \ \ \ \ \ \ \ ?\\\hline
{\small (48)}$\bullet${\small q}$_{b}^{\ast}${\small (9951)}$\overline
{q_{b}^{\ast}({\small 9951})}${\small =}$\Upsilon${\small (17805)} &
\ \ \ \ \ \ \ \ \ \ \ \ \ \ ? & \ \ \ \ \ \ \ \ \ \ ?\\\hline
{\small (45)}$\bullet${\small q}$_{b}^{\ast}${\small (9951)}$\overline
{q_{b}^{\ast}({\small 15811})}${\small =}$\Upsilon${\small (20210)} &
\ \ \ \ \ \ \ \ \ \ \ \ \ \ ? & \ \ \ \ \ \ \ \ \ \ ?\\\hline
{\small (48)}$\bullet${\small q}$_{b}^{\ast}${\small (15811)}$\overline
{q_{b}^{\ast}({\small 15811})}${\small =}$\Upsilon${\small (29597)} &
\ \ \ \ \ \ \ \ \ \ \ \ \ \ ? & \ \ \ \ \ \ \ \ \ \ ?\\\hline
\ \ \ \ \ \ \ \ \ \ \ \ \ \ \ \ \ \ \ \ ... & \ \ \ \ \ \ \ \ \ \ \ \ ... &
\ \ \ \ \ \ \ \ ...\\\hline
\end{tabular}

\newpage

\qquad\qquad\qquad\qquad\qquad\qquad{\LARGE Appendix III }

\qquad\qquad\qquad\qquad\ \ \ \ \ {\Large The High Isospin Mesons}

The BCC Quark Model predicts many high isospin mesons (I=1(S, or C, or b
$\neq0)$, I = 3/2, I = 2, I = 5/2, and I = 3). They are not predicted by the
Quark Model. They have not been discovered by experiment yet (except a meson
$\chi(1600)$ with I = 2).

First, we study the high isospin mesons with O(Meson)
$<$%
24.\ From (\ref{O(QiQj)-MAX}) and (\ref{O(M(I))}), we
have{\tiny \ \ \ \ \ \ \ \ \ \ \ \ \ \ }\ \ 

1). $\overline{q_{\Sigma}^{\ast}(m_{i})}q_{\Delta}^{\ast}(m_{j}),$\ O(K,
I=1/2) =17, O(F, I=3/2) =12, O(S, I=5/2) =7.

2). q$_{\Delta}^{\ast}(m_{i})\overline{q_{\Delta}^{\ast}(m_{i})},$ O($\eta$,
I=0) =32, O($\pi,$I=1) =24, O(T, I=2) =16, O(W, I=3) =8.

3). q$_{\Sigma}(m_{i})\overline{q_{\Sigma}^{\ast}(m_{i})},$ O($\eta$, I=0)=24,
O($\pi,$I=1)=16, O(T, I=2) =8.

4). $\overline{q_{b}^{\ast}(5531)}q_{\Sigma}^{\ast}(m),$ O($\pi_{b}$, I=1) = 16.

5). $\overline{q_{S}^{\ast}(1111)}q_{\Delta}^{\ast}(m),$ O(F, I=3/2) = 20.\ 

6). $\overline{q_{b}^{\ast}(5531)}q_{\Delta}^{\ast}(m),$ O(F$_{b}$, I=3/2) = 16.25.

7). q$_{C}^{\ast}(2271)\overline{q_{\Delta}^{\ast}(m)}\rightarrow$ F$_{C}%
$(I=3/2), O(F$_{C}$) = 23.75.

Above, the mesons with I $\geq$ 3/2 all have O(Meson)
$<$%
24. Thus, they cannot be observed now (\ref{O(I)<24}). Especially, the mesons
(W) with I = 3 (from q$_{\Delta}^{\ast}\overline{q_{\Delta}^{\ast}}$) have
O(W, I=3) = 8 (%
$<$%
$<$%
24). Thus, they cannot be observed now. The mesons with I = 5/2 (from
q$_{\Delta}^{\ast}\overline{q_{\Sigma}^{\ast}}$) have O(S, I=5/2) = 7 (%
$<$%
$<$%
24). Thus, they cannot be observed either.

Second, we study the high isospin mesons with O(Meson) $\geq$ 24. From
(\ref{O(QiQj)-MAX}) and (\ref{O(M(I))}), we
have{\tiny \ \ \ \ \ \ \ \ \ \ \ \ \ \ }\ \ 

1). For q$_{N}^{\ast}$(931)$\overline{q_{\Sigma}^{\ast}(m)}\rightarrow$
K(I=1/2), O(K) = 47; F(I=3/2), O(F) = 28.

2). For q$_{C}^{\ast}(2271)\overline{q_{\Sigma}^{\ast}(m)}\rightarrow$
$\pi_{C}$(I=1), O($\pi_{C}$) = 24.

3). For q$_{N}^{\ast}$(931)$\overline{q_{\Delta}^{\ast}(m)}\rightarrow\pi$(I =
1), O($\pi$) = 56; T(I=2), O(T) = 37.$\qquad\qquad\ \ \ \ \ \ \ \ \ $\ \ \ \ \ \ \ \ \ \ \ \ \ \ \ 

The mesons composed of $\overline{q_{N}^{\ast}(931)}q_{\Sigma}^{\ast}$(m), F(I
= 3/2), have O(q$_{i}^{\ast}\overline{q_{j}^{\ast}}$)= 28 (%
$>$%
24). Thus they should be discovered by today's experiments (\ref{O(I)<24}).
However, there are four members (I$_{z}$= 3/2, Q=1; I$_{z}$=1/2, Q=0; I$_{z}$=
-1/2, Q= -1; I$_{z}$= -3/2, Q= -2$)$ \ in the mesons. The member with Q = -2
will be composed of $\overline{q_{N}^{\ast}(931)^{+2/3}}$(Q=-2/3) and
$q_{\Sigma}^{\ast}$(Q=-4/3). The quark and the antiquark with the same kind of
electric charge will repel each other four times more strongly (2/3$\times
4/3=8/9)$ than the mesons with Q = 1 (1/3$\times2/3=2/9)$ exhibit. Therefore,
the member (Q = -2) will have (such as 0.75$\times$28 = 21) O(q$_{i}^{\ast
}\overline{q_{j}^{\ast}}$)
$<$%
24. They are difficult to find. Similarly, the mesons with I =2 (from
q$_{N}^{\ast}$(931)$\overline{q_{\Delta}^{\ast}(m)})$ have O(q$_{i}^{\ast
}\overline{q_{j}^{\ast}}$)= 37 (%
$>$%
24). They should be discovered by today's experiments (\ref{O(I)<24}).
However, there are five members (I$_{z}$=2, Q=2; I$_{z}$=1, Q=1; I$_{z}$=0,
Q=0; I$_{z}$=-1, Q=-1; I$_{z}$=-2, Q=-2) in the meson's family. Similarly to
the I=3/2 case, the members with I$_{z}$= $\pm$2, Q = $\pm$2 will be very
difficult to find. After taking off the members with Q=$\pm2$, the mesons
change into new mesons [I=1; Q= 1, 0, -1]. These new mesons may be observed as
the mesons K, D and $\pi$ as shown in the following:%

\begin{equation}%
\begin{tabular}
[c]{|l|l|l|}\hline
{\small Meson (before taking off member(Q=}$\pm${\small 2))} & {\small Meson
(after...)} & {\small Meson (may be)}\\\hline
{\small q}$_{N}^{\ast}${\small (931)}$\overline{q_{\Sigma}^{\ast
}{\small (1201)}}${\small =F(901)[3/2;2,1,0,-1]} & {\small K(901)[1;1,0,-1]} &
{\small K}$^{\pm}${\small (901),K}$^{0}${\small (901)}\\\hline
{\small q}$_{C}^{\ast}{\small (2271)}\overline{\text{q}_{\Sigma}^{\ast
}\text{{\small (1201)}}}$=$\pi_{C}${\small (2228)[1; 2,1,0]} &
{\small D(2228)[1;1,0]} & {\small (24)D(2228)}\\\hline
{\small q}$_{N}^{\ast}${\small (931)}$\overline{q_{\Delta}^{\ast}(1291)}%
${\small =T(960)[2;2,1,0,-1,-2]} & $\pi${\small (960)[1;1,0,-1]} & $\pi
${\small (960)[1;1,0,-1]}\\\hline
\end{tabular}
\end{equation}

Although the mesons with I = 2 are very difficut to find, our great
experimental physicists have already discovered one--$\chi(1600)$ [I$^{G}%
$(J$^{PC}$) = 2$^{+}$(2$^{++}$)] \cite{Quark Mass-2000}. If the result can be
confirmed, it will be a great discovery in physics. It will clearly show that
the Quark Model (6 elementary quarks and SU(N) symmetries) needs modification
and that the BCC Quark Model is a good modification since it predicts the
meson (I=2) $\chi$(1600) [q$_{N}^{\ast}$(931)$\overline{q_{\Delta}^{\ast
}(2011)}$ = $\chi$(1603)].

\end{document}